\documentclass{emulateapj}
\usepackage{amsmath}
\usepackage{amssymb}
\usepackage{epsfig}
\usepackage{subfigure}
\usepackage{natbib}
\usepackage{graphicx}
\usepackage{color}

\shorttitle{Untargeted Study of SN~Ibc Host Galaxies}
\shortauthors{Sanders et al.}

\def\asec{\char'175 }
\def\ksp{p_{\rm KS}}

\def\WHb{W_{{\rm H}\beta}}

\def\WNL{{\rm WNL}/({\rm WNL+O})}

\newcommand{\zPTFonezeroaavz}{0.063}

\newcommand{\ztwozerozeroninejf}{0.008}

\newcommand{\RAoneninenineoneR}{15:54:53.52}
\newcommand{\DEConeninenineoneR}{+19:00:43.9}

\newcommand{\RAtwozerozerotwoex}{22:09:00.79}
\newcommand{\DECtwozerozerotwoex}{--10:36:25.8}

\newcommand{\RAtwozerozerotwogz}{02:34:10.36}
\newcommand{\DECtwozerozerotwogz}{--00:53:18.2}

\newcommand{\RAtwozerozerothreeev}{13:10:31.80}
\newcommand{\DECtwozerozerothreeev}{--21:39:49.6}

\newcommand{\RAtwozerozerofourcf}{14:11:05.77}
\newcommand{\DECtwozerozerofourcf}{--11:44:09.4}

\newcommand{\RAtwozerozerofourib}{02:40:56.40}
\newcommand{\DECtwozerozerofourib}{--00:10:48.3}

\newcommand{\RAtwozerozerofivehm}{21:39:00.65}
\newcommand{\DECtwozerozerofivehm}{--01:01:38.7}

\newcommand{\RAtwozerozerofivenb}{12:13:37.61}
\newcommand{\DECtwozerozerofivenb}{+16:07:16.2}

\newcommand{\RAtwozerozerosixip}{23:48:31.68}
\newcommand{\DECtwozerozerosixip}{--02:08:57.3}

\newcommand{\RAtwozerozerosixir}{23:04:35.68}
\newcommand{\DECtwozerozerosixir}{+07:36:21.5}

\newcommand{\RAtwozerozerosixjo}{01:23:14.72}
\newcommand{\DECtwozerozerosixjo}{--00:19:46.7}

\newcommand{\RAtwozerozerosixlc}{22:44:24.48}
\newcommand{\DECtwozerozerosixlc}{--00:09:53.5}

\newcommand{\RAtwozerozerosixnx}{03:33:30.63}
\newcommand{\DECtwozerozerosixnx}{--00:40:38.2}

\newcommand{\RAtwozerozerosixtq}{02:10:00.70}
\newcommand{\DECtwozerozerosixtq}{+04:06:00.9}

\newcommand{\RAtwozerozerosevenI}{11:59:13.15}
\newcommand{\DECtwozerozerosevenI}{--01:36:18.9}

\newcommand{\RAtwozerozerosevenaz}{08:25:23.80}
\newcommand{\DECtwozerozerosevenaz}{+69:54:29.4}

\newcommand{\RAtwozerozerosevenbr}{11:15:39.93}
\newcommand{\DECtwozerozerosevenbr}{--04:22:47.8}

\newcommand{\RAtwozerozerosevence}{12:10:17.96}
\newcommand{\DECtwozerozerosevence}{+48:43:31.5}

\newcommand{\RAtwozerozerosevendb}{11:17:10.30}
\newcommand{\DECtwozerozerosevendb}{--06:11:48.6}

\newcommand{\RAtwozerozerosevenea}{15:53:46.27}
\newcommand{\DECtwozerozerosevenea}{--27:02:15.5}

\newcommand{\RAtwozerozerosevenff}{01:24:10.24}
\newcommand{\DECtwozerozerosevenff}{+09:00:40.5}

\newcommand{\RAtwozerozerosevengg}{00:28:12.51}
\newcommand{\DECtwozerozerosevengg}{+00:07:04.8}

\newcommand{\RAtwozerozerosevengl}{01:05:50.11}
\newcommand{\DECtwozerozerosevengl}{+00:08:41.3}

\newcommand{\RAtwozerozerosevenhb}{05:02:01.28}
\newcommand{\DECtwozerozerosevenhb}{--21:07:55.1}

\newcommand{\RAtwozerozerosevenhl}{20:50:07.76}
\newcommand{\DECtwozerozerosevenhl}{--01:58:36.4}

\newcommand{\RAtwozerozerosevenhn}{21:02:46.85}
\newcommand{\DECtwozerozerosevenhn}{--04:05:25.2}

\newcommand{\RAtwozerozeroeightao}{03:07:46.66}
\newcommand{\DECtwozerozeroeightao}{+38:22:06.2}

\newcommand{\RAtwozerozeroeightfi}{01:53:23.17}
\newcommand{\DECtwozerozeroeightfi}{+29:21:28.4}

\newcommand{\RAtwozerozeroeightgc}{02:10:36.63}
\newcommand{\DECtwozerozeroeightgc}{--53:45:59.5}

\newcommand{\RAtwozerozeroeightik}{03:36:09.54}
\newcommand{\DECtwozerozeroeightik}{--35:13:00.7}

\newcommand{\RAtwozerozeroeightim}{04:01:02.15}
\newcommand{\DECtwozerozeroeightim}{+74:05:48.5}

\newcommand{\RAtwozerozeroeightiu}{04:36:55.20}
\newcommand{\DECtwozerozeroeightiu}{--00:21:35.6}

\newcommand{\RAtwozerozeroninehu}{14:53:29.82}
\newcommand{\DECtwozerozeroninehu}{+18:35:31.1}

\newcommand{\RAtwozerozeroninejf}{23:04:52.98}
\newcommand{\DECtwozerozeroninejf}{+12:19:59.5}

\newcommand{\RAtwozerozeroninenl}{03:39:47.78}
\newcommand{\DECtwozerozeroninenl}{--11:13:25.0}

\newcommand{\RAtwozeroonezeroQ}{10:26:27.11}
\newcommand{\DECtwozeroonezeroQ}{+39:01:50.9}

\newcommand{\RAtwozeroonezeroah}{11:44:02.99}
\newcommand{\DECtwozeroonezeroah}{+55:41:27.6}

\newcommand{\RAtwozeroonezeroam}{09:33:01.75}
\newcommand{\DECtwozeroonezeroam}{+15:49:08.8}

\newcommand{\RAtwozeroonezeroay}{12:35:27.19}
\newcommand{\DECtwozeroonezeroay}{+27:04:02.8}

\newcommand{\RAtwozeroonezerocn}{11:04:06.57}
\newcommand{\DECtwozeroonezerocn}{+04:49:58.7}

\newcommand{\RAtwozerooneoneD}{03:02:14.58}
\newcommand{\DECtwozerooneoneD}{+17:20:58.9}

\newcommand{\RAtwozerooneoneV}{09:27:38.76}
\newcommand{\DECtwozerooneoneV}{+28:47:27.2}

\newcommand{\RAtwozerooneonebv}{13:02:53.57}
\newcommand{\DECtwozerooneonebv}{--04:02:36.0}

\newcommand{\RAtwozerooneonecs}{12:08:01.08}
\newcommand{\DECtwozerooneonecs}{+49:13:33.0}

\newcommand{\RAtwozerooneonegh}{03:16:54.20}
\newcommand{\DECtwozerooneonegh}{+25:54:14.6}

\newcommand{\RAtwozerooneonehw}{22:26:14.54}
\newcommand{\DECtwozerooneonehw}{+34:12:59.1}

\newcommand{\RAtwozerooneoneit}{22:02:44.45}
\newcommand{\DECtwozerooneoneit}{+31:41:49.1}

\newcommand{\RALSQoneoneJW}{02:04:47.40}
\newcommand{\DECLSQoneoneJW}{+00:50:06.0}

\newcommand{\RAPTFzeroninedfk}{23:09:13.42}
\newcommand{\DECPTFzeroninedfk}{+07:48:15.4}

\newcommand{\RAPTFzeroninedxv}{23:08:34.73}
\newcommand{\DECPTFzeroninedxv}{+18:56:13.7}

\newcommand{\RAPTFzeronineiqd}{02:35:23.23}
\newcommand{\DECPTFzeronineiqd}{+40:17:08.7}

\newcommand{\RAPTFzeronineq}{12:24:50.11}
\newcommand{\DECPTFzeronineq}{+08:25:58.8}

\newcommand{\RAPTFonezeroaavz}{11:20:13.36}
\newcommand{\DECPTFonezeroaavz}{+03:44:45.2}

\newcommand{\RAPTFonezerobip}{12:34:10.52}
\newcommand{\DECPTFonezerobip}{+08:21:48.5}

\newcommand{\RAPTFonezerovgv}{22:16:01.17}
\newcommand{\DECPTFonezerovgv}{+40:52:03.3}

\newcommand{\RAPTFoneonehyg}{23:27:57.34}
\newcommand{\DECPTFoneonehyg}{+08:46:38.0}

\newcommand{\PrevPublished}{15}

\newcommand{\NSanders}{58}

\newcommand{\NSandersIIb}{10}

\newcommand{\NSandersIb}{13}

\newcommand{\NSandersIbc}{3}

\newcommand{\NSandersIc}{24}

\newcommand{\NSandersIcBL}{8}
\newcommand{\NSandersAGN}{60}

\newcommand{\inGold}{48}
\newcommand{\inSilver}{10}
\newcommand{\inExp}{43}
\newcommand{\inNuc}{15}
\newcommand{\RatioIcIb}{1.8}
\newcommand{\fromBC}{23}
\newcommand{\fromLDSS}{25}
\newcommand{\fromIMACS}{9}
\newcommand{\medredall}{0.036}
\newcommand{\medredIcBL}{0.056}
\newcommand{\medredOther}{0.034}
\newcommand{\percentIcBL}{13}
\newcommand{\percentIIb}{17}
\newcommand{\KSModjazIbIc}{0.08^{+0.26}_{-0.06}}

\newcommand{\diagPPzerofourNtwomedunc}{0.09}
\newcommand{\diagPPzerofourNtwoN}{50}

\newcommand{\diagPPzerofourOthreeNtwoN}{31}

\newcommand{\diagPTzerofiveN}{13}

\newcommand{\diagZninefourN}{12}

\newcommand{\diagdirectN}{2}

\newcommand{\statZNIIbPPzerofourNtwo}{7}
\newcommand{\statZminIIbPPzerofourNtwo}{8.30}
\newcommand{\statZmaxIIbPPzerofourNtwo}{8.66}
\newcommand{\statZmedIIbPPzerofourNtwo}{8.46}
\newcommand{\statZstdIIbPPzerofourNtwo}{0.18}
\newcommand{\statZNIbPPzerofourNtwo}{12}
\newcommand{\statZminIbPPzerofourNtwo}{8.15}
\newcommand{\statZmaxIbPPzerofourNtwo}{8.79}
\newcommand{\statZmedIbPPzerofourNtwo}{8.48}
\newcommand{\statZstdIbPPzerofourNtwo}{0.16}

\newcommand{\statZNIcPPzerofourNtwo}{21}
\newcommand{\statZminIcPPzerofourNtwo}{8.14}
\newcommand{\statZmaxIcPPzerofourNtwo}{8.88}
\newcommand{\statZmedIcPPzerofourNtwo}{8.61}
\newcommand{\statZstdIcPPzerofourNtwo}{0.22}
\newcommand{\statZNIcBLPPzerofourNtwo}{7}
\newcommand{\statZminIcBLPPzerofourNtwo}{8.01}
\newcommand{\statZmaxIcBLPPzerofourNtwo}{8.53}
\newcommand{\statZmedIcBLPPzerofourNtwo}{8.34}
\newcommand{\statZstdIcBLPPzerofourNtwo}{0.21}

\newcommand{\statZNIIbEGPPzerofourNtwo}{5}

\newcommand{\statZmedDIFFIIbPPzerofourNtwo}{-0.04}
\newcommand{\statZNIbEGPPzerofourNtwo}{8}

\newcommand{\statZmedDIFFIbPPzerofourNtwo}{-0.05}

\newcommand{\statZNIcEGPPzerofourNtwo}{13}

\newcommand{\statZmedDIFFIcPPzerofourNtwo}{-0.01}
\newcommand{\statZNIcBLEGPPzerofourNtwo}{3}

\newcommand{\statZmedDIFFIcBLPPzerofourNtwo}{0.02}

\newcommand{\AVNIIb}{9}

\newcommand{\AVNIb}{12}

\newcommand{\AVNIc}{23}

\newcommand{\AVNIcBL}{6}

\newcommand{\KSAvIbIc}{0.85}
\newcommand{\AvIbIcStd}{0.2^{+3.2}_{-0.2}}

\newcommand{\KSPPzerofourNtwoIcIb}{0.10^{+0.20}_{-0.08}}
\newcommand{\KSPPzerofourNtwoIbIcBL}{0.21^{+0.31}_{-0.16}}

\newcommand{\KSPPzerofourNtwoIIbIcBL}{0.42^{+0.00}_{-0.29}}
\newcommand{\KSPPzerofourNtwoIbIIb}{0.75^{+0.22}_{-0.23}}
\newcommand{\KSPPzerofourNtwoIcIIb}{0.34^{+0.17}_{-0.27}}
\newcommand{\KSPPzerofourNtwoIcIcBL}{0.01^{+0.03}_{-0.01}}

\newcommand{\KSEGPPzerofourNtwoIcIb}{0.07^{+0.21}_{-0.05}}
\newcommand{\KSEGPPzerofourNtwoIbIcBL}{}

\newcommand{\KSEGPPzerofourNtwoIIbIcBL}{}
\newcommand{\KSEGPPzerofourNtwoIbIIb}{0.67^{+0.27}_{-0.30}}
\newcommand{\KSEGPPzerofourNtwoIcIIb}{0.16^{+0.28}_{-0.09}}
\newcommand{\KSEGPPzerofourNtwoIcIcBL}{}

\newcommand{\noexcIbmed}{8.46}
\newcommand{\noexcIcmed}{8.42}
\newcommand{\cNpNucIb}{19}
\newcommand{\cNpNucIc}{28}

\newcommand{\KSPPzerofourNtwoIIIc}{0.52}

\newcommand{\KSPPzerofourNtwoIIIb}{1\times10^{-3}}

\newcommand{\KSPPzerofourNtwoIIIcBL}{8\times10^{-4}}

\newcommand{\statZmedDIFFIIallPPzerofourNtwo}{0.13}
\newcommand{\KSPPzerofourNtwoIIall}{0.01}
\newcommand{\KSlgrbIcBL}{0.13}
\newcommand{\NlgrbEL}{7}
\newcommand{\combN}{171}
\newcommand{\combNE}{114}

\newcommand{\combNU}{75}

\newcommand{\combnosN}{133}
\newcommand{\combnosNE}{84}

\newcommand{\combnosNU}{37}

\newcommand{\ZmedallTPPzerofourNtwo}{8.64}
\newcommand{\ZmedallNTPPzerofourNtwo}{8.52}
\newcommand{\ZdifallNTPPzerofourNtwo}{24}
\newcommand{\ZallNTKS}{3\times10^{-4}}
\newcommand{\ZNTcomparewidth}{70}
\newcommand{\numTsol}{1.2}
\newcommand{\numThalfsol}{2.4}
\newcommand{\numTthirdsol}{5.1}
\newcommand{\ModjazTargSimKS}{5.0}
\newcommand{\ModjazTargSimMeanM}{3.3}
\newcommand{\ModjazTargSimMeanZ}{64.2}
\newcommand{\compexpN}{30}
\newcommand{\compexpmeddif}{0.08}
\newcommand{\compexprms}{0.13}
\newcommand{\compexpmederr}{0.68}
\newcommand{\compexpIbvIc}{0.03}
\newcommand{\KScAllAllIIbIb}{0.56^{+0.31}_{-0.35}}

\newcommand{\KSNcAllAllIIb}{25}

\newcommand{\KScAllAllIbIc}{0.08^{+0.21}_{-0.06}}

\newcommand{\KSNcAllAllIb}{47}
\newcommand{\MedcAllAllIb}{8.56}
\newcommand{\StdcAllAllIb}{0.16}

\newcommand{\KScAllAllIbcIcBL}{2.3^{+21.0}_{-2.1}\times10^{-4}}
\newcommand{\KSNcAllAllIbc}{127}

\newcommand{\KScAllAllIcIcBL}{1.3^{+9.8}_{-1.2}\times10^{-4}}
\newcommand{\KSNcAllAllIc}{68}
\newcommand{\MedcAllAllIc}{8.65}
\newcommand{\StdcAllAllIc}{0.13}
\newcommand{\KSNcAllAllIcBL}{18}
\newcommand{\MedcAllAllIcBL}{8.36}

\newcommand{\KScAllTarIIbIb}{0.21^{+0.39}_{-0.15}}

\newcommand{\KSNcAllTarIIb}{17}

\newcommand{\KScAllTarIbIc}{0.65^{+0.24}_{-0.43}}

\newcommand{\KSNcAllTarIb}{26}
\newcommand{\MedcAllTarIb}{8.66}

\newcommand{\KScAllTarIbcIcBL}{4.7^{+13.1}_{-4.0}\times10^{-2}}
\newcommand{\KSNcAllTarIbc}{73}

\newcommand{\KScAllTarIcIcBL}{4.4^{+16.1}_{-3.8}\times10^{-2}}
\newcommand{\KSNcAllTarIc}{39}
\newcommand{\MedcAllTarIc}{8.65}

\newcommand{\KSNcAllTarIcBL}{6}
\newcommand{\MedcAllTarIcBL}{8.45}

\newcommand{\KScAllUntIIbIb}{0.75^{+0.19}_{-0.32}}

\newcommand{\KSNcAllUntIIb}{8}

\newcommand{\KScAllUntIbIc}{2.0^{+6.8}_{-1.7}\times10^{-2}}

\newcommand{\KSNcAllUntIb}{21}
\newcommand{\MedcAllUntIb}{8.46}

\newcommand{\KScAllUntIbcIcBL}{3.1^{+12.9}_{-2.7}\times10^{-3}}
\newcommand{\KSNcAllUntIbc}{54}

\newcommand{\KScAllUntIcIcBL}{1.7^{+6.0}_{-1.5}\times10^{-3}}
\newcommand{\KSNcAllUntIc}{29}
\newcommand{\MedcAllUntIc}{8.61}

\newcommand{\KSNcAllUntIcBL}{12}
\newcommand{\MedcAllUntIcBL}{8.30}

\newcommand{\KScNucAllIIbIb}{0.60^{+0.31}_{-0.33}}

\newcommand{\KSNcNucAllIIb}{13}

\newcommand{\KScNucAllIbIc}{0.48^{+0.37}_{-0.34}}

\newcommand{\KSNcNucAllIb}{11}

\newcommand{\KScNucAllIbcIcBL}{}
\newcommand{\KSNcNucAllIbc}{41}

\newcommand{\KScNucAllIcIcBL}{}
\newcommand{\KSNcNucAllIc}{26}

\newcommand{\KSNcNucAllIcBL}{3}

\newcommand{\KScNucTarIIbIb}{0.15^{+0.18}_{-0.12}}

\newcommand{\KSNcNucTarIIb}{12}

\newcommand{\KScNucTarIbIc}{0.57^{+0.30}_{-0.35}}

\newcommand{\KSNcNucTarIb}{7}

\newcommand{\KScNucTarIbcIcBL}{}
\newcommand{\KSNcNucTarIbc}{29}

\newcommand{\KScNucTarIcIcBL}{}
\newcommand{\KSNcNucTarIc}{18}

\newcommand{\KSNcNucTarIcBL}{2}

\newcommand{\KScNucUntIIbIb}{}

\newcommand{\KSNcNucUntIIb}{1}

\newcommand{\KScNucUntIbIc}{0.38^{+0.36}_{-0.33}}

\newcommand{\KSNcNucUntIb}{4}

\newcommand{\KScNucUntIbcIcBL}{}
\newcommand{\KSNcNucUntIbc}{12}

\newcommand{\KScNucUntIcIcBL}{}
\newcommand{\KSNcNucUntIc}{8}

\newcommand{\KSNcNucUntIcBL}{1}

\newcommand{\KScEsiAllIIbIb}{0.31^{+0.40}_{-0.25}}

\newcommand{\KSNcEsiAllIIb}{12}

\newcommand{\KScEsiAllIbIc}{0.16^{+0.33}_{-0.13}}

\newcommand{\KSNcEsiAllIb}{36}

\newcommand{\KScEsiAllIbcIcBL}{4.5^{+19.5}_{-3.8}\times10^{-3}}
\newcommand{\KSNcEsiAllIbc}{86}

\newcommand{\KScEsiAllIcIcBL}{1.8^{+12.3}_{-1.5}\times10^{-3}}
\newcommand{\KSNcEsiAllIc}{42}

\newcommand{\KSNcEsiAllIcBL}{15}

\newcommand{\KScEsiTarIIbIb}{0.32^{+0.42}_{-0.24}}

\newcommand{\KSNcEsiTarIIb}{5}
\newcommand{\MedcEsiTarIIb}{8.55}

\newcommand{\KScEsiTarIbIc}{0.63^{+0.28}_{-0.38}}

\newcommand{\KSNcEsiTarIb}{19}
\newcommand{\MedcEsiTarIb}{8.62}

\newcommand{\KScEsiTarIbcIcBL}{0.39^{+0.38}_{-0.31}}
\newcommand{\KSNcEsiTarIbc}{44}

\newcommand{\KScEsiTarIcIcBL}{0.38^{+0.42}_{-0.29}}
\newcommand{\KSNcEsiTarIc}{21}
\newcommand{\MedcEsiTarIc}{8.62}

\newcommand{\KSNcEsiTarIcBL}{4}

\newcommand{\KScEsiUntIIbIb}{0.70^{+0.20}_{-0.35}}

\newcommand{\KSNcEsiUntIIb}{7}

\newcommand{\KScEsiUntIbIc}{0.06^{+0.12}_{-0.05}}

\newcommand{\KSNcEsiUntIb}{17}

\newcommand{\KScEsiUntIbcIcBL}{1.1^{+3.1}_{-0.9}\times10^{-2}}
\newcommand{\KSNcEsiUntIbc}{42}

\newcommand{\KScEsiUntIcIcBL}{7.2^{+12.9}_{-6.0}\times10^{-3}}
\newcommand{\KSNcEsiUntIc}{21}

\newcommand{\KSNcEsiUntIcBL}{11}

\newcommand{\MdotdiffIbIccombEU}{1.11}
\newcommand{\MdotdiffIbIcEG}{1.17}
\newcommand{\MdotdiffIbIccombEUVink}{1.20}
\newcommand{\MdotdiffIbIcEGVink}{1.32}
\newcommand{\combYdwarf}{6}
\newcommand{\combAll}{171}
\newcommand{\combPdwarf}{4}

\newcommand{\HBagemedIc}{7.4}
\newcommand{\HBagestdIc}{1.1}
\newcommand{\HBageNIc}{10}
\newcommand{\HBagemedIb}{8.1}
\newcommand{\HBagestdIb}{1.5}
\newcommand{\HBageNIb}{6}

\newcommand{\HBagemedIcBL}{5.6}
\newcommand{\HBagestdIcBL}{1.6}
\newcommand{\HBageNIcBL}{4}
\newcommand{\HBagemedIIb}{6.0}
\newcommand{\HBagestdIIb}{1.2}
\newcommand{\HBageNIIb}{5}
\newcommand{\HBageKSIcIcBL}{0.34}
\newcommand{\HBageKSIcIb}{0.21}
\newcommand{\HBageKSIcBLGRB}{0.48}
\newcommand{\compN}{18}
\newcommand{\comprms}{0.15}
\newcommand{\compmean}{0.06}
\newcommand{\compterr}{1.6}
\newcommand{\compEEN}{5}
\newcommand{\compEErms}{0.08}

\newcommand{\NgaussdistPone}{200}
\newcommand{\NgaussdistPtwo}{50}
\newcommand{\NgaussdistPoneM}{100}
\newcommand{\NgaussdistPtwoM}{20}

\newcommand{\statZmedIbPPzerofourNtwoZsol}{0.62}
\newcommand{\statZmedIcPPzerofourNtwoZsol}{0.83}
\newcommand{\statZmedIcBLPPzerofourNtwoZsol}{0.45}
\def\actaa{\rm{Acta Astron.}}

\begin{document}

\title{A Spectroscopic Study of Type I\lowercase{bc} Supernova Host Galaxies \\from Untargeted Surveys}
\author{
N. E. Sanders\altaffilmark{1},
A. M. Soderberg\altaffilmark{1},
E. M. Levesque\altaffilmark{2},
R. J. Foley\altaffilmark{1}\altaffilmark{3},
R. Chornock\altaffilmark{1},
D. Milisavljevic\altaffilmark{1},
R. Margutti\altaffilmark{1},
E. Berger\altaffilmark{1},
M. R. Drout\altaffilmark{1},
I. Czekala\altaffilmark{1},
J. A. Dittmann\altaffilmark{1},
}

\altaffiltext{1}{Harvard-Smithsonian Center for Astrophysics, 60 Garden Street, Cambridge, MA 02138 USA}
\altaffiltext{2}{CASA, Department of Astrophysical and Planetary Sciences, University of Colorado, 389-UCB, Boulder, CO 80309, USA}
\altaffiltext{3}{Clay fellow}

\email{nsanders@cfa.harvard.edu}

\begin{abstract}

We present the first spectroscopic study of the host environments of Type~Ibc supernovae (SN~Ibc) discovered exclusively by untargeted SN searches.  Past studies of SN~Ibc host environments have been biased towards high-mass, high-metallicity galaxies by focusing on SNe discovered in galaxy-targeted SN searches. Our new observations more than double the total number of spectroscopic stellar population age and metallicity measurements published for untargeted SN~Ibc host environments.  For the \statZNIbPPzerofourNtwo\ SNe Ib and \statZNIcPPzerofourNtwo\ SNe Ic in our metallicity sample, we find median metallicities of $\statZmedIbPPzerofourNtwoZsol~Z_\odot$ and $\statZmedIcPPzerofourNtwoZsol~Z_\odot$, respectively, but determine that the discrepancy in the full distribution of metallicities is not statistically significant.  This median difference would correspond to only a small difference in the mass loss via metal-line 
driven winds ($\lesssim30\%$), suggesting this does not play the dominant role in distinguishing SN~Ib and Ic progenitors.  However, the median metallicity of the \statZNIcBLPPzerofourNtwo\ broad-lined SN~Ic (SN~Ic-BL) in our sample is significantly lower, $\statZmedIcBLPPzerofourNtwoZsol~Z_\odot$.  The age of the young stellar population of SN~Ic-BL host environments also seems to be lower than for SN~Ib and Ic, but our age sample is small.  Combining all SN~Ibc host environment spectroscopy from the literature to date does not reveal a significant difference in SN~Ib and Ic metallicities, but reinforces the significance of the lower metallicities for SN~Ic-BL.  This combined sample demonstrates that galaxy-targeted SN searches introduce a significant bias for studies seeking to infer the metallicity distribution of SN progenitors, and we identify and discuss other systematic effects that play smaller roles.  We discuss the path forward for making progress on SN~Ibc progenitor studies in the LSST era.

This paper includes data gathered with the 6.5 m Magellan Telescopes located at Las Campanas Observatory, Chile.

\smallskip
\end{abstract}

\keywords{supernovae: general --- galaxies: abundances --- surveys}

\section{INTRODUCTION}
\label{sec:intro}

Core-collapse supernovae show a diversity of absorption features in their spectra near maximum light, reflecting a diversity in the composition of the outer envelope of their massive star progenitors at the ends of their lives \citep{Filippenko1997,Woosley02}.  In particular, some SNe show no evidence of hydrogen (Type Ib) or no evidence for either hydrogen or helium (Type Ic), suggesting extensive mass loss in the progenitor star sufficient to complete stripping of the H and He layers of its outer envelope \citep[][but see also \citealt{Dessart12}]{Elias85,Filippenko85,Wheeler85,Uomoto85,Clocchiatti96,Hachinger12}. Stellar evolutionary considerations point to two likely channels for these stripped-envelope core-collapse supernovae (Type Ibc supernovae).\footnote{Hereafter we use ``SN~Ibc'' to refer to the class of stripped-envelope core-collapse supernovae generally.  We define SN~Ibc to include SNe of subtypes Ib, IIb, Ic, and Ic-BL.  We use ``SN~Ib/c'' to refer to supernovae whose spectroscopic type 
is uncertain, but likely to be one of the SN~Ibc subtypes.}  These channels are: (i) high-mass Wolf-Rayet (WR) stars with strong metal line-driven winds with rotation likely playing an important role \citep{Woosley95,Georgy12}, and (ii) lower-mass helium-stars in close-binary systems who lose their envelopes via Roche lobe overflow or common envelope evolution \citep{Podsiadlowski92,Yoon10,Dessart11}.  Searches for the progenitor stars of SN~Ibc in pre-explosion imaging have yet to yield a progenitor detection, but have provided upper limits that challenge the hypothesis that their progenitors are massive WR stars like those seen in the Local Group \citep{Smartt09}.

By measuring the metallicity of the host environments of Type~Ibc SNe as a proxy for the metallicity of the progenitor stars, we may be able to distinguish between these two progenitor models.  Mass loss in WR stars is enhanced at high metallicity \citep{Vink05}.  If the primary SN~Ibc progenitor channel is single WR stars, then the rate of SN~Ibc relative to SNe that show H features (SN~II) would be enhanced at high metallicity (see e.g. \citealt{BP03}) and the ratio of SN~Ic to Ib should similarly be higher.  In binary progenitor systems, massive primary stars may still strip their envelopes primarily via WR winds, but there exists an additional channel for SN~Ic to be produced by relatively low mass ($\sim12~M_\odot$) stars which may dominate at low metallicity.  This channel calls for the mass transfer to occur while the star is in the core 
helium burning or later phases \citep{Yoon10}.  Because either channel calls for massive, short-lived stars to produce the explosions, the metallicity of the SN host environment should be an appropriate proxy for the metallicity of the progenitor star.

Additionally, a connection has emerged between long-duration Gamma Ray Bursts (long GRBs) and one particular subtype of SN~Ibc: broad-lined Type Ic SNe (Ic-BL; \citealt{kfw+98}; see \citealt{woosley06} for a review).  The broad and highly blueshifted absorption features of SNe Ic-BL indicate high photospheric expansion velocities, $v_{\rm ph}>2\times10^4$~km~s$^{-1}$ \citep{Iwamoto98}.  This GRB-SN connection can be explained by the gravitational collapse of a massive ($M\gtrsim 20~M_{\odot}$) progenitor star that produces a rapidly rotating and accreting compact object (central engine) that powers a relativistic outflow \citep[the collapsar model,][]{MWH01}.  However, radio observations demonstrate that only a small fraction ($\lesssim1/3$) of SNe Ic-BL harbor relativistic outflows \citep{skn+06,scp+10}.  Because angular momentum loss due to metal line-driven winds could prevent the compact remnant from rotating fast enough to produce a relativistic outflow, a metallicity threshold ($Z\lesssim0.3~Z_\odot$) has been proposed for collapsars \citep{WoosleyHeger}. 

Recent observations fundamentally challenge the role of metallicity in GRB production \citep{Fryer07}.  SN~2009bb provides an example of a SN~Ic-BL produced in a super-solar metallicity environment and harboring a central engine \citep{lsf+10,scp+10}, and LGRBs with relatively high-metallicity host environments have now been identified \citep{lkg+10,Graham09}. In contrast, SN~2010ay was a SN~Ic-BL with extreme explosion properties that occurred in a sub-solar metallicity environment and without evidence for a central engine (\citealt{nes2010ay}; see also SN~2007bg, \citealt{Young10}).

In the past few years, several observational studies have sought to measure the characteristics of the host environments of SN~Ibc.  \cite{Prieto08} used SDSS spectroscopy to study the metallicity distribution of 115 SNe (19 SN~Ibc) and found that SN~Ibc host environments are metal-enriched compared to those of SN~II. Extending the work of \cite{BP03}, \cite{BP09} reach a similar result using SDSS photometry to estimate metallicities for 701 SN (98 SN~Ibc) host galaxies.  \cite{Arcavi10} examine the host galaxies of core-collapse SNe discovered by the Palomar Transient Factory (PTF) and find that SN~Ic are more common in high-metallicity environments (``giant'' host galaxies, $M_r<-18$~mag), while SN~Ib, IIb and Ic-BL dominate in low-metallicity environments (``dwarf'' host galaxies).  \cite{Anderson10} and \cite{Leloudas11} perform spectroscopy to measure the metallicity of the host environments of 28 and 20 SN~Ibc, respectively, finding no statistically significant difference between the metallicity distribution of SN Ib and Ic.  Extending the work of \cite{Modjaz08}, \cite{Modjaz11} performed a similar spectroscopy study of 35 SN~Ibc host environments, finding that SN~Ic come from significantly higher-metallicity host environments than SN~Ib, with SN~Ic-BL falling in between.  Following \cite{Kelly08}, \cite{Kelly11} have examined SDSS spectroscopy of 519 SNe (67 SN~Ibc), finding that SN~Ic-BL preferentially occur in low-mass, low-metallicity host galaxies relative to other core-collapse SNe.

However, there is a prominent observational bias that affects the progenitor metallicity distribution inferred from observations of most known SN host galaxies.  The well-known relation between the luminosity and global-metallicity of star-forming galaxies \citep[the $L-Z$ relation, see e.g.][]{Tremonti04} indicates that a metallicity distribution measured only from objects found in targeted SN searches, which look for transients in fields centered on nearby and luminous galaxies, will be biased towards high metallicities.

Here we describe a new spectroscopic study of SN~Ibc host galaxy metallicities unbiased with respect to the $L-Z$ relation.  We have obtained spectra of \NSandersAGN\ host environments of SN~Ibc discovered only by untargeted transient searches (\NSandersIb\ SNe Ib, \NSandersIIb\ SNe IIb, \NSandersIc\ SNe Ic, \NSandersIcBL\ SNe Ic-BL, \NSandersIbc\ of indeterminate type SN~Ib/c, and 2 with AGN-dominated host environments, Section~\ref{sec:fluxmea}).  Previous spectroscopic studies of SN~Ibc host environments have included relatively few SNe discovered by untargeted searches, totaling $\lesssim40$ objects.  Our study doubles the existing sample of host environment spectroscopy for untargeted SN~Ibc, offering considerable constraining power for inferring the metallicity distribution of the parent population.

In Section~\ref{sec:obs} we describe the characteristics of this sample, our optical observations, and our spectroscopic methodology.  We present and analyze the host galaxy metallicities and other physical properties derived from these observations in Section~\ref{sec:res} and combine with previous spectroscopic surveys of SN~Ibc host environments in Section~\ref{sec:combsurv}.  In Section~\ref{sec:SE} we discuss possible systematics affecting our results.  In Section~\ref{sec:disc} we discuss our results in relation to SN~Ibc progenitor models and the SN-GRB connection and we suggest implications for future studies of SN~Ibc. We conclude in Section~\ref{sec:conc}.

\section{SAMPLE CONSTRUCTION}
\label{sec:obs}

\subsection{SN sample}
\label{sec:sample}

We have observed the host galaxies of \NSandersAGN\ SNe~Ibc reported in the International Astronomical Union Circulars (IAUCs)\footnote{f1.html} and/or Astronomer's Telegrams\footnote{http://www.astronomerstelegram.org/} between 1990-2011.  When transient searches operate by returning repeatedly to a pre-selected, typically bright, set of galaxies, we refer to the SNe found in those galaxies by those searches to be ``targeted.'' We refer to any SN discovered by other means as ``untargeted,'' including discoveries by wide-field optical surveys, SNe identified by targeted searches in anonymous background galaxies, and SNe discovered serendipitously during observations of unrelated objects.  We observed only untargeted discoveries and prioritized those SNe with reliable classifications, host galaxies which did not already have previously published metallicity measurements, and which were visible at low airmass during the time of our observations.  

In total, we present optical spectroscopy with S/N sufficient for metallicity measurements for \diagPPzerofourNtwoN\ host galaxies, with median redshift $z=\medredall$.  The discoverer of each SN in our sample is listed in Table~\ref{tab:prop}.
Spectroscopic metallicity estimates have been previously published for only \PrevPublished\ of these galaxies.  A comparison to previous spectroscopic studies of SN~Ibc host environments is presented in Section~\ref{sec:premaetal}; additionally, the host environment of SN\,2009jf was previously studied by \cite{Valenti11} \citep[see also][]{Sahu11}.

\begin{deluxetable*}{llllllll}
\tablecaption{Properties of SNe in Untargeted Sample\label{tab:prop}}
\tablewidth{0pt}
\tabletypesize{\scriptsize}
\tablehead{ \colhead{SN} & \colhead{Type\tablenotemark{a}} & \colhead{Sample\tablenotemark{b}} & \colhead{$z$} & \colhead{Slit width (kpc)\tablenotemark{c}} & \colhead{Discoverer\tablenotemark{d}}  & \colhead{Classification\tablenotemark{e}}}

\startdata
1991R &  Ibc &  S & 0.035 &  0.7 &  \cite{IAUC5237} &  \cite{IAUC5237S} \\
2002ex &  Ib &  G & 0.037 &  1.1 &  SNF &  S. \\
2002gz &  IIb-pec &  S & 0.085 &  1.1 &  SNF & \cite{IAUC8059}  \\
2003ev &  Ic &  G & 0.024 &  0.5 &  LOSS &  \cite{CBET8158} \\
2004cf &  Ib &  S & 0.248 &  5.8 &  \cite{IAUC8352} &  M. T. Botticella \\
2004ib &  Ic &  G & 0.056 &  1.1 &  SDSS &  G. Leloudas \\
2005hm &  Ib &  G & 0.034 &  1.0 &  SDSS &  \cite{Leloudas11} \\
2005nb &  Ic-BL &  G & 0.024 &  0.5 &  \cite{IAUC8657} &  \cite{Modjaz11} \\
2006ip &  Ic &  G & 0.031 &  0.6 &  SNF &  \cite{Modjaz11} \\
2006ir &  Ic &  G & 0.021 &  0.4 &  SNF &  \cite{Leloudas11} \\
2006jo &  Ib &  G & 0.077 &  1.4 &  SDSS &  \cite{Leloudas11} \\
2006lc &  Ib &  G & 0.016 &  0.3 &  SDSS &   \cite{CBET699}\\
2006nx &  Ic-BL &  G & 0.137 &  2.4 &  SDSS &  \cite{Modjaz11} \\
2006tq &  Ic &  S & 0.261 &  3.0 &  ESSENCE &  \cite{CBET830}\tablenotemark{f} \\
2007I &  Ic-BL &  G & 0.022 &  0.4 &  LOSS &  \cite{Modjaz11} \\
2007az &  Ib &  G & 0.035 &  1.0 &  LOSS &  \cite{CBET909} \\
2007br &  IIb &  G & 0.053 &  1.0 &  SNF &  R. Thomas \\
2007ce &  Ic-BL &  S & 0.046 &  0.9 &  \cite{CBET953} &  \cite{CBET953S} \\
2007db &  Ic &  G & 0.048 &  0.7 &  SNF &  R. Thomas \\
2007ea &  IIb &  G & 0.040 &  0.8 &  SNF &  R. Thomas \\
2007ff &  Ic &  S & 0.049 &  0.7 &  SNF &  R. Thomas \\
2007gg &  Ib &  G & 0.038 &  0.8 &  SNF &  R. Thomas \\
2007gl &  Ic &  G & 0.028 &  0.4 &  SNF &  R. Thomas \\
2007hb &  Ic &  G & 0.021 &  0.4 &  SNF &  R. Thomas \\
2007hl &  Ic &  S & 0.056 &  1.1 &  SNF &  R. Thomas \\
2007hn &  Ic &  G & 0.028 &  0.6 &  SNF &  \cite{Leloudas11} \\
2008ao &  Ic &  G & 0.015 &  0.3 &  \cite{CBET1269} &  \cite{CBET1275} \\
2008fi &  IIb &  G & 0.026 &  0.5 &  \cite{CBET1493} &  \cite{CBET1503} \\
2008gc &  Ib &  G & 0.049 &  0.7 &  CHASE &  \cite{CBET1540} \\
2008ik &  Ic &  G & 0.064 &  0.9 &  CHASE &  \cite{CBET1629} \\
2008im &  Ib &  G & 0.008 &  0.2 &  \cite{CBET1635} &  S. \\
2008iu &  Ic-BL &  S & 0.130 &  1.6 &  CRTS &  \cite{CBET1681} \\
2009hu &  Ib &  G & 0.117 &  2.1 &  \cite{CBET1894S} & \cite{CBET1894S}  \\
2009jf &  Ib &  G & 0.008 &  0.1 &  LOSS/PTF &  \cite{Valenti11} \\
2009nl &  Ic &  G & 0.113 &  1.8 &  CRTS &  A. Drake \\
2010Q &  Ic &  G & 0.054 &  1.0 &  CRTS &  A. Drake \\
2010ah &  Ic-BL &  G & 0.050 &  1.0 &  PTF &  \cite{Corsi11} \\
2010am &  IIb &  G & 0.020 &  0.4 &  CRTS & \cite{CBET22082C}  \\
2010ay &  Ic-BL &  G & 0.067 &  1.3 &  CRTS &  \cite{nes2010ay} \\
2010cn &  Ib/IIb-pec &  G & 0.026 &  0.5 &  CHASE & \tablenotemark{g}  \\
2010lz &  Ic &  G & 0.090 &  1.2 &  CRTS &  A. Drake \\
2011D &  IIb &  G & 0.023 &  0.3 &  CRTS &  \cite{CBET2627S} \\
2011V &  IIb &  G & 0.014 &  0.3 &  CRTS &  \cite{CBET2650} \\
2011bv &  IIb &  G & 0.072 &  1.0 &  CRTS &  \cite{CBET2704} \\
2011cs &  Ic &  G & 0.101 &  1.8 &  CRTS &  \cite{CBET2733} \\
2011gh &  Ib/c &  S & 0.018 &  0.4 &  CRTS & \cite{CBET2853}  \\
2011hw &  Ibn &  G & 0.021 &  0.4 &  \cite{CBET2906} &  \cite{Smith12} \\
2011ip &  Ic &  S & 0.051 &  1.0 &  \cite{CBET2932} &  S., S. Valenti \\
2011it &  Ic &  G & 0.016 &  0.3 &  \cite{CBET2938} &  S. \\
LSQ11JW &  Ib &  G & 0.020 &  0.4 &  LSQ &  S. \\
PTF09dfk &  Ib &  G & 0.016 &  0.2 &  PTF &  A. Gal-Yam \\
PTF09dxv &  IIb &  G & 0.032 &  0.4 &  PTF &  A. Gal-Yam \\
PTF09iqd &  Ic &  G & 0.034 &  0.7 &  PTF &  A. Gal-Yam \\
PTF09q &  Ic &  G & 0.090 &  1.7 &  PTF &  A. Gal-Yam \\
PTF10aavz &  Ic-BL &  G & 0.063 &  0.9 &  PTF &  S. \\
PTF10bip &  Ic &  G & 0.051 &  1.0 &  PTF &  A. Gal-Yam \\
PTF10vgv &  Ic &  G & 0.015 &  0.3 &  PTF &  \cite{Corsi12} \\
PTF11hyg &  Ic &  G & 0.028 &  0.6 &  PTF &  A. Gal-Yam
\enddata

\tablenotetext{a}{A detailed discussion of SN classification is given in Section~\ref{sec:typing}.}
\tablenotetext{b}{G indicates the Gold sample and S indicates the Silver sample, as defined according to security of spectroscopic classification in Section~\ref{sec:typing}.}
\tablenotetext{c}{The size of the spectroscopic slit (see Table~\ref{tab:obs}) in physical units at the distance of the SN.}
\tablenotetext{d}{Reference for the discovery of the supernova.  Acronyms for untargeted SN searches are as follows: Catalina Real-time Transient Survey \citep[CRTS; ][]{CSS}, the Equation of State: SupErNovae trace Cosmic Expansion program \citep[ESSENCE; ][]{ESSENCE}, the La Silla-QUEST Variability Survey \citep[LSQ; ][]{LSQ}, the Palomar Transient Factory \citep[PTF; ][]{PTF}, the Sloan Digital Sky Survey-II Supernova Survey \citep[SDSS; ][]{SDSSSS}, and the Nearby Supernova Factory  \citep[SNF, including supernovae discovered by NEAT; ][]{SNF}.  Some objects were discovered serendipitously in background galaxies during targeted SN searches or other wide-field surveys, including the CHilean Automatic Supernova sEarch \citep[CHASE; ][]{CHASE} and the Lick Observatory Supernova Search \citep[LOSS; ][]{Li11}.}
\tablenotetext{e}{Reference for spectroscopic classification of SN type.  Author names refer to private communications with observers who performed spectroscopy of the SN.  ``S.'' indicates our own spectroscopy.}
\tablenotetext{f}{Additionally, private communication with S. Blondin (2012) indicates that the classification spectrum for SN~2006tq is not of S/N sufficient for the Gold sample.}
\tablenotetext{g}{Varying spectroscopic classifications of SN~2010cn are reported by \cite{CBET2270,CBET2272,CBET2279}; we consider it a SN~IIb for the purposes on our anaalysis.}
\end{deluxetable*}

\begin{deluxetable*}{p{35pt}rrllllrll}
\tablecaption{SN Host Galaxy Sample and Observing Configurations\label{tab:obs}}
\tablewidth{0pt}
\tabletypesize{\scriptsize}

\tablehead{ \colhead{SN} & \colhead{SN $\alpha_{2000}$} & \colhead{SN $\delta_{2000}$} & \colhead{Date (UT)} & \colhead{Instrument} & \colhead{Disperser\tablenotemark{a}} &  \colhead{Exp. time (s)\tablenotemark{b}} & \colhead{Slit width} & \colhead{E/N\tablenotemark{c}}}

\startdata
1991D\tablenotemark{d}    & 13:41:13.58     & -14:38:47.6    & 2008 May 30  & LDSS3 & VPH\_All           & 600       & $1.0\asec$ & E\\                  
1991R    & \RAoneninenineoneR     & \DEConeninenineoneR    & 2008 May 30  & LDSS3 & VPH\_All           & 600       & $1.0\asec$ & E\\                  
2002ex   & \RAtwozerozerotwoex    & \DECtwozerozerotwoex   & 2008 June 29 & LDSS3 & VPH\_Blue/Red      & 1200/1020 & $1.5\asec$ & E\\                  
2002gz   & \RAtwozerozerotwogz    & \DECtwozerozerotwogz   & 2008 June 29 & IMACS & 300-17.5           & 1200      & $0.7\asec$ & E\\                  
2003ev   & \RAtwozerozerothreeev  & \DECtwozerozerothreeev & 2008 June 1  & LDSS3 & VPH\_All           & 900       & $1.0\asec$ & E\\                  
2003jp\tablenotemark{d}   & 23:26:03.28  & -08:59:22.7 & 2007 Dec. 17 & LDSS3 & VPH\_All           & 600       & $0.75\asec$& N \\                 
2004cf   & \RAtwozerozerofourcf   & \DECtwozerozerofourcf  & 2006 June 30 & LDSS3 & VPH\_Red           & 1800      & $1.5\asec$ & N\\                  
2004ib   & \RAtwozerozerofourib   & \DECtwozerozerofourib  & 2007 Dec. 16 & LDSS3 & VPH\_All           & 1200      & $1.0\asec$ & E\\                  
2005hm   & \RAtwozerozerofivehm   & \DECtwozerozerofivehm  & 2006 June 30 & LDSS3 & VPH\_Blue/Red      & 2100/1500 & $1.5\asec$ & E\\                  
2005nb   & \RAtwozerozerofivenb   & \DECtwozerozerofivenb  & 2008 May 31  & LDSS3 & VPH\_All           & 415       & $1.0\asec$ & E\\                  
2006ip   & \RAtwozerozerosixip    & \DECtwozerozerosixip   & 2007 Dec. 16 & LDSS3 & VPH\_All           & 1200      & $1.0\asec$ & N\\                  
2006ir   & \RAtwozerozerosixir    & \DECtwozerozerosixir   & 2007 Dec. 14 & LDSS3 & VPH\_All           & 1200      & $1.0\asec$ & E\\                  
2006jo   & \RAtwozerozerosixjo    & \DECtwozerozerosixjo   & 2006 Dec. 24 & LDSS3 & VPH\_Blue/Red      & 1000/600  & $1.0\asec$ & N\\                  
2006lc   & \RAtwozerozerosixlc    & \DECtwozerozerosixlc   & 2007 Dec. 14 & LDSS3 & VPH\_All           & 900       & $1.0\asec$ & N\\                  
2006nx   & \RAtwozerozerosixnx    & \DECtwozerozerosixnx   & 2006 Dec. 24 & LDSS3 & VPH\_Blue/Red      & 1500/750  & $1.0\asec$ & N\\                  
2006tq   & \RAtwozerozerosixtq    & \DECtwozerozerosixtq   & 2012 Jan. 18 & LDSS3 & VPH\_ALL           & 1200      & $0.75\asec$& N\\                  
2007I    & \RAtwozerozerosevenI   & \DECtwozerozerosevenI  & 2008 Jan. 17 & LDSS3 & VPH\_All           & 1800      & $1.0\asec$ & N\\                  
2007az   & \RAtwozerozerosevenaz  & \DECtwozerozerosevenaz & 2011 Dec. 24 & BC    & 300GPM             & 1200      & $1.5\asec$ & E\\                  
2007br   & \RAtwozerozerosevenbr  & \DECtwozerozerosevenbr & 2007 Dec. 14 & LDSS3 & VPH\_All           & 1200      & $1.0\asec$ & N\\                  
2007ce   & \RAtwozerozerosevence  & \DECtwozerozerosevence & 2012 Jan. 01 & BC    & 300GPM             & 1800      & $1.0\asec$ & E\\                  
2007db   & \RAtwozerozerosevendb  & \DECtwozerozerosevendb & 2007 Dec. 15 & LDSS3 & VPH\_All           & 900       & $0.75\asec$& N \\                 
2007ea   & \RAtwozerozerosevenea  & \DECtwozerozerosevenea & 2008 May 30  & LDSS3 & VPH\_All           & 1200      & $1.0\asec$ & E\\             
2007ff   & \RAtwozerozerosevenff  & \DECtwozerozerosevenff & 2007 Dec. 17 & LDSS3 & VPH\_All           & 1200      & $0.75\asec$& N\\                  
2007gg   & \RAtwozerozerosevengg  & \DECtwozerozerosevengg & 2011 Sep. 06 & BC    & 300GPM             & 1200      & $1.0\asec$ & E\\                  
2007gl   & \RAtwozerozerosevengl  & \DECtwozerozerosevengl & 2011 Sep, 21 & IMACS & 300-17.5           & 1200      & $0.7\asec$ & E\\                  
2007hb   & \RAtwozerozerosevenhb  & \DECtwozerozerosevenhb & 2011 Nov. 19 & IMACS & 300-17.5           & 600       & $0.9\asec$ & E\\                  
2007hl   & \RAtwozerozerosevenhl  & \DECtwozerozerosevenhl & 2008 May 30  & LDSS3 & VPH\_All           & 900       & $1.0\asec$ & E\\                  
2007hn   & \RAtwozerozerosevenhn  & \DECtwozerozerosevenhn & 2008 May 30  & LDSS3 & VPH\_All           & 600       & $1.0\asec$ & N\\                  
2008ao   & \RAtwozerozeroeightao  & \DECtwozerozeroeightao & 2012 Jan. 01 & BC    & 300GPM             & 1800      & $1.0\asec$ & E\\                  
2008fi   & \RAtwozerozeroeightfi  & \DECtwozerozeroeightfi & 2012 Jan. 19 & BC    & 300GPM             & 1200      & $1.0\asec$ & E\\                  
2008gc   & \RAtwozerozeroeightgc  & \DECtwozerozeroeightgc & 2012 Jan. 18 & LDSS3 & VPH\_All           & 600       & $0.75\asec$& E\\                  
2008ik   & \RAtwozerozeroeightik  & \DECtwozerozeroeightik & 2012 Jan. 20 & LDSS3 & VPH\_All           & 600       & $0.75\asec$& E\\                  
2008im   & \RAtwozerozeroeightim  & \DECtwozerozeroeightim & 2011 Dec. 24 & BC    & 300GPM             & 500       & $1.5\asec$ & E\\                  
2008iu   & \RAtwozerozeroeightiu  & \DECtwozerozeroeightiu & 2011 Sep. 21 & IMACS & 300-17.5           & 1200      & $0.7\asec$ & E\\                  
2009hu   & \RAtwozerozeroninehu   & \DECtwozerozeroninehu  & 2012 Jan. 01 & BC    & 300GPM             & 1800      & $1.0\asec$ & E\\                  
2009jf   & \RAtwozerozeroninejf   & \DECtwozerozeroninejf  & 2011 Nov. 18 & IMACS & 300-17.5           & 600       & $0.7\asec$ & E\\                  
2009nl   & \RAtwozerozeroninenl   & \DECtwozerozeroninenl  & 2011 Sep. 20 & IMACS & 300-17.5           & 1200      & $0.9\asec$ & E\\                  
2010Q    & \RAtwozeroonezeroQ     & \DECtwozeroonezeroQ    & 2012 Jan. 18 & BC    & 300GPM             & 1800      & $1.0\asec$ & E\\                  
2010ah   & \RAtwozeroonezeroah    & \DECtwozeroonezeroah   & \tablenotemark{$\dagger$} & BC  & 300GPM  & 1500+1200 & $1.0\asec$ & E\\                  
2010am   & \RAtwozeroonezeroam    & \DECtwozeroonezeroam   & 2012 Jan. 01 & BC    & 300GPM             & 1800      & $1.0\asec$ & E\\                  
2010ay   & \RAtwozeroonezeroay    & \DECtwozeroonezeroay   & 2010 Apr. 11 & GMOS  & R400              & 1800 & $1.0\asec$& E\\                        
2010cn   & \RAtwozeroonezerocn    & \DECtwozeroonezerocn   & 2012 Jan. 18 & BC    & 300GPM             & 1200      & $1.0\asec$ & E\\             
2010lz   & 01:50:20.32            & -21:44:31.9            & 2012 Jan. 19 & LDSS3 & VPH\_All           & 1500      & $0.75\asec$ & E\\                 
2011D    & \RAtwozerooneoneD      & \DECtwozerooneoneD     & 2011 Nov. 29 & IMACS & 300-17.5           & 500       & $0.7\asec$ & E\\                  
2011V    & \RAtwozerooneoneV      & \DECtwozerooneoneV     & 2012 Jan. 01 & BC    & 300GPM             & 1800      & $1.0\asec$ & E\\                  
2011bv   & \RAtwozerooneonebv     & \DECtwozerooneonebv    & 2012 Jan. 19 & LDSS3 & VPH\_All           & 1200      & $0.75\asec$& E\\                  
2011cs   & \RAtwozerooneonecs     & \DECtwozerooneonecs    & 2012 Jan. 19 & BC    & 300GPM             & 1500      & $1.0\asec$ & E\\                  
2011gh   & \RAtwozerooneonegh     & \DECtwozerooneonegh    & 2012 Jan. 18 & BC    & 300GPM             & 600       & $1.0\asec$ & E\\                  
2011hw   & \RAtwozerooneonehw     & \DECtwozerooneonehw    & 2011 Nov. 29 & BC    & 300GPM             & 600       & $1.0\asec$ & E\\                  
2011ip   & 1:13:47.59             & -12:41:06.0            & 2011 Dec. 31 & BC    & 300GPM             & 1800      & $1.0\asec$ & E\\                  
2011it   & \RAtwozerooneoneit     & \DECtwozerooneoneit    & 2011 Dec. 31 & BC    & 300GPM             & 600       & $1.0\asec$ & E\\                  
LSQ11JW  & \RALSQoneoneJW         & \DECLSQoneoneJW        & 2011 Dec. 31 & BC    & 300GPM             & 1200      & $1.0\asec$ & E\\                  
PTF09q   & \RAPTFzeronineq        & \DECPTFzeronineq       & 2012 Jan. 18 & BC    & 300GPM             & 600       & $1.0\asec$ & E\\                  
PTF09dfk & \RAPTFzeroninedfk      & \DECPTFzeroninedfk     & 2011 Sep. 21 & IMACS & 300-17.5           & 900       & $0.7\asec$ & E\\                  
PTF09dxv & \RAPTFzeroninedxv      & \DECPTFzeroninedxv     & 2011 Nov. 18 & IMACS & 300-17.5           & 500       & $0.7\asec$ & N\\                  
PTF09iqd & \RAPTFzeronineiqd      & \DECPTFzeronineiqd     & 2012 Jan. 18 & BC    & 300GPM             & 500       & $1.0\asec$ & N\\                  
PTF10aavz& \RAPTFonezeroaavz      & \DECPTFonezeroaavz     & 2011 Jan. 13 & LDSS3 & VPH\_All           & 1200      & $0.75\asec$& E\\                  
PTF10bip & \RAPTFonezerobip       & \DECPTFonezerobip      & 2012 Jan. 01 & BC    & 300GPM             & 1800      & $1.0\asec$ & E\\                  
PTF10vgv & \RAPTFonezerovgv       & \DECPTFonezerovgv      & 2011 Sep. 06 & BC    & 300GPM             & 1200      & $1.0\asec$ & E\\                  
PTF11hyg & \RAPTFoneonehyg        & \DECPTFoneonehyg       & 2011 Sep. 06 & BC    & 300GPM             & 1200      & $1.0\asec$ & E\\                  
\enddata

\tablenotetext{a}{The grism or grating used in the spectroscopic observation. When multiple configurations were used, we list both, separated with a slash.}
\tablenotetext{b}{The exposure time of the spectroscopic observations.  When two times are given separated by a slash, they correspond to two different dispersers.} 
\tablenotetext{c}{E indicates a slit position on the SN explosion site, N indicates slit placement at the galaxy center (see Section~\ref{sec:spectra})}
\tablenotetext{d}{Objects excluded from the sample due to AGN contamination; see Section~\ref{sec:fluxmea}.}
\tablenotetext{$\dagger$}{Spectra acquired on 2011 May 06 and 2012 Jan. 18 were coadded.}
\end{deluxetable*}

\label{sec:typing}

For those SNe in our sample whose classification spectroscopy is not well discussed in the circulars or the literature, we confirm the spectroscopic classification by contacting the authors of the classification circular to re-evaluate their original spectra using e.g., the Supernova Identification tool \citep{Blondin07} (see private communications in Table~\ref{tab:prop}).  We construct a ``Gold'' sample of \inGold{} SNe for which the classification spectrum suggests a clear subtype (but see Section~\ref{sec:classeffect}); a ``Silver'' sample of \inSilver{} SN where the classification is based on a spectrum with poor S/N.  SNe where we are not able to distinguish between two sub-types are listed as "Undetermined Ibc."  For the purposes of this study, we consider the peculiar object SN~2011hw (classified as Type Ibn; \citealt{Smith12}) to be of indeterminate type, we consider SN~2006lc (whose He lines were weak) a Type~Ib \citep{Leloudas11}, and we group the peculiar Type~IIb SNe~2002gz and 2010cn together with the other Type IIb SNe.  

\subsection{Spectroscopic Observations}
\label{sec:spectra}

\begin{figure*}
\plotone{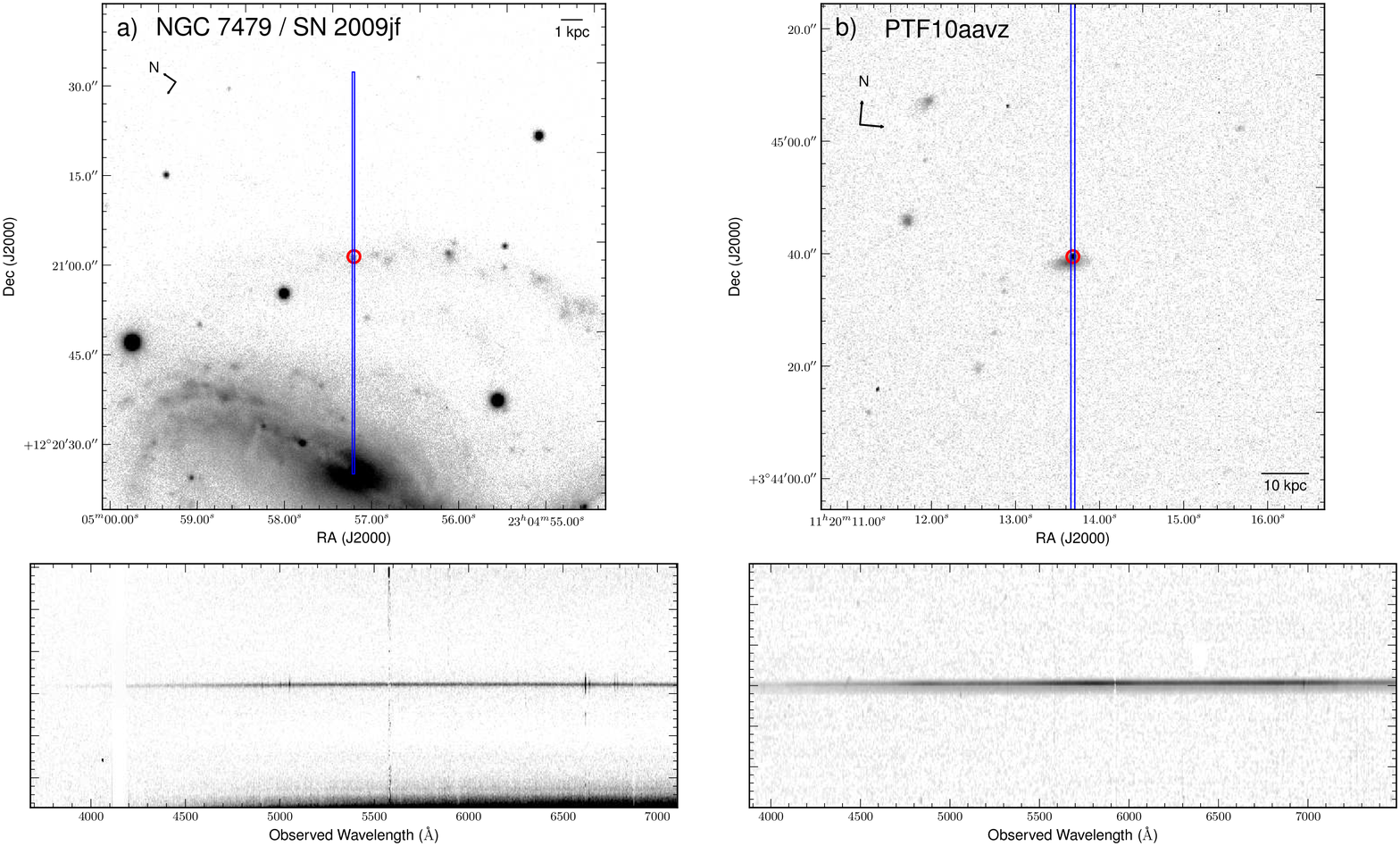}
\caption{\label{fig:twod}Example of optical spectroscopy procedure for two SN~Ibc host galaxies in our sample.  Shown are the $r$-band acquisition images (top) and 2D-spectrum (bottom) obtained using Magellan/IMACS and LDSS3 (see Table~\ref{tab:obs}) for a) the Type~Ib SN~2009jf host environment in the galaxy NGC~7479 ($z=\ztwozerozeroninejf$) and b) the Type~Ic-BL PTF10aavz ($z=\zPTFonezeroaavz$).  The slit (blue rectangle) was aligned with the SN explosion site (red circle).  The IMACS chip gap is visible in the spectrum at $\sim4100$~\AA.}
\end{figure*}

We obtained acquisition images in $r$-band and long-slit spectra ($\sim3600-8500$~\AA) of \fromLDSS\ SN~Ibc host galaxies using the Low Dispersion Survey Spectrograph 3 (LDSS3) instrument on the Magellan-Clay Telescope, \fromBC\ using the BlueChannel Spectrograph (BC) on the MMT (for which there is no associated imager), \fromIMACS\ using the Inamori-Magellan Areal Camera \& Spectrograph (IMACS) instrument on the Magellan-Baade Telescope, and 1 using the Gemini Multi-Object Spectrograph (GMOS) instrument on the Gemini-North Telescope.  The host galaxies were observed at the parallactic angle, except for some IMACS spectra which were obtained with the atmospheric dispersion compensator.

When possible, the spectrum was extracted at the location of the SN explosion site within the host galaxy.  We consider the spectrum to sample the explosion site (``E''; \inExp{} SNe) when we extract at the position of the explosion site and the slit width corresponds to a physical size $\lesssim2$~kpc ($z\leq0.11$ for a 1\asec\ slit), and otherwise consider the spectrum nuclear (``N''; \inNuc{} SNe, see Table~\ref{tab:obs}).  Explosion site spectroscopy represents a luminosity-weighted average of the physical properties of the star-forming gas within the extracted region.  However, because our SN sample consists of events discovered by galaxy-impartial surveys, the host galaxies are typically smaller and more distant than those observed in previous studies.  In some cases, the explosion sites of SNe in the outskirts of their host galaxies did not have sufficient flux for spectroscopy and instead we extract at the galaxy nucleus.  Some intrinsically-dim host galaxies have apparent sizes so small that a significant fraction of all the galaxy light will fall in the slit, even at relatively-low redshift (see Figure~\ref{fig:twod}).  For comparison, the explosion site spectroscopy of \cite{Modjaz11} includes host galaxies at a maximum redshift of $z=0.183$ using a $1\asec$ slit ($\sim3$~kpc) and the studies of \cite{Prieto08} and \cite{Kelly11} employ $3\asec$ SDSS fiber spectroscopy for SN~Ibc at $z\lesssim0.04$ ($\sim2$~kpc) and 0.07 ($\sim4$~kpc), respectively.

\subsection{Spectroscopic analysis}
\label{sec:specreduce}

\begin{figure*}
\plotone{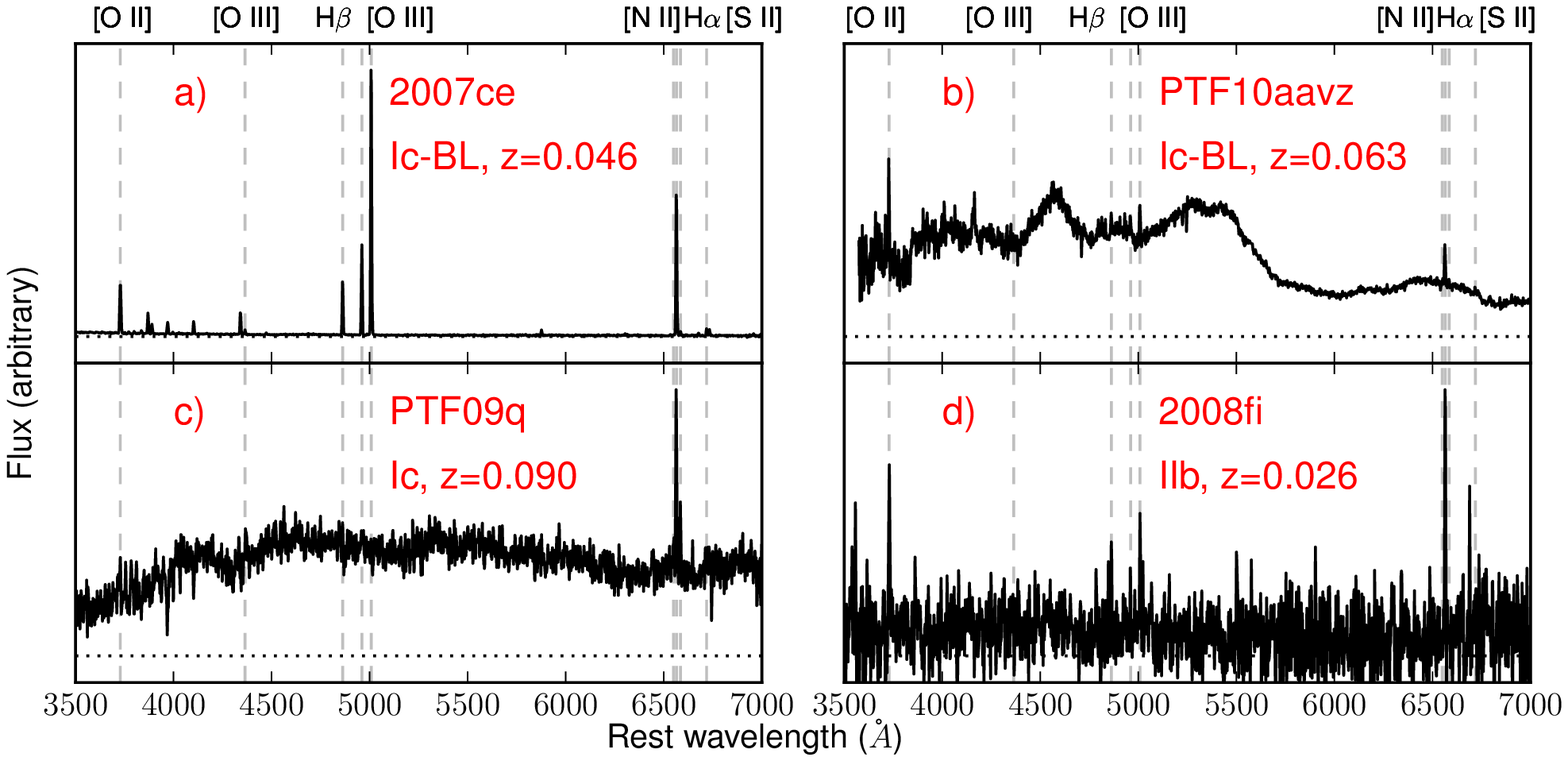}
\caption{\label{fig:onedfour}Examples of spectra from our SN~Ibc host environment observations.  The panels are labeled with the names, classifications, and redshifts of the SNe and represent a) a high-S/N spectrum dominated by nebular emission from the host environment, b) a spectrum dominated by SN flux, c) a spectrum dominated by stellar continuum, and d) a low S/N spectrum dominated by nebular emission from the host environment.  The gray vertical lines illustrate the position of the major nebular emission features used in the metallicity analysis.  The continuum level can be judged against the dashed horizontal line, illustrating zero flux.}
\end{figure*}

We employed standard two-dimensional long-slit image reduction and spectral extraction routines in IRAF\footnote{IRAF is distributed by the National Optical Astronomy Observatory, which is operated by the Association of Universities for Research in Astronomy (AURA) under cooperative agreement with the National Science Foundation.}. Examples of reduced spectra are displayed in Figure~\ref{fig:onedfour}.  

\label{sec:fluxmea}

The line fluxes of prominent nebular emission lines (H$\alpha$ and H$\beta$; [\ion{O}{2}]~$\lambda3727$\footnote{The [\ion{O}{2}] $\lambda\lambda3726,3729$ doublet is not resolved in these spectra, and we refer to the sum we effectively measure as [\ion{O}{2}] $\lambda3727$.}, [\ion{O}{3}]~$\lambda4363$, [\ion{O}{3}]~$\lambda4959$, [\ion{O}{3}]~$\lambda5007$, [\ion{N}{2}]~$\lambda\lambda6548$, 6584; and [\ion{S}{2}]~$\lambda\lambda6717$, 6731) were measured by fitting Gaussian functions to their profiles using a Markov Chain Monte Carlo (MCMC) technique \citep{pymc}. We fit the profiles to a wavelength range $20$ \AA{} in width centered on the rest frame wavelength of each line.  We fit a linear continuum to $20$~\AA{} regions of the spectra off the wings of each line and use this continuum measurement to calculate equivalent widths for spectral lines.  We constrain the amplitude of the Gaussian to be positive and fit a single redshift and line width for the set of Balmer lines and an independent redshift and width for the forbidden lines.  We adopt the fitted redshift of the Balmer lines as the redshift of the host galaxy.  We obtain estimates of the uncertainty in the line flux from the MCMC trace and require a detection confidence of 99\%\footnote{Assuming a normal distribution, the 99\% confidence interval corresponds to the requirement that the median value be $2.576\sigma$ greater than zero.  Here we calculate $\sigma$ as the difference between the 50th and 16th percentile values of the amplitude distribution in the MCMC trace.}.

The line fluxes measured for each host environment are presented in Table~\ref{tab:lines}.  We use observations of standard stars from the same night as the host environment observations to achieve relative flux calibration.  No correction for underlying stellar absorption has been made, which could potentially affect the measurement of the Balmer line fluxes (particularly H$\beta$) for galaxies with significant stellar continuum flux.  Only 5 of \NSanders\ spectra (those of SN~2002ex, 2006ip, 2007I, 2007ff, and PTF10vgv) have stellar continuum levels that may indicate significant underlying absorption, and we therefore do not measure quantities that depend on H$\beta$ for these objects, including extinction ($A_V$).

We tested for AGN contamination using a \cite{Baldwin81} excitation-mechanism diagnostic diagram and the classification scheme of \citep{Kauffmann03}.  We found emission lines from the host galaxies to be consistent with typical star-forming galaxies with few exceptions.  The host environments of SN~1991D and 2003jp were excluded from our sample (and all analysis below) due to significant AGN contamination that would bias certain metallicity diagnostics \citep{KE08}.  We neglect the potential effects of AGN contamination for SN~2006jo (Type~Ib) and PTF10xla (Ib/c), whose host environments show evidence for a composite classification.  Some objects have supernova flux contamination (2010ay, 2011bv, 2011gh, 2011ip, LSQ11JW, PTF10aav), although the broad SN absorption features do not affect the measurement of the flux in the narrow galaxy emission lines.

\section{HOST ENVIRONMENT PROPERTIES}
\label{sec:res}

From our optical spectra we measure the magnitude of dust extinction, metallicity, young stellar population ages, and Wolf-Rayet star populations of the host environments of the SN~Ibc in our sample.  We discuss each of these in the following sections.

\subsection{Dust extinction}
\label{sec:AVdist}

We estimate the line-of-sight extinction for each SN host galaxy spectrum (including both Galactic and intrinsic reddening) from the Balmer flux decrement. We assume $\mbox{H}\alpha/\mbox{H}\beta=2.85$, which corresponds to $T=10,000$~K and $n_e=10^4~\mbox{cm}^{-3}$ for Case B recombination \citep{oandf}.  The extinction curve of \cite{cardelli89} was applied to correct individual line fluxes for reddening, assuming $R_V=3.1$.  The value of the extinction derived for each host galaxy is presented in Table~\ref{tab:lines}.

\begin{deluxetable*}{p{33pt}lllllllllll}
\tablecaption{Nebular Emission Line Fluxes From SN Host Galaxies\label{tab:lines}}
\tabletypesize{\scriptsize}
\tablehead{\colhead{SN} & \colhead{$A_V$} & \colhead{[\ion{O}{2}]} & \colhead{[\ion{O}{3}]} & \colhead{H$\beta$}  & \colhead{[\ion{O}{3}]} & \colhead{[\ion{O}{3}]} & \colhead{[\ion{N}{2}]} & \colhead{H$\alpha$} & \colhead{[\ion{N}{2}]} & \colhead{[\ion{S}{2}]} & \colhead{[\ion{S}{2}]} \\ 
      && \colhead{$\lambda$ 3727} & \colhead{$\lambda$ 4363} & \colhead{$\lambda$ 4861} & \colhead{$\lambda$ 4959} & \colhead{$\lambda$ 5007} & \colhead{$\lambda$ 6548} & \colhead{$\lambda$ 6562} & \colhead{$\lambda$ 6584} & \colhead{$\lambda$ 6717} & \colhead{$\lambda$ 6731}}
\startdata
\multicolumn{11}{l}{\textit{SN Type: IIb}} \\
2007br & \nodata & $170\pm30$ & \nodata & \nodata & \nodata & $30\pm10$ & \nodata & $100\pm30$ & $38\pm8$ & $45\pm8$ & $34\pm7$\\
2007ea & $0.8^{+0.2}_{-0.3}$ & $350\pm6$ & \nodata & $100\pm2$ & $84\pm5$ & $250\pm10$ & $17\pm1$ & $360\pm20$ & $51\pm3$ & $54\pm3$ & $36\pm2$\\
2010am & $0.0^{+0.0}_{-0.0}$ & $373\pm3$ & \nodata & $100\pm2$ & $104\pm7$ & $290\pm20$ & $11\pm2$ & $220\pm20$ & $35\pm3$ & $49\pm4$ & $32\pm3$\\
2010cn & $0.0^{+0.2}_{-0.0}$ & $300\pm10$ & \nodata & $100\pm7$ & $104\pm10$ & $290\pm30$ & \nodata & $270\pm30$ & $27\pm5$ & $45\pm6$ & $23\pm5$\\
2011D & $5.6^{+1.1}_{-1.0}$ & \nodata & \nodata & $100\pm30$ & $60\pm20$ & $200\pm50$ & $120\pm30$ & $1500\pm400$ & $130\pm40$ & $220\pm70$ & $140\pm40$\\
2011V & $0.0^{+0.6}_{-0.0}$ & $140\pm20$ & \nodata & $100\pm40$ & \nodata & \nodata & $80\pm20$ & $300\pm100$ & $100\pm20$ & $80\pm20$ & $80\pm20$\\
PTF09dxv & $5.0^{+0.8}_{-0.7}$ & \nodata & \nodata & $100\pm40$ & \nodata & \nodata & $140\pm30$ & $1200\pm400$ & $430\pm60$ & $320\pm50$ & $170\pm30$\\
\hline\\
\sidehead{\textit{SN Type: Ib}}
2002ex & \nodata & \nodata & \nodata & \nodata & $5.8\pm0.2$ & $20.4\pm0.3$ & $7.2\pm0.6$ & $100\pm10$ & $22\pm2$ & $25\pm2$ & $18\pm1$\\
2004cf & $1.3^{+0.5}_{-0.6}$ & \nodata & \nodata & $100\pm10$ & $110\pm10$ & $300\pm20$ & \nodata & $410\pm70$ & $59\pm8$ & \nodata & $150\pm20$\\
2005hm & $0.0^{+0.5}_{-0.0}$ & \nodata & $20\pm6$ & $100\pm4$ & $180\pm20$ & $590\pm50$ & \nodata & $290\pm50$ & $27\pm8$ & $74\pm8$ & $47\pm7$\\
2006jo & $0.2^{+0.4}_{-0.2}$ & $450\pm10$ & \nodata & $100\pm5$ & $110\pm10$ & $200\pm20$ & $55\pm6$ & $310\pm40$ & $190\pm20$ & $56\pm6$ & $46\pm5$\\
2006lc & $3.1^{+0.3}_{-0.3}$ & \nodata & \nodata & $100\pm20$ & \nodata & \nodata & $60\pm10$ & $700\pm100$ & $200\pm20$ & $76\pm6$ & $58\pm5$\\
2007az & $0.0^{+0.0}_{-0.0}$ & $215\pm3$ & \nodata & $100\pm2$ & $126\pm7$ & $400\pm20$ & $6\pm1$ & $260\pm20$ & $12\pm2$ & $35\pm3$ & $17\pm2$\\
2007gg & $0.7^{+0.6}_{-0.6}$ & $310\pm10$ & \nodata & $100\pm20$ & $100\pm20$ & $240\pm30$ & \nodata & $350\pm50$ & $80\pm10$ & $120\pm20$ & $70\pm10$\\
2008gc & $0.4^{+0.3}_{-0.3}$ & $290\pm20$ & \nodata & $100\pm5$ & $80\pm7$ & $220\pm20$ & $26\pm3$ & $320\pm30$ & $30\pm3$ & $52\pm4$ & $37\pm4$\\
2009hu & \nodata & $12\pm5$ & \nodata & \nodata & \nodata & \nodata & $22\pm7$ & $100\pm20$ & $34\pm9$ & \nodata & \nodata\\
2009jf & $5.1^{+0.7}_{-0.6}$ & \nodata & \nodata & $100\pm10$ & $150\pm20$ & $190\pm20$ & $160\pm30$ & $1300\pm200$ & $310\pm40$ & $110\pm20$ & $140\pm30$\\
LSQ11JW & $0.0^{+0.1}_{-0.0}$ & $300\pm8$ & \nodata & $100\pm8$ & $50\pm6$ & $150\pm10$ & \nodata & $260\pm30$ & $44\pm6$ & $40\pm6$ & \nodata\\
PTF09dfk & $3.5^{+0.5}_{-0.5}$ & $79\pm6$ & \nodata & $100\pm4$ & $110\pm10$ & $350\pm30$ & $30\pm4$ & $800\pm100$ & $110\pm10$ & $170\pm20$ & $130\pm10$\\
\hline\\
\sidehead{\textit{SN Type: Ic}}
2003ev & \nodata & \nodata & \nodata & \nodata & \nodata & $30\pm7$ & $15\pm5$ & $100\pm30$ & $62\pm7$ & $36\pm5$ & $22\pm5$\\
2004ib & $2.4^{+0.3}_{-0.3}$ & $630\pm20$ & \nodata & $100\pm6$ & $73\pm7$ & $230\pm10$ & $25\pm5$ & $570\pm40$ & $108\pm8$ & $128\pm9$ & $93\pm6$\\
2006ip & \nodata & $42\pm2$ & \nodata & \nodata & $2.4\pm0.8$ & $3.9\pm0.9$ & $9.8\pm0.9$ & $100\pm10$ & $32\pm2$ & $17\pm1$ & $14\pm1$\\
2006ir & $0.8^{+0.2}_{-0.3}$ & $190\pm10$ & \nodata & $100\pm3$ & $79\pm6$ & $260\pm20$ & $12\pm3$ & $360\pm20$ & $39\pm3$ & $54\pm4$ & $34\pm3$\\
2006tq & \nodata & $160\pm10$ & \nodata & \nodata & $50\pm10$ & $100\pm20$ & \nodata & $100\pm20$ & $80\pm30$ & \nodata & $90\pm30$\\
2007db & $0.9^{+0.4}_{-0.4}$ & $600\pm20$ & \nodata & $100\pm5$ & $76\pm7$ & $220\pm10$ & $34\pm6$ & $370\pm40$ & $63\pm6$ & $92\pm7$ & $66\pm6$\\
2007ff & \nodata & $30\pm10$ & \nodata & \nodata & \nodata & \nodata & \nodata & $100\pm20$ & $63\pm8$ & $13\pm3$ & $13\pm3$\\
2007gl & \nodata & \nodata & \nodata & \nodata & \nodata & \nodata & $20\pm10$ & $100\pm100$ & $90\pm20$ & $30\pm10$ & $40\pm10$\\
2007hb & $2.1^{+0.4}_{-0.4}$ & \nodata & \nodata & $100\pm20$ & $17\pm4$ & $17\pm4$ & $44\pm9$ & $500\pm100$ & $160\pm20$ & \nodata & \nodata\\
2007hl & $2.5^{+0.2}_{-0.2}$ & $460\pm10$ & \nodata & $100\pm5$ & $41\pm4$ & $91\pm5$ & $59\pm4$ & $600\pm40$ & $159\pm9$ & $137\pm7$ & $93\pm5$\\
2007hn & \nodata & $140\pm40$ & \nodata & \nodata & \nodata & \nodata & $33\pm8$ & $100\pm40$ & $91\pm9$ & $57\pm6$ & $36\pm5$\\
2008ao & $1.5^{+0.3}_{-0.3}$ & $97\pm1$ & \nodata & $100\pm3$ & $12\pm1$ & $37\pm2$ & $50\pm4$ & $450\pm40$ & $150\pm10$ & $74\pm6$ & $58\pm5$\\
2008ik & $1.9^{+0.6}_{-0.5}$ & $370\pm10$ & \nodata & $100\pm20$ & \nodata & \nodata & $90\pm10$ & $500\pm100$ & $370\pm30$ & $170\pm20$ & $130\pm10$\\
2009nl & $3.8^{+0.8}_{-0.7}$ & \nodata & \nodata & $100\pm20$ & \nodata & \nodata & \nodata & $900\pm200$ & $120\pm40$ & $150\pm30$ & \nodata\\
2010Q & $0.0^{+0.1}_{-0.0}$ & $212\pm5$ & $10\pm3$ & $100\pm4$ & $160\pm10$ & $480\pm40$ & $10\pm2$ & $260\pm30$ & $12\pm3$ & $55\pm5$ & $18\pm2$\\
2011it & $0.7^{+0.5}_{-0.5}$ & $260\pm10$ & \nodata & $100\pm9$ & $33\pm6$ & $110\pm10$ & $45\pm6$ & $350\pm50$ & $100\pm10$ & $96\pm10$ & $69\pm8$\\
PTF09iqd & \nodata & $140\pm20$ & \nodata & \nodata & \nodata & \nodata & $40\pm10$ & $100\pm30$ & $50\pm20$ & $70\pm20$ & $40\pm10$\\
PTF09q & \nodata & $10\pm3$ & \nodata & \nodata & \nodata & \nodata & $10\pm2$ & $100\pm10$ & $21\pm3$ & \nodata & \nodata\\
PTF10bip & $0.6^{+0.3}_{-0.4}$ & $393\pm5$ & \nodata & $100\pm4$ & $85\pm7$ & $250\pm20$ & $18\pm3$ & $340\pm30$ & $34\pm4$ & $72\pm6$ & $46\pm5$\\
PTF10vgv & \nodata & $35.5\pm0.8$ & \nodata & \nodata & $8\pm1$ & $22\pm2$ & $9\pm1$ & $100\pm10$ & $30\pm3$ & $24\pm2$ & $19\pm2$\\
PTF11hyg & $1.2^{+0.6}_{-0.6}$ & $144\pm10$ & \nodata & $100\pm30$ & \nodata & $50\pm10$ & $80\pm10$ & $400\pm100$ & $200\pm20$ & $70\pm10$ & $51\pm9$\\
\hline\\
\sidehead{\textit{SN Type: Ic-BL}}
2005nb & $1.0^{+0.5}_{-0.4}$ & \nodata & \nodata & $100\pm10$ & $50\pm10$ & $140\pm20$ & \nodata & $380\pm40$ & $70\pm10$ & $70\pm10$ & $50\pm10$\\
2006nx & $0.1^{+0.6}_{-0.1}$ & $600\pm20$ & \nodata & $100\pm4$ & $120\pm10$ & $350\pm40$ & $17\pm5$ & $290\pm60$ & $38\pm7$ & $80\pm10$ & $33\pm6$\\
2007I & \nodata & $130\pm10$ & \nodata & \nodata & $7\pm3$ & $29\pm4$ & \nodata & $100\pm10$ & $23\pm3$ & $41\pm4$ & $20\pm3$\\
2007ce & $0.0^{+0.2}_{-0.0}$ & $119\pm1$ & $9.8\pm0.8$ & $100.0\pm1.0$ & $200\pm10$ & $550\pm20$ & $2.5\pm0.4$ & $270\pm30$ & $7.4\pm0.7$ & $14\pm1$ & $12\pm1$\\
2008iu & $0.0^{+0.0}_{-0.0}$ & $110\pm7$ & $23\pm4$ & $100\pm2$ & $210\pm20$ & $640\pm70$ & $13\pm2$ & $230\pm40$ & $11\pm1$ & $12\pm2$ & \nodata\\
2010ah & $0.7^{+0.5}_{-0.5}$ & $380\pm20$ & \nodata & $100\pm10$ & $60\pm10$ & $180\pm20$ & \nodata & $350\pm40$ & $33\pm9$ & $70\pm10$ & $43\pm10$\\
2010ay & $0.0^{+0.0}_{-0.0}$ & $233\pm7$ & $4.6\pm0.6$ & $100.0\pm0.6$ & $170\pm20$ & $582\pm4$ & $7.7\pm0.6$ & $250\pm30$ & $25\pm2$ & $29\pm3$ & $22\pm2$\\
\hline\\
\sidehead{\textit{SN Type: Undetermined Ibc}}
1991R & $1.6^{+0.3}_{-0.3}$ & $560\pm30$ & \nodata & $100\pm6$ & $49\pm6$ & $180\pm10$ & $41\pm5$ & $460\pm30$ & $106\pm8$ & $77\pm6$ & $54\pm6$\\
2011gh & $2.0^{+0.4}_{-0.4}$ & $193\pm7$ & \nodata & $100\pm10$ & $12\pm4$ & $48\pm6$ & $56\pm7$ & $520\pm80$ & $190\pm20$ & $93\pm9$ & $72\pm7$\\
2011hw & \nodata & $40\pm10$ & \nodata & \nodata & \nodata & \nodata & $23\pm7$ & $100\pm20$ & $61\pm9$ & $40\pm10$ & $50\pm10$\\
\enddata

\tablecomments{$A_V$ has been derived from the Balmer decrement as described in the test, but fluxes reported here have not been dereddened.  Fluxes are reported relative to H$\beta=100$ when possible; H$\alpha$ or oxygen lines are used for normalization when necessary.}
\end{deluxetable*}

\begin{figure}
\plotone{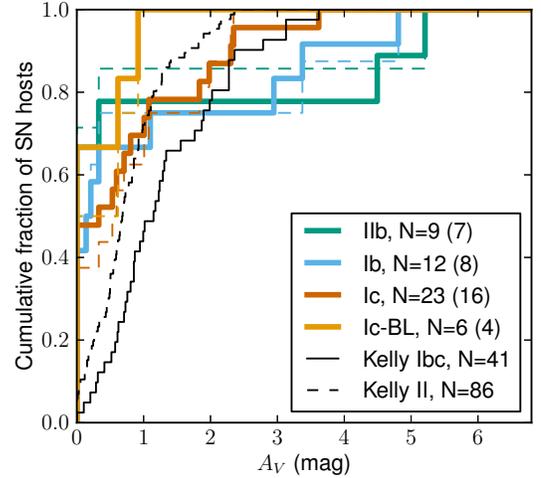}
\caption{\label{fig:AVdist}Cumulative distribution of visual-band extinction ($A_V$) measured for host environments of SN of different types from the spectra collected in our sample.  The dashed lines illustrate the same distributions, for only explosion-site spectroscopy.  The number of objects included in each sample is given in the legend, with explosion-site only numbers in parentheses.  The distributions measured by \cite{Kelly11} for SDSS fiber spectra are shown in black.}
\end{figure}

In Figure~\ref{fig:AVdist}, we show the distribution of visual-band extinction ($A_V$) derived for the SN host environments from the Balmer decrement, for all spectra (solid lines) and for only explosion-site spectroscopy (dashed lines).  Galactic extinction has been subtracted using the infrared dust map of \cite{SF11}.  The difference introduced by restricting the sample to explosion sites is small.

Using the Kolmogorov-Smirnov (KS) test, we do not identify a statistical difference between the extinction distributions of SN~Ib and Ic (with KS $p$-value, $\ksp=\KSAvIbIc$).  Combining the \AVNIb\ SN~Ib and \AVNIc\ SN~Ic with $A_V$ measurements and subtracting Galactic extinction, the median and 14th and 86th percentile values ($1\sigma$) are $A_V=\AvIbIcStd$~mag.  The median extinction values for the SN~IIb and Ic-BL are consistent with 0, but given the small sample sizes ($N=\AVNIIb$ and \AVNIcBL, respectively), the difference from the SN~Ib+Ic distribution is not significant ($\ksp\gg0.1$). 

In Figure~\ref{fig:AVdist} we also show the $A_V$ distribution measured by \cite{Kelly11} for SN~II and for the combination of SN~Ib and Ic.  A consistent extinction distribution was estimated for 19 SN~Ibc by \cite{Drout11} based on light curve colors.  For SN~Ibc, \cite{Kelly11} find median and $1\sigma$ values of $A_V=1.1^{+1.1}_{-0.6}$~mag, with smaller values for SN~II.  We find $A_V$ values with a more extreme range, with some SNe having no measurable redenning and some as large as $A_V\approx5$~mag.  This discrepancy is likely due to methodological differences.    Because they employ $3\asec$ SDSS fibers, \cite{Kelly11} probe gas in the nuclear region of the host galaxy, whereas most of our spectra are from the SN explosion site where the line of sight could be significantly different.  However, in some cases the discovery magnitudes reported for the SNe do not allow for extinctions as large as we measure for the host environment\footnote{e.g. SN~2011D, discovered at $m=18.2$~mag by \citealt{CBET2627}, implying $M<-20$ adopting the Balmer decrement extinction}.  This implies that the line of sight to the SN may be substantially different than that represented by the integrated light from the star forming gas we observe.

\subsection{Metallicity Estimation}
\label{sec:metalmea}

Host galaxy oxygen abundances were calculated from the extinction-corrected line flux ratios according to a number of independently calibrated abundance diagnostics, described below.  We report the abundance derived from each method in Table~\ref{tab:diag}.

\begin{deluxetable*}{llllllll}
\tablecaption{Metallicities and Ages of SN Host Environments\label{tab:diag}}
\tabletypesize{\scriptsize}
\tablehead{\colhead{SN Host Galaxy} & \multicolumn{6}{c}{log(O/H)+12} & \colhead{Age\tablenotemark{a}}\\
 & \colhead{Direct} & \colhead{Z94} & \colhead{KD02} & \colhead{PP04N2} & \colhead{PP04O3N2} & \colhead{PT05} & \colhead{(Myr)}}

\startdata
\multicolumn{7}{l}{\textit{SN Type: IIb}} \\
2007br & \nodata &\nodata &\nodata &$8.66\pm0.09$ &\nodata &\nodata & \nodata\\
2007ea & \nodata &$8.46\pm0.04$ &$8.55\pm0.07$ &$8.42\pm0.02$ &$8.33\pm0.01$ &\nodata & $5.2\pm0.4$\\
2010am & \nodata &$8.47\pm0.03$ &$8.58\pm0.03$ &$8.45\pm0.03$ &$8.33\pm0.02$ &$8.16\pm0.02$ & $6.0\pm0.5$\\
2010cn & \nodata &$8.54\pm0.07$ &$8.54\pm0.08$ &$8.33\pm0.05$ &$8.26\pm0.03$ &$8.26\pm0.06$ & $6.3\pm0.4$\\
2011D & \nodata &\nodata &\nodata &$8.30\pm0.10$ &$8.33\pm0.06$ &\nodata & $5.4\pm0.3$\\
2011V & \nodata &\nodata &$9.04\pm0.07$ &$8.6\pm0.1$ &\nodata &\nodata & $8.4\pm0.9$\\
PTF09dxv & \nodata &\nodata &\nodata &$8.64\pm0.09$ &\nodata &\nodata & $8.3\pm0.9$\\
\hline\\
\sidehead{\textit{SN Type: Ib}}
2002ex & \nodata &\nodata &\nodata &$8.52\pm0.04$ &\nodata &\nodata & \nodata\\
2004cf & \nodata &\nodata &\nodata &$8.42\pm0.05$ &$8.31\pm0.03$ &\nodata & $8.1\pm0.6$\\
2005hm & \nodata &\nodata &\nodata &$8.32\pm0.09$ &$8.16\pm0.05$ &\nodata & $7.3\pm0.4$\\
2006jo & \nodata &\nodata &$8.93\pm0.05$ &$8.79\pm0.04$ &$8.57\pm0.02$ &\nodata & $8.4\pm0.9$\\
2006lc & \nodata &\nodata &\nodata &$8.59\pm0.05$ &\nodata &\nodata & $8.2\pm0.8$\\
2007az & \nodata &$8.50\pm0.03$ &\nodata &$8.15\pm0.04$ &$8.12\pm0.02$ &$8.3\pm0.2$ & $5.4\pm0.2$\\
2007gg & \nodata &\nodata &$8.7\pm0.1$ &$8.52\pm0.05$ &$8.40\pm0.03$ &\nodata & $9.0\pm0.8$\\
2008gc & \nodata &$8.62\pm0.07$ &$8.51\pm0.08$ &$8.31\pm0.03$ &$8.29\pm0.02$ &$8.25\pm0.08$ & $6.0\pm0.3$\\
2009hu & \nodata &\nodata &\nodata &$8.64\pm0.09$ &\nodata &\nodata & \nodata\\
2009jf & \nodata &\nodata &\nodata &$8.55\pm0.06$ &$8.47\pm0.03$ &\nodata & $9.0\pm0.8$\\
LSQ11JW & \nodata &$8.78\pm0.06$ &$8.71\pm0.04$ &$8.46\pm0.04$ &$8.43\pm0.03$ &$8.30\pm0.05$ & \nodata\\
PTF09dfk & \nodata &$8.58\pm0.08$ &$8.7\pm0.1$ &$8.42\pm0.04$ &$8.30\pm0.02$ &$8.32\pm0.08$ & $9.0\pm0.6$\\
\hline\\
\sidehead{\textit{SN Type: Ic}}
2003ev & \nodata &\nodata &\nodata &$8.78\pm0.08$ &\nodata &\nodata & \nodata\\
2004ib & \nodata &\nodata &\nodata &$8.49\pm0.03$ &$8.40\pm0.02$ &\nodata & $8.7\pm0.7$\\
2006ip & \nodata &\nodata &\nodata &$8.62\pm0.04$ &\nodata &\nodata & \nodata\\
2006ir & \nodata &$8.69\pm0.05$ &$8.66\pm0.06$ &$8.35\pm0.03$ &$8.29\pm0.02$ &$8.37\pm0.05$ & $5.7\pm0.3$\\
2006tq & \nodata &\nodata &\nodata &$8.9\pm0.1$ &\nodata &\nodata & \nodata\\
2007db & \nodata &\nodata &\nodata &$8.46\pm0.04$ &$8.38\pm0.02$ &\nodata & $7.9\pm0.6$\\
2007ff & \nodata &\nodata &\nodata &$8.78\pm0.07$ &\nodata &\nodata & \nodata\\
2007gl & \nodata &\nodata &\nodata &$8.9\pm0.2$ &\nodata &\nodata & \nodata\\
2007hb & \nodata &\nodata &\nodata &$8.61\pm0.06$ &$8.82\pm0.04$ &\nodata & $6.3\pm0.7$\\
2007hl & \nodata &\nodata &$8.43\pm0.09$ &$8.57\pm0.02$ &$8.57\pm0.01$ &\nodata & $8.2\pm0.8$\\
2007hn & \nodata &\nodata &\nodata &$8.88\pm0.10$ &\nodata &\nodata & \nodata\\
2008ao & \nodata &$9.13\pm0.02$ &$9.02\pm0.04$ &$8.63\pm0.03$ &$8.73\pm0.01$ &$8.44\pm0.04$ & $6.6\pm0.7$\\
2008ik & \nodata &\nodata &$8.87\pm0.09$ &$8.83\pm0.06$ &\nodata &\nodata & $8.4\pm0.9$\\
2009nl & \nodata &\nodata &\nodata &$8.4\pm0.1$ &\nodata &\nodata & $8.1\pm0.5$\\
2010Q & \nodata &\nodata &\nodata &$8.14\pm0.07$ &$8.09\pm0.04$ &$8.2\pm0.1$ & $5.3\pm0.2$\\
2011it & \nodata &$8.82\pm0.09$ &$8.85\pm0.08$ &$8.60\pm0.04$ &$8.55\pm0.02$ &$8.26\pm0.10$ & $8.1\pm0.8$\\
PTF09iqd & \nodata &\nodata &\nodata &$8.7\pm0.1$ &\nodata &\nodata & \nodata\\
PTF09q & \nodata &\nodata &\nodata &$8.52\pm0.05$ &\nodata &\nodata & \nodata\\
PTF10bip & \nodata &\nodata &\nodata &$8.33\pm0.04$ &$8.29\pm0.02$ &\nodata & $6.9\pm0.4$\\
PTF10vgv & \nodata &\nodata &\nodata &$8.61\pm0.04$ &\nodata &\nodata & \nodata\\
PTF11hyg & \nodata &\nodata &$9.03\pm0.08$ &$8.72\pm0.07$ &$8.72\pm0.04$ &\nodata & $7.9\pm0.9$\\
\hline\\
\sidehead{\textit{SN Type: Ic-BL}}
2005nb & \nodata &\nodata &\nodata &$8.46\pm0.05$ &$8.44\pm0.03$ &\nodata & $6.4\pm0.5$\\
2006nx & \nodata &\nodata &\nodata &$8.39\pm0.07$ &$8.28\pm0.04$ &\nodata & $6.7\pm0.5$\\
2007I & \nodata &\nodata &\nodata &$8.53\pm0.05$ &\nodata &\nodata & \nodata\\
2007ce & $7.92\pm0.05$ &\nodata &\nodata &$8.01\pm0.03$ &$7.99\pm0.02$ &$8.3\pm0.2$ & $3.8\pm0.1$\\
2008iu & \nodata &\nodata &$8.58\pm0.05$ &$8.14\pm0.05$ &$8.05\pm0.03$ &$8.27\pm0.04$ & $4.7\pm0.2$\\
2010ah & \nodata &$8.6\pm0.1$ &\nodata &$8.31\pm0.08$ &$8.32\pm0.05$ &\nodata & $8.0\pm0.4$\\
2010ay & $8.35\pm0.07$ &\nodata &$8.62\pm0.03$ &$8.33\pm0.04$ &$8.16\pm0.02$ &$8.18\pm0.01$ & \nodata\\
\hline\\
\sidehead{\textit{SN Type: Undetermined Ibc}}
1991R & \nodata &\nodata &$8.4\pm0.1$ &$8.54\pm0.02$ &$8.46\pm0.02$ &\nodata & $8.1\pm0.8$\\
2011gh & \nodata &$8.85\pm0.10$ &$8.85\pm0.06$ &$8.65\pm0.04$ &$8.70\pm0.02$ &\nodata & \nodata\\
2011hw & \nodata &\nodata &\nodata &$8.78\pm0.07$ &\nodata &\nodata & \nodata\\
\enddata

\tablecomments{The oxygen abundance, log(O/H)$+12$, is estimated using the following diagnostics, as described in the text: Direct (based on [\ion{O}{3}] $\lambda 4363$), Z94 \citep{Zaritsky94}, KD02 \citep{kewley02}, PP04N2 and PP04O3N2 \citep{PP04}, and PT05 \citep{PT05}.}
\tablenotetext{a}{The age of the young stellar population has been estimated based on the rest-frame equivalent width of the H$\beta$ line, following \cite{Levesque10} (see Section~\ref{sec:Yage}).}
\end{deluxetable*}

\label{sec:metalmeadiag}

We derive ``direct'' oxygen abundance estimates by estimating the electron temperature of the gas' dominant excitation zone, which is only possible if the [\ion{O}{3}] $\lambda4363$ line is detected.  Following \cite{Levesque10}, the electron temperatures are estimated using IRAF's five-level nebular modeling package \texttt{nebular} \citep{shaw94}.  The \texttt{nebular} task \texttt{temden} is first applied to iteratively estimate the O$^{++}$ temperature ($T_e(\rm{O}^{++})$) and density ($n_e$) of the nebula from the [\ion{O}{3}] and [\ion{S}{2}] line ratios, respectively.  If the measured line ratios correspond to unphysical conditions (outside the range for which \texttt{temden} is calibrated, $500<T_e(\rm{O}^{++})<10^5$~K and $1<n_e<10^8$~cm$^{-3}$), we do not calculate the direct abundance.  The O$^+$ temperature is then estimated using the linear empirical relation of \cite{Garnett92}.  The \ion{O}{2} and \ion{O}{3} abundances are then estimated using the density, ionic temperatures, and [\ion{O}{2}] and [\ion{O}{3}] line ratios following the ionization correction factor (ICF) prescription of \cite{Shi06}.  The total oxygen abundance is taken to be the sum of these two ionic abundances.  

We employ several strong line diagnostics chosen to represent each of the major classes calibrated in the literature: $R_{23}$, N2O2, N2, O3N2, and $P$ \citep[see][ for a recent review]{lopez2010}.  First, we apply the $R_{23}$ oxygen abundance calibration of \cite{Zaritsky94}, an average of three earlier methods, hereafter referred to as ``Z94.''  Z94 is only calibrated for the higher-metallicity upper branch of the $R_{23}$-abundance degeneracy ($8.4<$log(O/H)$+12<9.6$); in every case where we apply Z94, the [\ion{N}{2}]/[\ion{O}{2}] ratio suggests an upper branch solution.  Second, we apply the [\ion{N}{2}]/[\ion{O}{2}] oxygen abundance calibration of \cite{kewley02} (as updated by \citealt{KE08}), hereafter referred to as ``KD02.''  \cite{kewley02} synthesize a variety of modern photoionization models and observational calibrations to produce recommendations for producing an abundance estimate given different permutations of available emission lines and uses the [\ion{N}{2}]/[\ion{O}{2}] ratio to break the degeneracy between the upper and lower branches of $R_{23}$.  Third, we apply the empirical [\ion{N}{2}]/H$\alpha$ (``N2'') and [\ion{O}{3}]/[\ion{N}{2}] (``O3N2'')  oxygen abundance calibrations of \cite{PP04}, hereafter referred to as ``PP04.''  Fourth, we apply the excitation parameter (``$P$ method'') oxygen abundance calibration of \cite{PT05}, hereafter referred to as ``PT05.''  $P$ is calculated from the ratio of [\ion{O}{3}] to ([\ion{O}{2}]+[\ion{O}{3}]) \citep{P01}, and the [\ion{N}{2}]/[\ion{O}{2}] ratio is used to break the $R_{23}$ degeneracy.

\label{sec:metalmeacompare}

There are well-known offsets between the diagnostics which are particularly large between empirically and theoretically-calibrated diagnostics \citep[see e.g.][]{Stas02}.    However, the relative metallicity difference measured between a given pair of galaxies in different diagnostics is consistent with an rms scatter typically $\sim0.07$~dex, and $0.15$~dex between the most discrepant diagnostics \citep{KE08}. Comparing the metallicities we measure for the same host galaxies in different diagnostics, we find discrepancies consistent with the rms scatter reported by \cite{KE08}.  Hereafter we refer to this uncertainty intrinsic to the diagnostics as the ``systematic uncertainty;'' we do not factor the systematic uncertainty into the metallicities reported in Table~\ref{tab:diag}, but we do consider this systematic uncertainty in our statistical analysis (Section~\ref{sec:stat}).  The systematic uncertainty is typically as large as the ``statistical uncertainty'' associated with the line flux and $A_V$ measurement errors; for example, the median statistical uncertainty associated with our PP04N2 measurements is \diagPPzerofourNtwomedunc~dex.

The galaxy sample varies with the diagnostic chosen.  The PP04N2 diagnostic can be applied to nearly all galaxies in our sample ($N=\diagPPzerofourNtwoN$), while the similar PP04O3N2 diagnostic can only be applied to the \diagPPzerofourOthreeNtwoN{} galaxies which have measurements of the potentially fainter [\ion{O}{3}] and H$\beta$ lines.  The [PT05,Z94] diagnostics can only be applied to [\diagPTzerofiveN{},\diagZninefourN{}] galaxies, due to their strict dependence on the full complement of [\ion{O}{2}] and [\ion{O}{3}] lines.  We can apply the direct diagnostic to only \diagdirectN{} galaxies due to its reliance on the auroral line which is weak at the metallicity regime probed here.  However, the N2 diagnostic produces relative metallicity estimates consistent with the other strong line methods \cite{KE08}.  N2 is also less sensitive to systematic effects: it employs lines with a very small wavelength separation (so extinction correction may be neglected) and the effect of underlying absorption is less important (because the absorption equivalent width of H$\alpha$ is typically equal to or less than H$\beta$, despite the $\sim3$~times stronger flux, e.g. \citealt{Brinchmann04}).  However, in N2 diagnostics no correction is made for ionization parameter and the N2 ratio can saturate at high metallicities when nitrogen becomes the dominant coolant \citep{kewley02}.

\subsection{Metallicity Distribution of SN~Ibc Progenitor Environments}
\label{sec:metaldist}

\begin{figure*}
\plotone{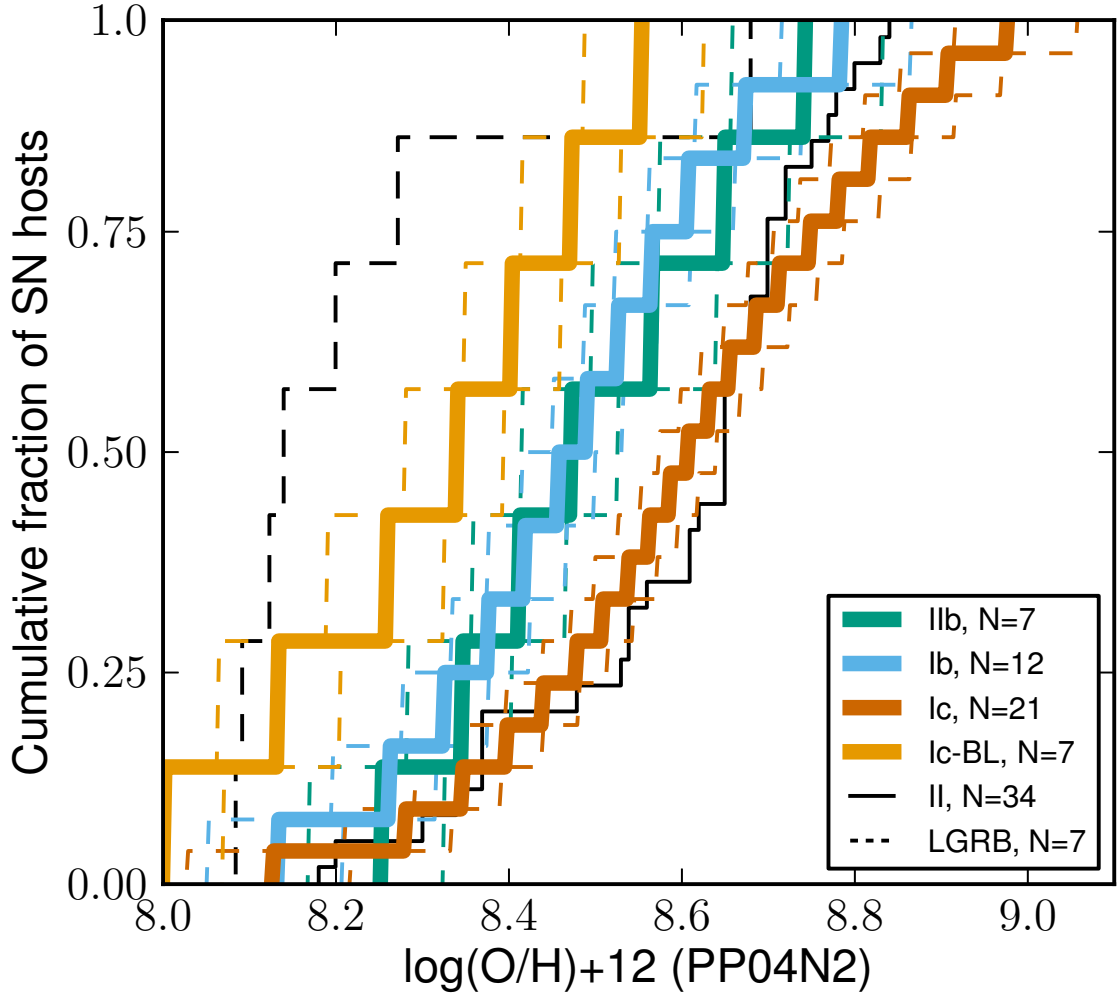}
\caption{\label{fig:CDFpp}Cumulative metallicity functions for SN host galaxies from this study, using the PP04N2 diagnostic.  The 16th and 84th percentiles ($1\sigma$ limits) of the probability distribution for the CDFs are illustrated by the dashed lines, as computed by propagation of the uncertainties in the line flux measurements and the $0.07$~dex diagnostic systematic error.  For comparison, we show the distribution of PP04N2 explosion site metallicities of SN~II discovered by the untargeted PTF survey as measured by \cite{Stoll12}, and that of low-redshift ($z<0.3$) LGRBs from  \cite{Levesque10}, \cite{lsf+10}, and \cite{Chornock2010bh}.}
\end{figure*}

Using the PP04N2 abundance diagnostic, we measure metallicities for \statZNIbPPzerofourNtwo{} SN~Ib host galaxies and find values ranging from log(O/H)$+12=\statZminIbPPzerofourNtwo - \statZmaxIbPPzerofourNtwo$, with a median abundance and standard deviation of log(O/H)$+12=\statZmedIbPPzerofourNtwo$ and $\statZstdIbPPzerofourNtwo$~dex, respectively (Figure~\ref{fig:CDFpp}).  The characteristics of the \statZNIcPPzerofourNtwo{} SN~Ic host galaxies in this diagnostic are similar, with metallicities ranging from log(O/H)$+12=\statZminIcPPzerofourNtwo - \statZmaxIcPPzerofourNtwo$ and median and standard deviation of log(O/H)$+12=\statZmedIcPPzerofourNtwo$ and $\statZstdIcPPzerofourNtwo$~dex.  Among the \statZNIIbPPzerofourNtwo{} SNe~IIb for which we measure PP04N2 metallicities, we find a range of log(O/H)$+12=\statZminIIbPPzerofourNtwo-\statZmaxIIbPPzerofourNtwo$ with median log(O/H)$+12=\statZmedIIbPPzerofourNtwo$ and the standard deviation is \statZstdIIbPPzerofourNtwo~dex~dex, similar to the SN~Ib.  In contrast, the \statZNIcBLPPzerofourNtwo{} SN~Ic-BL host environments typically have lower metallicities, with median log(O/H)$+12=\statZmedIcBLPPzerofourNtwo$ and standard deviation of \statZstdIcBLPPzerofourNtwo~dex.  The minimum metallicity for SNe Ic-BL host galaxies (log(O/H)$+12=\statZminIcBLPPzerofourNtwo$) is only $\sim0.1$~dex lower that of SN~Ib and Ic host galaxies, but the highest metallicity measured for an SN~Ic-BL host galaxy (log(O/H)$+12=\statZmaxIcBLPPzerofourNtwo$) is similar to the median for SN~Ib and Ic.

We report the results from the other diagnostics in Table~\ref{tab:diststat}.  We also note that \cite{Stoll12} have developed an emperical, linear conversion between oxygen and iron abundance.  Applying this to our host environment metallicities yields $\rm{[Fe/H]}=[-0.6\pm0.2,-0.4\pm0.3,-0.6\pm0.2,-0.8\pm0.3]$ using our full sample of SNe~[Ib,Ic,IIb,Ic-BL].

Table~\ref{tab:diststat} illustrates that the effect of restricting our sample to only those objects with Gold classifications and explosion site spectroscopy is modest.  Among the \statZNIcPPzerofourNtwo{} SNe Ic host galaxies in our sample for which the PP04N2 diagnostic was applied, \statZNIcEGPPzerofourNtwo{} have Gold classifications and explosion site spectroscopy.  The difference between the median metallicity measured between this subsample and the full sample is $\statZmedDIFFIcPPzerofourNtwo$~dex.  For SNe [Ib,IIb,Ic-BL], the subsample fraction is [$\statZNIbEGPPzerofourNtwo{}/\statZNIbPPzerofourNtwo{},\statZNIIbEGPPzerofourNtwo{}/\statZNIIbPPzerofourNtwo{},\statZNIcBLEGPPzerofourNtwo{}/\statZNIcBLPPzerofourNtwo{}$] and the difference in the median is similarly small at [$\statZmedDIFFIbPPzerofourNtwo,\statZmedDIFFIIbPPzerofourNtwo,\statZmedDIFFIcBLPPzerofourNtwo$]~dex.  These differences, smaller than the systematic uncertainty associated with the abundance diagnostics (Section~\ref{sec:metalmea}), generally reflect the lower metallicities of the SN explosion sites as compared to the galaxy nuclei.

\begin{deluxetable*}{llllll}
\tablecaption{Statistics of Measured Host Galaxy Metallicity (log(O/H)$+12$) Distribution\label{tab:diststat}}
\tabletypesize{\scriptsize}
\tablehead{\colhead{Diagnostic} & \colhead{N} & \colhead{Minimum} & \colhead{Maximum} & \colhead{Median} & \colhead{$\sigma$}}

\startdata
\sidehead{SNe IIb} \\
direct & 0 (0) & \nodata (\nodata) & \nodata (\nodata) & \nodata (\nodata) & \nodata (\nodata) \\
Z94 & 3 (3) & 8.46 (8.46) & 8.54 (8.54) & 8.48 (8.48) & 0.06 (0.06) \\
KD02 & 4 (4) & 8.54 (8.54) & 9.04 (9.04) & 8.58 (8.58) & 0.25 (0.25) \\
PP04N2 & 7 (5) & 8.30 (8.30) & 8.66 (8.65) & 8.46 (8.42) & 0.18 (0.13) \\
PP04O3N2 & 4 (4) & 8.26 (8.26) & 8.33 (8.33) & 8.32 (8.32) & 0.04 (0.04) \\
PT05 & 2 (2) & 8.16 (8.16) & 8.26 (8.26) & 8.17 (8.17) & 0.07 (0.07) \\
\sidehead{SNe Ib}
direct & 0 (0) & \nodata (\nodata) & \nodata (\nodata) & \nodata (\nodata) & \nodata (\nodata) \\
Z94 & 4 (4) & 8.50 (8.50) & 8.78 (8.78) & 8.57 (8.60) & 0.13 (0.13) \\
KD02 & 5 (4) & 8.51 (8.51) & 8.93 (8.73) & 8.69 (8.65) & 0.21 (0.16) \\
PP04N2 & 12 (8) & 8.15 (8.15) & 8.79 (8.55) & 8.48 (8.43) & 0.16 (0.14) \\
PP04O3N2 & 9 (7) & 8.12 (8.12) & 8.57 (8.47) & 8.32 (8.31) & 0.16 (0.16) \\
PT05 & 4 (4) & 8.25 (8.25) & 8.32 (8.32) & 8.29 (8.30) & 0.10 (0.09) \\
\sidehead{SNe Ib/c}
direct & 0 (0) & \nodata (\nodata) & \nodata (\nodata) & \nodata (\nodata) & \nodata (\nodata) \\
Z94 & 1 (0) & \nodata (\nodata) & \nodata (\nodata) & \nodata (\nodata) & \nodata (\nodata) \\
KD02 & 2 (0) & 8.44 (\nodata) & 8.85 (\nodata) & 8.78 (\nodata) & 0.24 (\nodata) \\
PP04N2 & 3 (1) & 8.54 (\nodata) & 8.78 (\nodata) & 8.65 (\nodata) & 0.12 (\nodata) \\
PP04O3N2 & 2 (0) & 8.46 (\nodata) & 8.70 (\nodata) & 8.57 (\nodata) & 0.13 (\nodata) \\
PT05 & 0 (0) & \nodata (\nodata) & \nodata (\nodata) & \nodata (\nodata) & \nodata (\nodata) \\
\sidehead{SNe Ic}
direct & 0 (0) & \nodata (\nodata) & \nodata (\nodata) & \nodata (\nodata) & \nodata (\nodata) \\
Z94 & 3 (3) & 8.69 (8.69) & 9.13 (9.13) & 8.72 (8.72) & 0.33 (0.33) \\
KD02 & 6 (5) & 8.43 (8.66) & 9.03 (9.03) & 8.74 (8.83) & 0.29 (0.28) \\
PP04N2 & 21 (13) & 8.14 (8.14) & 8.88 (8.86) & 8.61 (8.60) & 0.22 (0.22) \\
PP04O3N2 & 10 (8) & 8.09 (8.09) & 8.82 (8.82) & 8.45 (8.45) & 0.23 (0.24) \\
PT05 & 4 (4) & 8.23 (8.23) & 8.44 (8.44) & 8.31 (8.31) & 0.15 (0.15) \\
\sidehead{SNe Ic-BL}
direct & 2 (1) & 7.92 (\nodata) & 8.35 (\nodata) & 7.98 (\nodata) & 0.25 (\nodata) \\
Z94 & 1 (1) & \nodata (\nodata) & \nodata (\nodata) & \nodata (\nodata) & \nodata (\nodata) \\
KD02 & 2 (1) & 8.58 (\nodata) & 8.62 (\nodata) & 8.55 (\nodata) & 0.12 (\nodata) \\
PP04N2 & 7 (3) & 8.01 (8.31) & 8.53 (8.46) & 8.34 (8.36) & 0.21 (0.09) \\
PP04O3N2 & 6 (3) & 7.99 (8.16) & 8.44 (8.44) & 8.20 (8.32) & 0.19 (0.14) \\
PT05 & 3 (1) & 8.18 (\nodata) & 8.31 (\nodata) & 8.20 (\nodata) & 0.12 (\nodata)
\enddata

\tablecomments{The statistical properties of the distribution of metallicities measured for the host galaxies, divided by SN type.  The first values listed represent all the SNe in the sample; the values in parenthesis reflect only those SNe with secure typing (Gold sample; Section~\ref{sec:typing}) and explosion site spectroscopy (Section~\ref{sec:spectra}).  The oxygen abundance diagnostics applied are described in Section~\ref{sec:metalmea}.  SNe of type ``Ib/c'' have uncertain typing.  The medians and standard deviations ($\sigma$) have been calculated from the MCMC samplings.}
\end{deluxetable*}

Finally, there are several SN host environments at which we performed spectroscopy, but could not measure metallicities.  For 7 objects, we detect narrow emission lines in the host galaxy, but they are not sufficient to estimate the metallicity in any of the strong line diagnostics: 2002gz (IIb), 2008fi (IIb), 2008im (Ib), 2011bv (IIb), 2011cs (Ic), 2011ip (Ic), and PTF10aavz (Ic-BL).  For 2 objects, we did not detect any narrow emission lines in our host galaxy spectrum: 2004ai (Ib) and 2010lz (Ic).  With the exception of SN~2004ai, these host galaxies are not at exceptionally high redshifts with respect to the remainder of the sample, and poor S/N in the spectroscopy is due to the intrinsic low luminosity of the host galaxies.  

We place upper limits on the host environment metallicity of these SNe using archival photometry combined with the $L-Z$ relation of \cite{Tremonti04}\footnote{Converting the \cite{Tremonti04} $L-Z$ relation from the T04 to the PP04N2 scale and adopting log(O/H)$_\odot+12=8.9$ on the T04 scale \citep{Delahaye06}, a galaxy with $M_B\sim-19.8$~mag should have solar metallicity.} and, when H$\alpha$ is detected, the the 3$\sigma$ upper limit measured for the [\ion{N}{2}] flux combined with the PP04N2 diagnostic.  We summarize this investigation in Table~\ref{tab:noZ} and note a few special cases here, but we do not use these metallicity limits in our figures or statistics except where explicitly noted.  SN~2008im occurred $\sim8$~kpc from the nucleus of the Sb galaxy UGC 02906 \citep[with $z=0.008$]{CBET1635}.  In our spectrum, light from an older stellar population dominates over any signature of star formation.  Given that the host galaxy has an absolute magnitude $M_B\approx-19$~mag, the explosion site is likely to be of sub-solar metallicity.  The only one of these nine host environments for which we cannot place useful constraints on metallicity is that of SN~2004ai, reported as a SN~Ic by \cite{IAUC8296}, but revised to SN~Ib (M. T. Botticella, private communication).  The supernova spectrum indicates $z\sim0.59$, making it by far the most distant SN in our sample.  H$\alpha$ is redshifted out of our spectral range and nothing is visible in DSS images at this position, from which we infer that $M_B>-23$~mag, which does not allow us to distinguish between sub- or super-solar metallicities.  The remaining seven host environments are constrained to be at sub-solar metallicities, although for SN~2008im and 2010lz we do not have sufficient S/N in H$\alpha$ to place spectroscopic limits on the metallicity.

\begin{deluxetable*}{lllrll}
\tablecaption{Host environments without metallicity measurements\label{tab:noZ}}
\tablehead{ \colhead{SN} & \colhead{SN Type} & \colhead{$z$} & \colhead{$M_B$~(mag)} & \colhead{$Z$ (phot.)\tablenotemark{a}} & \colhead{$Z$ (spec.)\tablenotemark{b}}}

\startdata
2002gz    & IIb-pec (G) & 0.085 & $>-18$ &                           $\lesssim8.4$      & $\lesssim8.5$ \\
2004ai    & Ib  (S) & 0.590 & $>-23$ &                           $\lesssim8.9$          & \nodata\\
2008fi    & IIb (G) & 0.026 & $>-15$ &                           $\lesssim8.2$          & $\lesssim8.2$\\
2008im    & Ib  (G) & 0.008 & -19    &                           $\sim8.5$          & \nodata\\
2010lz    & Ic  (G) & 0.090\tablenotemark{c} & $>-12$ &          $<8.0\tablenotemark{d}$         & \nodata\\
2011bv    & IIb (G) & 0.072 & $>-17$ &                      $\lesssim8.3$          & $\lesssim8.3$\\
2011cs    & Ic  (G) & 0.101 & $>-18$ &                           $\lesssim8.4$          & $\lesssim8.1$ \\
2011ip    & Ic  (G) & 0.051 & $>-17$ &                           $\lesssim8.3$          & $\lesssim8.5$ \\
PTF10aavz & Ic-BL (G) & 0.063   & $-16.5$  &                $\sim8.2$          & $\lesssim8.1$  
\enddata
\tablenotetext{a}{Metallicity (log(O/H)~$+12$) limit implied by the $L-Z$ relation using the photometry listed, on the PP04 scale.}
\tablenotetext{a}{Metallicity (log(O/H)~$+12$) limit implied by the PP04N2 diagnostic using the $3\sigma$ upper limit measured for the [\ion{N}{2}] flux, stated when H$\alpha$ is detected.}
\tablenotetext{c}{\cite{CBET2645}}
\tablenotetext{d}{The absolute magnitude limit places this host galaxy below the range over which the $L-Z$ relation of \cite{Tremonti04} is calibrated ($\sim8.0-9.2$ on the T04 scale).}
\tablecomments{The security of the spectral classification (Silver, S, or Gold, G; see Section~\ref{sec:typing}) is indicated in parenthesis.  Unless otherwise noted, we retrieve photometry for these host galaxies from NED, the NASA/IPAC Extragalactic Database operated by the Jet Propulsion Laboratory, California Institute of Technology, under contract with the National Aeronautics and Space Administration.  When photometry is not available in NED (not detected in DSS), we assume $m_B>20$~mag.  Redshifts are taken from NED or our own spectroscopy, except where noted.}
\end{deluxetable*}

\subsection{Statistical tests on metallicity distributions}
\label{sec:stat}

\label{sec:statunt}

We apply the KS test to our metallicity measurements (Table~\ref{tab:ksNES}) and interpret $\ksp<0.05$ to indicate statistically significant evidence for a difference in the parent populations of the two sets being compared.   When Monte Carlo simulations\footnote{Following \cite{Leloudas11}, we incorporate the uncertainty in the individual metallicity estimates (but not the diagnostic systematic uncertainty) by repeating the KS test through Monte Carlo simulations where we sample from the full probability distribution for the metallicity of each host galaxy.} indicate that $\ksp$ rises above this threshold ($>0.05$) in at least 14\% of trials, we refer to this as ``marginal'' evidence of statistical significance.

We find no significant difference between the metallicity distribution of SN~Ib and Ic, with $\ksp=\KSPPzerofourNtwoIcIb{}$ using the PP04N2 diagnostic ($\ksp=\KSEGPPzerofourNtwoIcIb{}$ if restricted to Gold classifications and explosion site metallicity measurements).  This contrasts with the finding of \cite{Modjaz11}, that the distributions disagree at the $\ksp=0.01$ level using the equivalent PP04O3N2 diagnostic, but is consistent with the null result of \cite{Anderson10,Kelly11,Leloudas11}.  We note that when we apply the KS test to the sample of SN~Ib and Ic from \cite{Modjaz11} using our methodology, we find $\ksp=\KSModjazIbIc$.  The difference is that we separate SN~IIb and Ibn from Ib, which changes the metallicity distribution and reduces the SN~Ib sample size by 5~objects.

\begin{deluxetable}{lrrrr}
\tablecaption{KS test $p$-values for SNe Ibc\label{tab:ksNES}}
\tablehead{ & \colhead{IIb} & \colhead{Ib} & \colhead{Ic} & \colhead{Ic-BL}}

\startdata
IIb & \nodata & $\KSPPzerofourNtwoIbIIb$  & $\KSPPzerofourNtwoIcIIb$  & $\KSPPzerofourNtwoIIbIcBL$  \\ 
Ib & $ \KSEGPPzerofourNtwoIbIIb$ & \nodata & $\KSPPzerofourNtwoIcIb$  & $\KSPPzerofourNtwoIbIcBL$  \\ 
Ic & $ \KSEGPPzerofourNtwoIcIIb$ & $\KSEGPPzerofourNtwoIcIb$  & \nodata & $\KSPPzerofourNtwoIcIcBL$  \\ 
Ic-BL & $\KSEGPPzerofourNtwoIIbIcBL$ & $\KSEGPPzerofourNtwoIbIcBL$  & $\KSEGPPzerofourNtwoIcIcBL$  & \nodata 
\enddata

\tablecomments{The values in the table are KS test $p$-values ($\ksp$) for the probability that the measured host galaxy metallicities of the SNe of the two indicated types were drawn from the same parent populations.  Above the diagonal, all objects in our sample are considered; below the diagonal, only objects with Gold spectroscopic classification and explosion site spectroscopy are considered.  The PP04N2 diagnostic is used for metallicity estimation.  The KS test is performed only if metallicity measurements are available for $\geq4$ SN of each type.  The upper and lower limits listed are the 16th and 84th percentile values of the results of the KS test Monte Carlo simulations.}
\end{deluxetable}

We find different metallicity distributions for SN~Ic and Ic-BL, with $\ksp=\KSPPzerofourNtwoIcIcBL$ using the PP04N2 diagnostic.  Given the smaller sample size and lower median metallicity of the SN~Ib, the evidence for a difference in the Ic-BL and Ib metallicity distributions is not significant ($\ksp=\KSPPzerofourNtwoIbIcBL$).  We find no evidence for a significant difference between the SN~Ib and IIb populations ($\ksp=\KSPPzerofourNtwoIbIIb$).

\label{sec:statprev}

\subsection{Young stellar population ages}
\label{sec:Yage}

We estimate the age of the young stellar population using the method of \cite{Levesque10}, assuming an instantaneous-burst star formation history \citep[for a review see ][]{Stas96}.  The age estimate is based on the rest frame equivalent width of the H$\beta$ line ($\WHb$) and evolutionary synthesis models for starburst galaxies based on the Geneva HIGH evolutionary tracks \citep{Meynet94,Schaerer98}.  We use the PP04N2 diagnostic to break the metallicity degeneracy, restrict the sample to spectra obtained at the explosion site, and exclude objects with supernova flux contamination (SNe~2010ay, 2011bv, 2011gh, 2011ip, LSQ11JW, PTF10aav).

\begin{figure}
\plotone{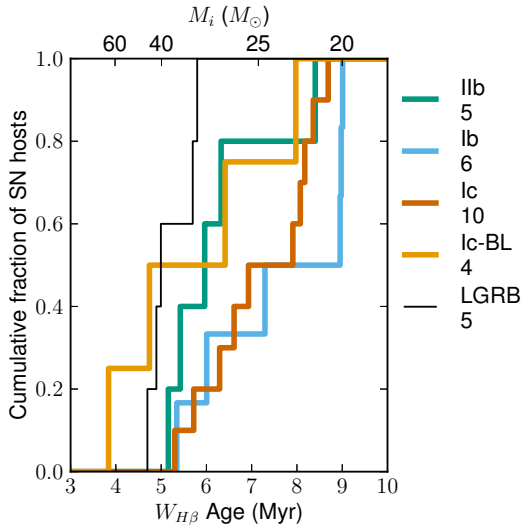}
\caption{\label{fig:age}Cumulative young stellar population age distribution for the SN~Ibc host galaxies, based on the rest frame equivalent width of the H$\beta$ line ($\WHb$) measured from explosion-site spectroscopy only.  The color coding and number of objects of each type are listed on the right.  The low-redshift ($z<0.3$) LGRB host environment ages come from \cite{Levesque10}.  The top axis illustrates the initial masses of rotating Wolf-Rayet stars with lifetimes equivalent to the stellar population ages, as predicted by \cite{Meynet94}.}
\end{figure}

As illustrated in Figure~\ref{fig:age}, we find that the ages of young stellar populations in Type~Ib and Ic SN host environments are not discrepant, with median ages of $[\HBagemedIb,\HBagemedIc]$~Myr and standard deviations of $[\HBagestdIb,\HBagestdIc]$~Myr for $N=[\HBageNIb,\HBageNIc]$ SN~[Ib,Ic] with explosion site spectra, and $\ksp=\HBageKSIcIb$.  However, the ages of SN~Ic-BL environments ($N=\HBageNIcBL$) are somewhat lower, with a median of $\HBagemedIcBL$~Myr and standard deviation of $\HBagestdIcBL$~Myr. Given the small sample size, the KS test cannot confirm statistical significance, $\ksp=\HBageKSIcIcBL$.  Similarly, the age of SN~IIb environments seems to be lower than that of SN~Ib and Ic, with median age $\HBagemedIIb$~Myr and standard deviation of $\HBagestdIIb$~Myr, but the sample size is small ($N=\HBageNIIb$).  

We note that the distribution of ages for SN~Ic-BL host environments in our sample are similar to that of the low-redshift ($z<0.3$) LGRB hosts studied in \cite{Levesque10}, which have a median age of $\sim5$~Myr.  The KS test does not suggest that the LGRB and SN~Ic-BL age distributions are significantly different ($\ksp=\HBageKSIcBLGRB$), but with only 5 low-$z$ LGRBs the sample sizes are small.  Moreover, the observation that SN~Ic-BL and IIb occur in younger stellar populations is consistent with the finding of \cite{Kelly11} that these types of SNe have bluer explosion site $u^\prime-z^\prime$ colors. 

With some caveats, we can interpret the young stellar population ages in terms of the lifetime of massive stars.  For consistency with the $\WHb$ models of \cite{Levesque10}, we use the stellar lifetimes from the ``Geneva HIGH'' (high mass loss) solar-metallicity ($Z=0.02$) evolutionary tracks computed by \cite{Meynet94} (shown in Figure~\ref{fig:age}).  The measured ages therefore imply progenitor stars with initial masses $M_i\sim20-40~M_\odot$ for SN~Ib and Ic.  This is similar to the mass range for SN~Ib/c progenitors expected from stellar evolutionary theory and indicated by the relative rate of SN~Ibc and SN~II \citep{BP09}, but may be discrepant with progenitor non-detection upper limits from pre-explosion imaging \citep[for a review, see][]{Smartt09}.  For SN~IIb and Ic-BL, the young stellar population ages imply progenitor stars that are somewhat more massive, $M_i\lesssim60~M_\odot$.  While stars of a given mass are longer-lived at lower metallicity in the Geneva models, and the SN Ic-BL progenitor stars are found in lower metallicity environment, the effect is small because the lifetimes vary by $\sim10\%$ over a factor of $\sim3$ in metallicity.  However, because the ages are estimated assuming an instantaneous burst of star formation and neglecting any ongoing star formation, the possibility of younger progenitor stars within the population is not precluded.  Further assumptions inherent to this $\WHb$ diagnostic include complete absorption of ionizing photons, spatially uniform dust extinction in both nebular and stellar emission regions, and it is based on evolutionary tracks that do not include the effects of stellar rotation.

\subsection{Wolf-Rayet star populations}
\label{sec:WR}

Broad Wolf-Rayet (WR) features (``bumps'') in galaxy spectra, reflecting the existence of evolved, massive stars ($M\gtrsim25~M_\odot$), can be used to characterize the nature of ongoing star formation in the galaxy \citep{Allen76,Kunth81,Schaerer98,Schaerer99}.  In particular, the ``blue bump'' which is primarily due to the \ion{He}{2}~$\lambda4686$ line is an indicator of late-type WN Wolf-Rayet (WNL) stars.  A visual inspection of the spectra of our SN~Ibc host environments does not reveal any recognizable blue bumps.  We estimate the $3\sigma$ upper limit of the flux in the blue bump feature as $3$ times the rms of the continuum flux in a $40$~\AA\ window at its location.

We follow the method of \cite{Lopez10WR} to place limits on the fraction of WNL stars in the young stellar population, $\WNL$, based on the ratio of the upper limit flux of the blue bump to H$\beta$.  For WR galaxies in their dataset, this fraction ranges from $\WNL\sim0.03-0.3$, smaller than we can constrain with most of our spectra due to the continuum S/N.  However, there are 3 host environments for which we can rule out WNL populations at that level (2007az, 2008iu, 2010Q; all explosion site spectra) and one for which we can rule out $\WNL>0.1$ (2007ce).  Significantly higher S/N spectroscopy or narrow-band imaging could provide stricter constraints for typical SN host galaxies.  This analysis demonstrates that investigations of the WR populations of SN host environments requires a significantly different observing strategy than a study designed for strong line metallicity measurements, and will likely be limited to nearby SN host galaxies.

\section{Combined SN~I\lowercase{bc} Dataset}
\label{sec:combsurv}

Next we combine our dataset with those of previous/concurrent spectroscopic studies of SNe~Ibc (\citealt{Anderson10}, \citealt{Kelly11}, \citealt{Leloudas11}, \citealt{Modjaz11}, and this work).  
We summarize the characteristics of these surveys in Table~\ref{tab:tcount}.

\subsection{Criteria of the Combined SN~I\lowercase{bc} dataset}
\label{sec:premaetal}

\begin{deluxetable}{lrrrrrr}
\tablecaption{Host Galaxy Samples by type\label{tab:tcount}}
\tabletypesize{\scriptsize}
\tablehead{ \colhead{SN Type} & \colhead{A+10} & \colhead{K+11} & \colhead{L+11} & \colhead{M+11} & \colhead{S+12} & \colhead{TW}}

\startdata
\sidehead{Targeted SNe} \\
IIb & 1 & 13 & 2 & 3 & 0 & 0\\
Ib & 10 & 10 & 3 & 7 & 0 & 0\\
Ib/c & 3 & 2 & 2 & 1 & 0 & 0\\
Ic & 14 & 23 & 1 & 10 & 0 & 0\\
Ic-BL & 0 & 5 & 0 & 6 & 0 & 0\\
$ z$  & 0.005 & 0.011 & 0.016 & 0.012 & \nodata & \nodata\\
\sidehead{Untargeted SNe}
IIb & 0 & 1 & 0 & 1 & 2 & 10\\
Ib & 0 & 3 & 6 & 6 & 3 & 13\\
Ib/c & 0 & 1 & 1 & 0 & 0 & 3\\
Ic & 0 & 5 & 4 & 4 & 3 & 24\\
Ic-BL & 0 & 4 & 0 & 9 & 1 & 8\\
$ z$  & \nodata & 0.034 & 0.037 & 0.037 & 0.036 & 0.036\\
\sidehead{All SNe}
IIb & 1 & 14 & 2 & 4 & 2 & 10\\
Ib & 10 & 13 & 9 & 13 & 3 & 13\\
Ib/c & 3 & 3 & 3 & 1 & 0 & 3\\
Ic & 14 & 28 & 5 & 14 & 3 & 24\\
Ic-BL & 0 & 9 & 0 & 15 & 1 & 8\\
$ z$  & 0.005 & 0.015 & 0.022 & 0.017 & 0.036 & 0.036
\enddata

\tablecomments{The number of host galaxies in each of the literature samples divided by SN type, and median redshifts, $z$.  The numbers are given for SNe discovered by untargeted (targeted) surveys.  SNe of type ``Ib/c'' have uncertain typing.  The samples come from the following references: A+10, \cite{Anderson10}; K+11, \cite{Kelly11} (only objects with metallicities measured); L+11, \cite{Leloudas11}; M+11, \cite{Modjaz11}; S+12, \cite{Stoll12}; TW, this work.}
\end{deluxetable}

\begin{deluxetable}{lrrrrrr}
\tablecaption{Overlap between SN~Ibc host galaxy samples\label{tab:overlap}}
\tabletypesize{\scriptsize}
\tablehead{ & \colhead{A+10} & \colhead{K+11} & \colhead{L+11} & \colhead{M+11} & \colhead{S+12} & \colhead{TW}}

\startdata
A+10 & 28 & 7 & 0 & 0 & 0 & 0\\
K+11 &  & 67 & 3 & 12 & 3 & 4\\
L+11 &  &  & 20 & 5 & 0 & 5\\
M+11 &  &  &  & 47 & 0 & 5\\
S+12 &  &  &  &  & 9 & 4\\
TW &  &  &  &  &  & 58\\
\enddata

\tablecomments{The numbers on the diagonal represent the total number of SNe Ibc in each sample; the numbers above the diagonal represent the intersection of the samples.  The samples are the same as in Table~\ref{tab:tcount}.}
\end{deluxetable}

For the purposes of assembling a statistical sample, we consider only measurements made on the PP04 scale (see Section~\ref{sec:metalmeacompare})\footnote{For our observations and those of \cite{Anderson10} and \cite{Leloudas11}, we employed measurements made using the PP04N2 diagnostic; for \cite{Kelly11} and \cite{Modjaz11}, who do not report PP04N2 measurements, we instead employ the PP04O3N2 measurements and apply the small transformation from \cite{KE08}.}.  Among these five samples, there are metallicity measurements for \combN{} unique SN~Ibc host galaxies, of which \combnosN{} had metallicity measurements predating our study.  To ensure consistency between authors when propagating uncertainties, we add in quadrature a representative systematic uncertainty of 0.07~dex (see Section~\ref{sec:metalmeacompare}) to the metallicity estimates from our study, \cite{Leloudas11}, and \cite{
Modjaz11}; for \cite{Kelly11} and \cite{Stoll12}, who do not report metallicity uncertainties, we assume this is the sole uncertainty.

Following the authors' own evaluations, we consider all observations from \cite{Modjaz11} and all but three from \cite{Leloudas11} to be at the explosion site.  If we were to apply our own criteria (Section~\ref{sec:spectra}), 4 of the observations from \cite{Modjaz11} would not qualify as explosion site due to their high redshift (SNe~2007jy, 2007qw, 2005kr, and 2006nx).  For \cite{Anderson10}, following Section~\ref{sec:spectra}, we consider the $25/28$ observations where the spectrum was extracted $<2$~kpc from the SN position to be at the explosion site.  For \cite{Kelly11}, who employ $3\asec$ fiber spectroscopy from the SDSS, we consider all of the observations to be nuclear rather than explosion site measurements, as they sample the global properties of the host galaxy.  We have investigated the discrepancies between those host galaxies whose metallicity measurements were performed by multiple authors and they are typically $\lesssim0.15$~dex.  When SNe have metallicities reported by multiple authors ( see Table~\ref{tab:overlap}), we adopt the average and use only explosion site metallicities where possible.  We neglect any systematic uncertainty introduced by differences in instrumental characteristics and spectroscopic analysis between the samples.  

Of this combined sample, \combNU{} of the SNe were discovered by untargeted searches and \combNE{} of the metallicity measurements come from explosion site spectroscopy. Before our observations, these sample sizes were only \combnosNU{} and \combnosNE{}, respectively.  Our results therefore approximately double the number of untargeted SN~Ibc for which host galaxy metallicity measurements have been published, and significantly increase the number of explosion-site metallicity measurements.

\begin{figure}
\plotone{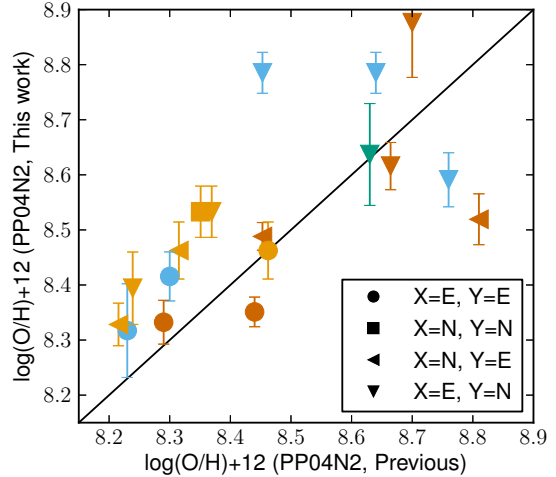}
\caption{\label{fig:compold}Comparison of SN host galaxy metallicity measurements this work and previous studies in the combined sample.  The overlap between these samples is described in Table~\ref{tab:overlap}.  The error bars correspond to the statistical error estimated in this work, from propagation of the line flux errors.  The marker shapes defined in the legend as ``X, Y'' correspond to the spectroscopy type (explosion site, ``E'' and nuclear, ``N'') of the previous work (X) and this work (Y).  The colors correspond to SN types as in Figure~\ref{fig:CDFpp}.}
\end{figure}

Figure~\ref{fig:compold} compares the metallicity measurements made in this work to \compN\ previous measurements of the same host galaxies from the combined sample.  Generally, discrepancies are small, with a mean residual of \compmean~dex and rms of \comprms~dex.  In the median, the discrepancy is $\compterr\times$ the statistical error in our measurement.  The largest outliers can be explained by clear differences in spectroscopic methodology between our sample and the previous works.  For example, for PTF09q we measured $\log(\rm{O/H})+12=8.52$ in our explosion site spectrum, whereas \cite{Stoll12} measure a higher $\log(\rm{O/H})+12=8.81$ from their SDSS spectrum due to the nuclear placement of the Sloan fiber ($\sim5$~kpc from the SN site). There are \compEEN\ cases where we can compare an explosion site measurement from our sample to an equivalent measurement from the literature, and in those cases the rms discrepancy is 
only half as large (\compEErms~dex).

\begin{figure}
\plotone{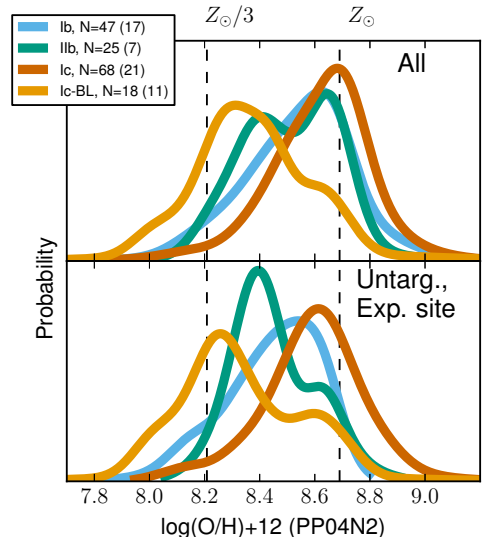}
\caption{\label{fig:ideogram}Continuous probability distribution of SN~Ibc host environment metallicities from the combined sample.  The curves are sums of Gaussians defined by the metallicity measurement and their associated uncertainties.  Top: All SN~Ibc (sample size N noted in legend); Bottom: Only SN~Ibc from untargeted surveys with explosion site metallicity measurements (sample size in parenthesis in legend).}
\end{figure}

\begin{figure*}
\plotone{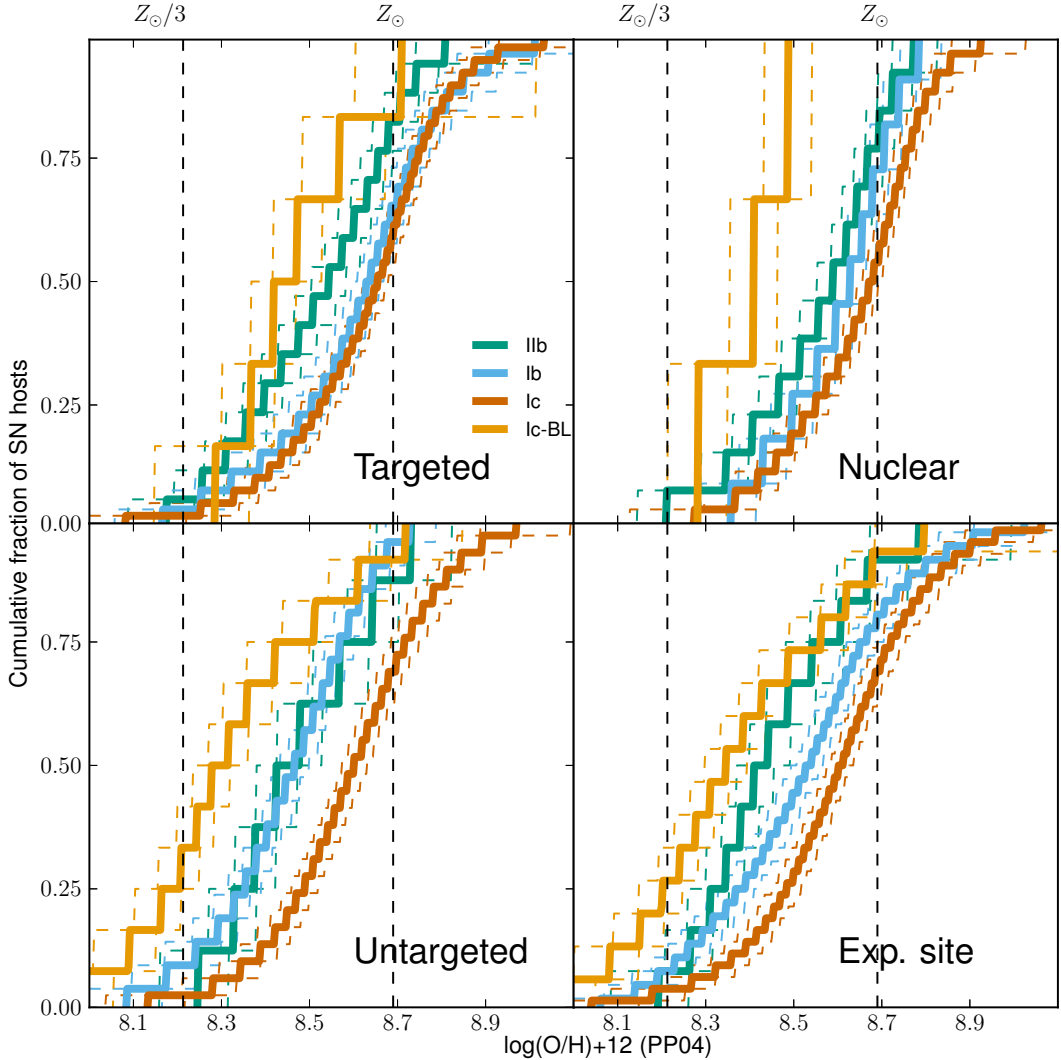}
\caption{\label{fig:CDFppC}Cumulative metallicity functions for SN host environments from the combined sample, using the PP04 diagnostic.  The colors corresponding to SN types are given in the first panel; the sample sizes are given in Table~\ref{tab:KScomb}.  As labeled, the panels correspond to different cuts on the discovery mode of the supernova (galaxy-targeted or untargeted) and the method of spectroscopy (nuclear or explosion site).  The dashed colored lines are $1\sigma$ limits for the CDFs as in Figure~\ref{fig:CDFpp}, but incorporating the systematic uncertainty of $0.07$~dex (Section~\ref{sec:metalmeacompare}).  The dashed vertical lines illustrate $Z_\odot$ and $Z_\odot/3$.}
\end{figure*}

\subsection{SN Ib vs. Ic metallicities}
\label{sec:combIbIc}
\label{sec:combdist}

Figure~\ref{fig:ideogram} illustrates the metallicity distribution for all the SN~Ibc host galaxies from the combined sample.  First we consider the distributions with no selection criteria for SN discovery or spectroscopy methodology.  With sample sizes of [\KSNcAllAllIb,\KSNcAllAllIc], we find the median metallicity of SN~[Ib,Ic] to be log(O/H)$+12=[\MedcAllAllIb,\MedcAllAllIc]$.  The difference between the distributions is not statistically significant ($\ksp=\KScAllAllIbIc$) and the difference in the medians is small relative to the width of the distributions ($[\StdcAllAllIb,\StdcAllAllIc]$~dex standard deviation).

If instead we only consider SNe discovered by targeted SN searches, we find SN host environments with systematically higher metallicities (as quantified in Section~\ref{sec:targeffect}) and we find that the differences between the metallicity distributions for SNe of different types are reduced (Figure~\ref{fig:CDFppC}).  The median metallicity of the $N=[\KSNcAllTarIb,\KSNcAllTarIc]$ SN~[Ib,Ic] from the targeted searches is log(O/H)$+12=[\MedcAllTarIb,\MedcAllTarIc]$ ($\ksp=\KScAllTarIbIc$).  Among the subsample with spectroscopy at the SN explosion site ($N=[\KSNcEsiTarIb,\KSNcEsiTarIc]$ SN~[Ib,Ic]) we find median metallicities of log(O/H)$+12=[\MedcEsiTarIb,\MedcEsiTarIc]$ ($\ksp=\KScEsiTarIbIc$).  

Looking exclusively to untargeted SNe, for $N=[\KSNcAllUntIb,\KSNcAllUntIc]$ SN~[Ib,Ic] we find median metallicities of log(O/H)$+12=[\MedcAllUntIb,\MedcAllUntIc]$ (Figure~\ref{fig:CDFppC}).  While this difference is marginally statistically significant ($\ksp=\KScAllUntIbIc$), it is biased by an unequal numbers of galaxy-nucleus versus explosion site spectroscopy in the SN~Ib and Ic samples (\cNpNucIb\ and \cNpNucIc\% nuclear spectroscopy, respectively).  This sample construction bias raises the metallicities of SN~Ic relative to SN~Ib (see Section~\ref{sec:esiteeffect}).  If the sample is restricted to only explosion site measurements, the median difference is similar ($\sim0.15$~dex), but the difference in the full distribution is not significant ($\ksp=\KScEsiUntIbIc$).  The difference in the distributions is most apparent at the high-metallicity end, where very few SN~Ib discovered by untargeted surveys are found at super-solar metallicities.  However, the explosion site spectroscopy in Figure~\ref{fig:CDFppC} illustrates that SN~Ib discovered by targeted searches do occur in super-solar metallicity environments, and therefore their absence among the untargeted objects must be attributed to small sample size.

\begin{figure}
\plotone{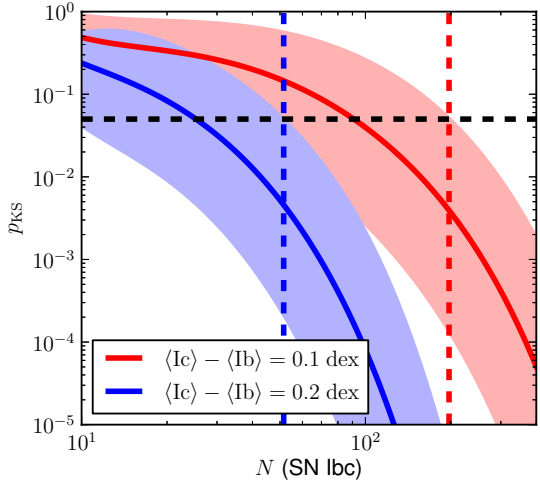}
\caption{\label{fig:kssim}Simulations demonstrating the sample size $N$ of SN~Ibc necessary to distinguish a difference in the median metallicity of SN~Ib and Ic of $[0.1,0.2]$~dex ([red,blue]) assuming an underlying Gaussian distribution with $\sigma=0.2$~dex.  The shaded regions illustrate the $1\sigma$ range in $\ksp$ over random draws from the distribution.  The horizontal dashed line illustrates the threshold for statistical significant, $\ksp=0.05$ at the $1\sigma$ level.  The vertical lines mark the sample size necessary to distinguish the difference in the metallicity distribution to this threshold.}
\end{figure}

How many observations would be required to distinguish between a true difference in the underlying distribution of progenitor metallicities for SN~Ib and Ic?  If we assume that the metallicity distribution of SN~Ic progenitors is enriched by $0.1$~dex with respect to SN~Ib progenitors (as we find in Section~\ref{sec:metaldist}) and that both distributions are Gaussians with standard deviation $\sigma=0.2$~dex, then we can randomly sample from these distributions to investigate the value of $\ksp$ we would infer from studies of different sizes (see Figure~\ref{fig:kssim}).  We take statistical significance to be indicated by the KS test when $\ksp<0.05$ and we assume a sample ratio $N_{\rm Ic}/N_{\rm Ib}=1.6$.  

We find that, in the absence of systematics, only $N\sim\NgaussdistPtwo$ ($\NgaussdistPtwoM$) SN~Ibc would be required to distinguish the discrepancy in 85\% (50\%) of trials given a difference in the median metallicity of $0.2$~dex \citep[as suggested by][]{Modjaz11}.  The combined sample includes $N=[\KSNcAllAllIb,\KSNcAllAllIc]$ SN~[Ib,Ic] metallicity measurements; $N=[\KSNcAllUntIb,\KSNcAllUntIc]$ of these come from untargeted SN searches and therefore have substantially reduced systematics (see Section~\ref{sec:targeffect}).  Therefore, the observations to date should be sufficient to distinguish a median metallicity difference of $0.2$~dex between SN~Ib and Ic (even among exclusively-untargeted surveys), which is not supported by the data in the combined sample.  However, a much larger sample of $N\sim\NgaussdistPone$ ($\NgaussdistPoneM$) observations of SN~Ib and Ic would be necessary to distinguish a median difference of $0.1$~dex in the progenitor distribution in 84\% (50\%) of trials.  Therefore a sample $\gtrsim2\times$ as large as the combined sample would be required to unambiguously distinguish a relatively subtle discrepancy of $0.1$~dex.

The host environment metallicity measurements made for 34 untargeted Type~II SNe by \cite{Stoll12} (shown in Figure~\ref{fig:CDFpp}) constitute an interesting proxy for the metallicity distribution of the massive star progenitors of core-collapse SNe.  Applying the KS test using all the untargeted objects in the combined sample, we do not find a significant difference between the metallicities of SN~Ic and SN~II ($\ksp=\KSPPzerofourNtwoIIIc$), although there is a significant difference between SN~II and [Ib,Ic-BL] ($\ksp=[\KSPPzerofourNtwoIIIb,\KSPPzerofourNtwoIIIcBL]$).  Combining all the SN~Ibc in our sample, we find that they have a median metallicity $\statZmedDIFFIIallPPzerofourNtwo$~dex lower than the SN~II and the difference in the full distribution is significant at $\ksp=\KSPPzerofourNtwoIIall$.  A higher median metallicity for SN~II than Ibc would conflict with several previous findings (e.g. \citealt{Prieto08,BP09,Anderson10,Kelly11}) and would not be consistent with metal-line driven winds stripping the progenitors of SN~Ibc, unless combined with a significantly more bottom-heavy IMF at low metallicities.  This discrepancy may instead indicate differences in sample construction.  The relatively bright SN~Ic-BL may be over-represented in the combined sample relative to a volume limited survey (see Section~\ref{sec:selfoleffect}), biasing the SN~Ibc metallicity downward, and some low-luminosity SN~II and Ibc host galaxies do not have metallicity measurements (see Section~\ref{sec:deptheffect} in this work and Section~3 of \citealt{Stoll12}).

\subsection{SN Ic vs. Ic-BL metallicities}
\label{sec:combIcBL}

When no cut is placed on SN discovery or spectroscopy characteristics, the SN~Ic-BL in the combined sample ($N=\KSNcAllAllIcBL$) have an appreciably smaller ($\gtrsim0.2$~dex) median metallicity (log(O/H)$+12=\MedcAllAllIcBL$) than other SN~Ibc, and the difference relative to the SN~Ic distribution is significant with $\ksp=\KScAllAllIcIcBL$ (see Figure~\ref{fig:CDFppC}).  Restricting our scope to SNe discovered by targeted searches, the sample of targeted SN~Ic-BL is limited ($N=\KSNcAllTarIcBL$).  The SN~Ic-BL again show a median metallicity (log(O/H)$+12=[\MedcAllTarIcBL]$) $\sim0.2$~dex smaller than other SN~Ibc, but the difference compared to the SN~Ic has only marginal statistical significance ($\ksp=\KScAllTarIcIcBL$).  However, for $N=\KSNcAllUntIcBL$ SN~Ic-BL from untargeted surveys (primarily from this work and \cite{Modjaz11}), we find a median metallicity (log(O/H)$+12=\MedcAllUntIcBL$) that is $\sim0.15$~dex lower than from the targeted SN searches.  This median metallicity is significantly different\footnote{Statistical significance is verified despite a smaller sample size than in the SN Ib versus Ic comparison because the difference in the distributions is much larger.  Simulations of the type illustrated by Figure~\ref{fig:kssim} indicate that a sample of $\sim30$ SN~Ic+Ic-BL is sufficient to expose a $0.3$~dex median difference at the $\ksp<0.05$ level in $\geq84\%$ of trials.} from the distribution of either SN~Ic ($\sim0.3$~dex lower median, $\ksp=\KScAllUntIcIcBL$) or the combination of the $N=\KSNcAllUntIbc$ SN~Ib, Ic, and Ib/c ($\ksp=\KScAllUntIbcIcBL$).  As the SN~Ic-BL metallicities derive almost exclusively from explosion site spectroscopy, this difference is not dependent on spectroscopic methodology ($\ksp=\KScEsiUntIcIcBL$ for SN~Ic vs SN~Ic-BL with explosion-site spectroscopy only).

It has been suggested that the metallicity distribution of SNe~Ic-BL may be bimodal \citep[see e.g.][]{Modjaz08}.  This question is entangled with SN search methodology, because some SN~Ic-BL are discovered via associated GRBs (by untargeted gamma-ray searches), while many SN~Ic-BL without GRBs have been found by targeted SN searches.  While we have shown SN~Ic-BL are preferentially found at lower metallicity than other SN~Ibc, Figure~\ref{fig:ideogram} illustrates that the metallicity distribution of SN~Ic-BL is broad, extending to super-solar metallicities.  The Figure is visually suggestive of a bimodality, even among objects only from untargeted SN searches and with explosion-site metallicity measurements, but given the sample size the distribution is not significantly different than a Gaussian.  Further studies comparing SN~Ic-BL explosion properties (optical luminosity, photospheric velocity, etc.) with explosion site metallicity may inform the discussion of progenitor (sub)classes for these objects.  

\subsection{SN IIb vs. Ibc metallicities}
\label{sec:combIIb}

\cite{Arcavi10} reported that SN~IIb preferentially occur in low-luminosity, likely low-metallicity, host environments.  However, when no cut is placed on SN discovery or spectroscopy characteristics, the difference between the combined sample distributions for SN~Ib ($N=\KSNcAllAllIb$) and IIb ($N=\KSNcAllAllIIb$) is not significant, with $\ksp=\KScAllAllIIbIb$ (see Figure~\ref{fig:CDFppC}).  Among only objects with explosion site spectroscopy ($N_{\rm{IIb}}=\KSNcEsiTarIIb$ SN~IIb), the SN~IIb median log(O/H)$+12=\MedcEsiTarIIb$, $\sim0.1$~dex lower than SN~Ib, but this difference is not significant ($\ksp=\KScEsiTarIIbIb$).  Among SNe discovered by untargeted searches ($N_{\rm IIb}=\KSNcAllUntIIb$, mostly from this work), there is no suggestion of a difference between the host environment metallicity of SN~IIb and SN~Ib, with $\ksp=\KScAllUntIIbIb$.  As with SN~Ic-BL, Figure~\ref{fig:ideogram} points to a bimodality in the SN~IIb metallicity distribution that cannot be statistically verified with the present sample size.

\begin{deluxetable*}{lrrr}
\tablecaption{Statistics of the combined SN~Ibc sample\label{tab:KScomb}}
\tabletypesize{\scriptsize}
\tablehead{&\colhead{All} & \colhead{Nuclear} & \colhead{Exp. Site}}

\startdata
\cutinhead{SN~Ib vs Ic}
All             &       $\KSNcAllAllIb;\KSNcAllAllIc;\KScAllAllIbIc$  &       $\KSNcNucAllIb;\KSNcNucAllIc;\KScNucAllIbIc$  &       $\KSNcEsiAllIb;\KSNcEsiAllIc;\KScEsiAllIbIc$ \\
Targeted         &       $\KSNcAllTarIb;\KSNcAllTarIc;\KScAllTarIbIc$  &       $\KSNcNucTarIb;\KSNcNucTarIc;\KScNucTarIbIc$  &       $\KSNcEsiTarIb;\KSNcEsiTarIc;\KScEsiTarIbIc$ \\
Untargeted       &       $\KSNcAllUntIb;\KSNcAllUntIc;\KScAllUntIbIc$  &       $\KSNcNucUntIb;\KSNcNucUntIc;\KScNucUntIbIc$  &       $\KSNcEsiUntIb;\KSNcEsiUntIc;\KScEsiUntIbIc$ \\

\cutinhead{SN~IIb vs Ib}
All             &       $\KSNcAllAllIIb;\KSNcAllAllIb;\KScAllAllIIbIb$  &       $\KSNcNucAllIIb;\KSNcNucAllIb;\KScNucAllIIbIb$  &       $\KSNcEsiAllIIb;\KSNcEsiAllIb;\KScEsiAllIIbIb$ \\
Targeted         &       $\KSNcAllTarIIb;\KSNcAllTarIb;\KScAllTarIIbIb$  &       $\KSNcNucTarIIb;\KSNcNucTarIb;\KScNucTarIIbIb$  &       $\KSNcEsiTarIIb;\KSNcEsiTarIb;\KScEsiTarIIbIb$ \\
Untargeted       &       $\KSNcAllUntIIb;\KSNcAllUntIb;\KScAllUntIIbIb$  &       $\KSNcNucUntIIb;\KSNcNucUntIb;\KScNucUntIIbIb$  &       $\KSNcEsiUntIIb;\KSNcEsiUntIb;\KScEsiUntIIbIb$ \\

\cutinhead{SN~Ic-BL vs Ic}
All             &       $\KSNcAllAllIcBL;\KSNcAllAllIc;\KScAllAllIcIcBL$  &       $\KSNcNucAllIcBL;\KSNcNucAllIc;\KScNucAllIcIcBL$  &       $\KSNcEsiAllIcBL;\KSNcEsiAllIc;\KScEsiAllIcIcBL$ \\
Targeted         &       $\KSNcAllTarIcBL;\KSNcAllTarIc;\KScAllTarIcIcBL$  &       $\KSNcNucTarIcBL;\KSNcNucTarIc;\KScNucTarIcIcBL$  &       $\KSNcEsiTarIcBL;\KSNcEsiTarIc;\KScEsiTarIcIcBL$ \\
Untargeted       &       $\KSNcAllUntIcBL;\KSNcAllUntIc;\KScAllUntIcIcBL$  &       $\KSNcNucUntIcBL;\KSNcNucUntIc;\KScNucUntIcIcBL$  &       $\KSNcEsiUntIcBL;\KSNcEsiUntIc;\KScEsiUntIcIcBL$ \\

\cutinhead{SN~Ic-BL vs Ibc}
All             &       $\KSNcAllAllIcBL;\KSNcAllAllIbc;\KScAllAllIbcIcBL$  &       $\KSNcNucAllIcBL;\KSNcNucAllIbc;\KScNucAllIbcIcBL$  &       $\KSNcEsiAllIcBL;\KSNcEsiAllIbc;\KScEsiAllIbcIcBL$ \\
Targeted         &       $\KSNcAllTarIcBL;\KSNcAllTarIbc;\KScAllTarIbcIcBL$  &       $\KSNcNucTarIcBL;\KSNcNucTarIbc;\KScNucTarIbcIcBL$  &       $\KSNcEsiTarIcBL;\KSNcEsiTarIbc;\KScEsiTarIbcIcBL$ \\
Untargeted       &       $\KSNcAllUntIcBL;\KSNcAllUntIbc;\KScAllUntIbcIcBL$  &       $\KSNcNucUntIcBL;\KSNcNucUntIbc;\KScNucUntIbcIcBL$  &       $\KSNcEsiUntIcBL;\KSNcEsiUntIbc;\KScEsiUntIbcIcBL$ \\

\enddata

\tablecomments{This table lists the KS test $p$-value ($\ksp$) and sample size ($N$) for SN~Ibc metallicity distributions (PP04 scale) in the combined sample.  Each table entry reflects a different cut on the SN discovery (galaxy-targeted or untargeted) and spectroscopic methods (nuclear or explosion site; see also Figure~\ref{fig:CDFppC}).  Each entry is given as $N_1;N_2;\ksp$.  ``SN~Ibc'' is a combination of SN~Ib, Ic, and Ib/c.}
\end{deluxetable*}

\subsection{Type Ic supernovae from dwarf host galaxies}
\label{sec:Iclow}

Investigating SNe discovered in the first year of operation of PTF, \cite{Arcavi10} found that SN~IIb and Ic-BL occur preferentially in dwarf host galaxies (defined by $M_r>-18$~mag), while they found SN~Ic only in giant host galaxies ($M_r<-18$~mag).  They interpret this as evidence for a dependence of the SN explosion properties on the metallicity of the explosion site.  Transforming the $L-Z$ relation of \cite{Tremonti04} to the $r$ band, they find that this threshold corresponds to a characteristic metallicity of $0.35~Z_\odot$, or log(O/H)$+12=8.23$ on the PP04 scale. However, this photometric approach has several limitations (see Section~3.1 of \citealt{Arcavi10}).  In particular, the statistics are limiting, with only 6 SN~Ibc in dwarf galaxies, and the absolute magnitude of the host galaxy does not necessarily reflect the metallicity of the explosion site accurately (Section~\ref{sec:esiteeffect}).  Moreover, while the mass-metallicity relation for star-forming galaxies is relatively tight (\citealt{Tremonti04} find $\sigma=0.10$~dex), the scatter is larger when luminosity is used as a proxy for mass (\citealt{Tremonti04} find $\sigma=0.16$~dex for $M_B$) and the scatter increases by a factor of $\gtrsim2$ at low-luminosities/masses in the dwarf regime \citep[see e.g.][]{KE08,Mannucci11}.  Additional scatter is introduced in the survey of \cite{Arcavi10} because they use a statistical transformation to convert their $r$-band magnitudes to the $B$-band and because they do not correct for extinction of the host galaxies, which \cite{Tremonti04} find is typically $A_B\approx0.3$~mag ($\sim0.05$~dex in $Z$).  Finally, the statistical analysis of \cite{Arcavi10} depends on the arbitrary choice of the luminosity threshold ($M_r=-18$~mag).

Using the spectroscopic metallicity threshold log(O/H)$+12\leq8.23$, \combPdwarf\% of the combined sample of SN~Ibc host galaxy observations we consider here are in dwarf hosts (\combYdwarf/\combAll\ with PP04 metallicity measurements).  Of these six, two are SN~Ic-BL discovered by untargeted surveys (2007ce and 2008iu, both with explosion-site metallicity measurements presented by this work), one is an untargeted SN~Ib (2007az; this work), one is a targeted SN~IIb (2008ax; nuclear metallicity from \citealt{Kelly11}), and two are SN~Ic.  The SN~Ic are SN\,2002jz, discovered by a targeted survey with an explosion site metallicity measured by \cite{Anderson10}, and SN\,2010Q, discovered by the untargeted CRTS survey and with explosion site metallicity measured as a member of our Gold spectroscopic sample.  However, the spectral classification of SN\,2002jz is somewhat uncertain, as \cite{2002jzIAUC} suggest that it possibly displays H$\alpha$ absorption reflective of Type IIb SNe.  Additionally, there are 3 SN~Ic (all Gold classifications) in our sample for which we can not measure metallicities using strong line diagnostics, but are likely to be in sub-solar metallicity hosts and at least one of these (SN~2010lz) is in an exceptionally low-luminosity host galaxy (see Section~\ref{sec:SE}).  Finally, we note SN\,2005kf, which \cite{Modjaz11} report as a Type~Ic SN from a $M_B=-17$~mag dwarf host galaxy with super-solar metallicity (log(O/H)$+12\approx8.8$).

In summary, spectroscopic metallicity measurements of the host environments of SN~Ibc do not support the conclusion of \cite{Arcavi10} that SN~Ic do not occur in low-metallicity host galaxies.  This discrepancy may be due to the small number of objects included in the sample of \cite{Arcavi10}.  Moreover, because secondary metallicity estimates made using host galaxy photometry introduce additional scatter, their role in detecting the subtle difference that could exist between the metallicity distributions of SN~Ib and Ic may be limited.  In Section~\ref{sec:combIbIc} we show that existing observations demonstrate that the difference in the median metallicity of SN~Ib and Ic host galaxies is almost certainly $<0.2$~dex, while the systematic uncertainty in photometric metallicity measurements is $\sim0.2$~dex.

\section{Systematic Effects}
\label{sec:SE}

\subsection{Targeted vs. Untargeted SN Searches}
\label{sec:targeffect}

\begin{figure}
\plotone{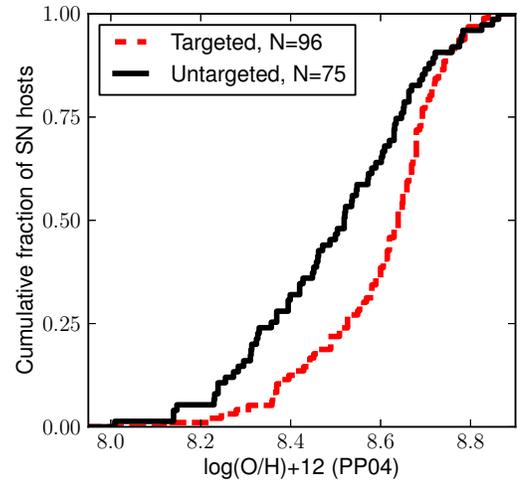}
\caption{\label{fig:comptarg}Cumulative metallicity distribution for the host galaxies SN~Ibc (of all subtypes) discovered by targeted and untargeted searches, from the combined dataset (Section~\ref{sec:combsurv}).  SNe with and without explosion site spectroscopy are shown (adopting only the explosion site measurement when both are available); if we restrict the sample to objects with explosion site spectroscopy, the effect is to shift to lower metallicities by $\ll0.05$~dex.}
\end{figure}

Galaxy-targeted supernova searches can bias SN host environment studies towards higher metallicities due to the galaxy $L-Z$ relation, as illustrated in Figure~\ref{fig:comptarg}.   The median metallicity measured for SNe discovered by targeted searches is log(O/H)
$+12=\ZmedallTPPzerofourNtwo$, while for untargeted searches it is log(O/H)$+12=\ZmedallNTPPzerofourNtwo$ ($\sim$\ZdifallNTPPzerofourNtwo\% lower).  The difference between the distributions is statistically significant ($\ksp=\ZallNTKS$).  The fraction of galaxies with metallicity $<1/3~Z_\odot$ is larger by a factor of $N_U/N_T\approx\numTthirdsol$ in untargeted surveys, meaning that low-metallicity galaxies are strongly underrepresented in targeted surveys.  This ratio is still appreciable for $1/2~Z_\odot$ ($N_U/N_T\approx\numThalfsol$), and even galaxies of solar metallicity are somewhat under-represented in targeted searches ($N_U/N_T\approx\numTsol$).

Consequently, galaxy-targeted SN searches offer a smaller baseline over which to probe for differences in metallicity distributions.  The $1\sigma$ spread in the metallicity distribution of SN~Ibc discovered by targeted SN searches in the combined sample is only $\ZNTcomparewidth\%$ as large as that of untargeted surveys.  As a result, the metallicity distributions of SNe are compressed and differences are reduced, as illustrated by Figure~\ref{fig:CDFppC}.

This $L-Z$ bias has important consequences for studies which combine observations of SNe discovered by both targeted and untargeted surveys.  To illustrate this, we take the assumption that SN~Ib and Ic share the same metallicity distribution and randomly draw samples from the targeted and untargeted SN~Ibc metallicity distributions shown in Figure~\ref{fig:comptarg}.  We produce simulated samples constructed identically to that of \cite{Modjaz11} with respect to SN discovery characteristics: [6,11] targeted and [6,4] untargeted SN~[Ib,Ic].  We find the SN~Ic in the simulated samples to have higher average metallicities than the SN~Ib in \ModjazTargSimMeanZ\% of simulated trials, due solely to the $L-Z$ bias.  The average difference in the SN~Ic and SN~Ib metallicities reported by \cite{Modjaz11} ($\geq0.14$~dex, depending on the diagnostic scale) is reproduced or exceeded in \ModjazTargSimMeanM\% of the trials.  Similarly, we find that in \ModjazTargSimKS\% of trials, a KS test on the simulated observations would indicate that SN~Ic host environments are significantly different ($\ksp\leq0.05$) from Ib environments.  This represents a systematic effect biasing the results of the study, above and beyond the statistical ambiguity indicated by the KS test $p$-value.  We conclude that differences in sample construction alone (the ratio of supernovae from targeted versus untargeted searches) can lead to erroneous differences in the metallicity distribution measured for the host environments of different SN types.

\subsection{Isolating the SN Explosion Site}
\label{sec:esiteeffect}

Relative to nearby galaxies observed in targeted surveys, host galaxies in our exclusively-untargeted sample will typically have smaller intrinsic radii and luminosities and are typically found at greater distances.  As a result, we are not able to resolve and measure the local metallicity at the explosion site of every host galaxy in our sample.  As we discuss in Section~\ref{sec:spectra}, many of our measurements reflect integrated galaxy flux or only light from the brightest region of the galaxy (the nucleus).   The situation is similar for SNe hosted by edge-on galaxies, where the host environment spectrum is necessarily integrated over the full line-of-sight.

It has been shown that SN~Ic preferentially occur in the innermost regions of galaxies, moreso than SN~Ib or IIb \citep{Prieto08,Kelly08,Anderson09,Habergham12}.  In cases where we resort to measuring the galaxy nuclear metallicity, this implies that we would overestimate the metallicities of SN~Ib relative to Ic in the presence of a strong gradient (e.g. \citealt{Zaritsky94}).  We would therefore be less sensitive to the scenario where SN~Ib come from lower metallicity environments than SN~Ic within galaxies of similar nuclear characteristics.

However, the role of gradients is not clear cut and the ability to resolve explosion environments is limited, even for surveys of local galaxies.  Individual SN~Ibc host environments (HII regions) can only be resolved from nearby emission nebulae for the most local events (e.g. $z\lesssim0.005$ for a 100~pc HII region and a 1\asec\ slit).  Moreover, long-slit spectroscopy of nearby spiral galaxies illustrates that the oxygen abundance difference between the galaxy nucleus and disk outskirts is modest (very rarely as large as $0.3$~dex; \citealt{Moran12}).  Metallicity gradients in galaxies should play an even smaller role for explosion sites close to the galaxy nucleus, where SNe Ibc usually occur \citep{Kelly08}.  Moreover, nearby spiral galaxies show significant intrinsic scatter in metallicity gradients (e.g. \citealt{Fabbiano04}, \citealt{Rosolowsky08}, \citealt{SandersM31}; but see also \citealt{Bresolin11}).

\begin{figure*}
\plotone{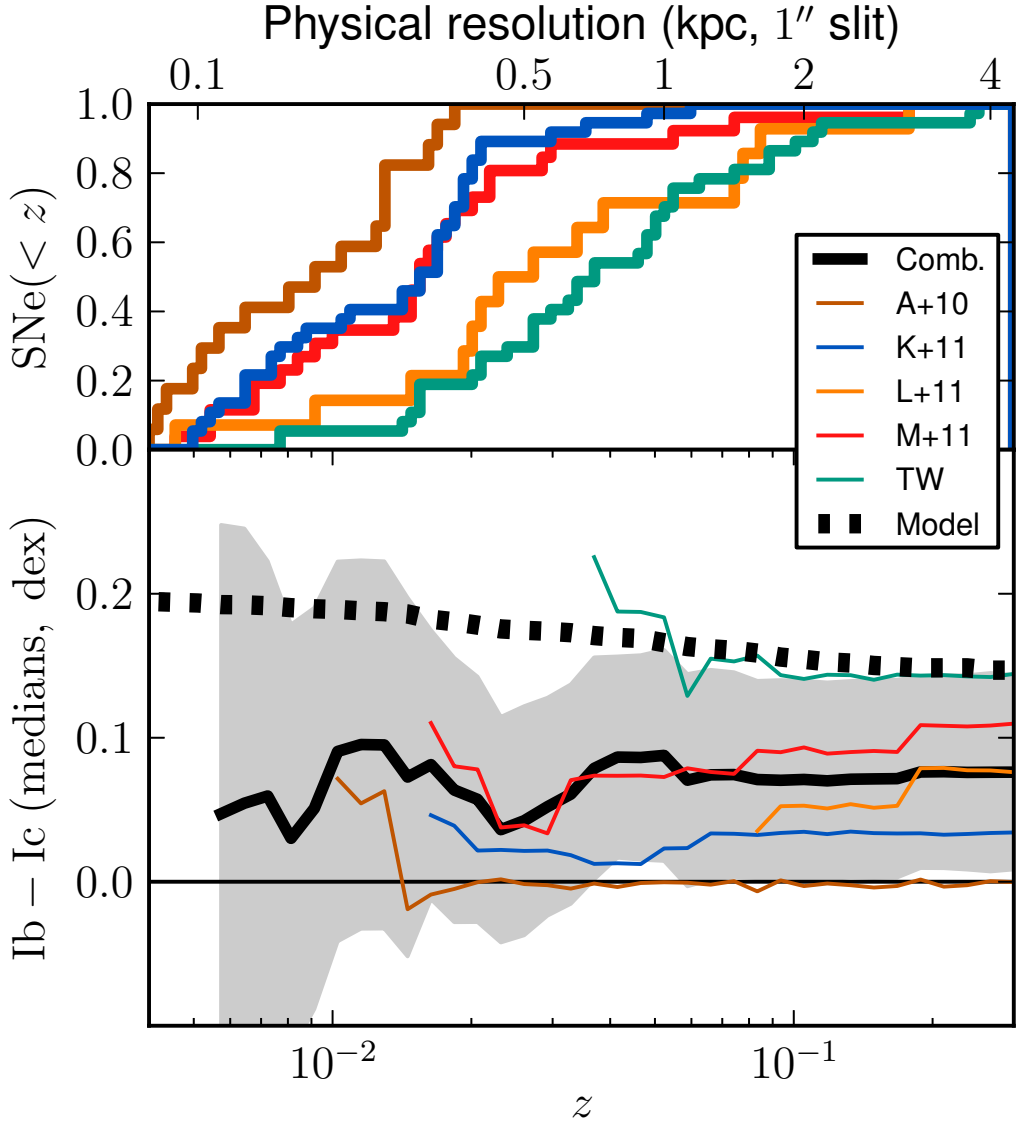}
\caption{\label{fig:meddiffZ}Bottom: the cumulative difference measured between the median metallicity of SN~Ib and Ic host environments at redshift~$<z$ where $>4$ SN~Ib and Ic are available, using log(O/H) as estimated from the PP04 diagnostic.  The black dashed line illustrates the model for the effect of physical resolution on the measurement of a $0.2$~dex difference in the median, described in the text.  Except for the K+11 sample, only objects with explosion site spectroscopy claimed are considered (see Section~\ref{sec:premaetal}).  The shaded region illustrates the $1\sigma$ bootstrap errors on the median difference in the combined sample.  Top: redshift distribution of SN~Ib and Ic in the individual studies of the combined sample (see Table~\ref{tab:tcount}).}
\end{figure*}

We investigate the effect of physical resolution on the median metallicity difference measured between SN~Ib and Ic host environments in Figure~\ref{fig:meddiffZ}.  Monte Carlo simulations incorporating the uncertainties in the individual metallicity measurements demonstrate that the difference in the median metallicity of SN~Ib and Ic is never significantly larger than the uncertainty in the median for any sample.  For comparison, we show a model where the true median difference of $0.2$~dex that can be measured accurately at $z=0$, but is diminished linearly with increasing physical resolution until it is apparently 0~dex at a modest redshift ($z\approx0.1$, resolution of $\sim2$~kpc).  The model is shown cumulatively for a sample with the redshift distribution of the combined sample.

Comparing the model to the existing datasets in Figure~\ref{fig:meddiffZ} suggests that physical resolution is not the limiting factor obscuring a difference in the median metallicity of SN~Ib and Ic.  The median difference in the combined sample is not greater than $0.1$~dex at any redshift (and always consistent with 0~dex at $1~\sigma$), even though the corresponding model suggests that a $0.15-0.2$~dex metallicity difference would be preserved in the sample out to $z\gtrsim0.2$.  We conclude that limitations due to physical resolution are not sufficient to mask a significant difference between the metallicity distributions of the SN~Ib and Ic in the combined sample.

\begin{figure}
\plotone{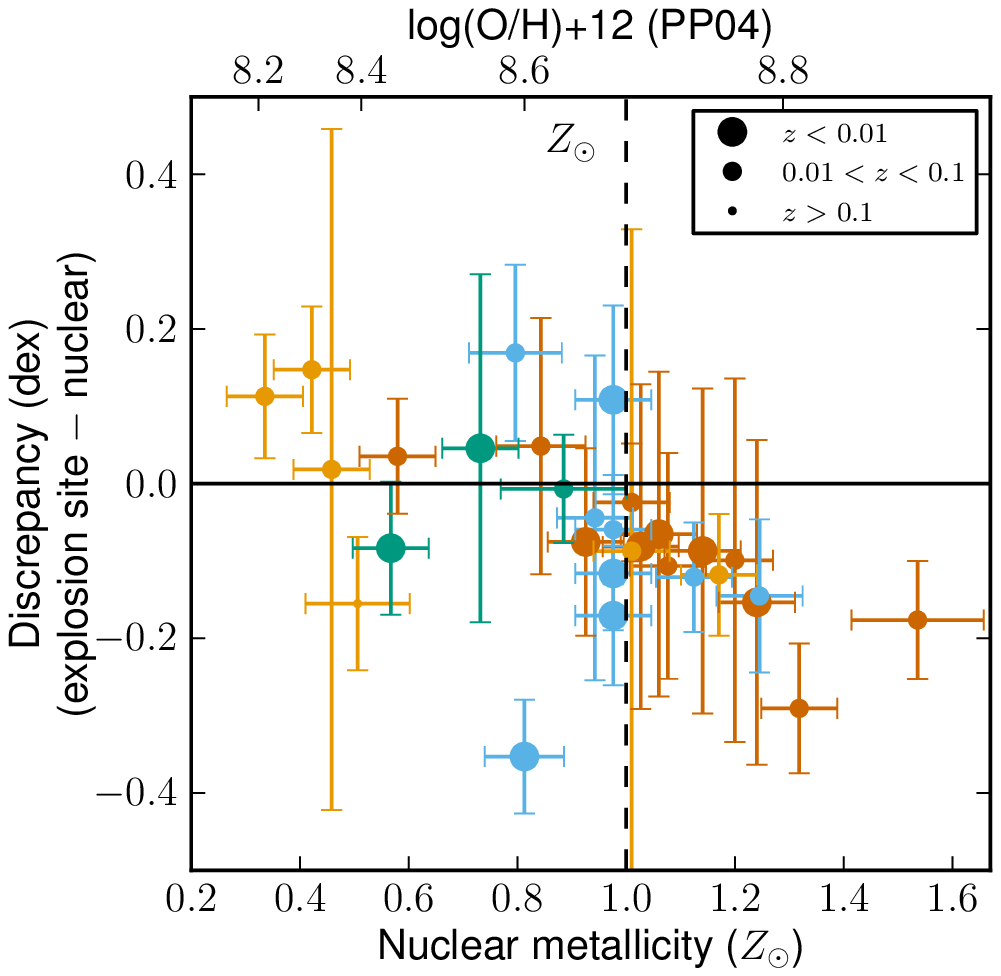}
\caption{\label{fig:compesite}Comparison of PP04 metallicity measurements for \compexpN{} SN~Ibc host environments from galaxy nucleus and explosion site spectroscopy, from the combined dataset.  The horizontal line marks equality and the dashed vertical line marks the solar metallicity.  The colors reflect SN type as in Figure~\ref{fig:CDFppC} and the sizes encode the redshift of the object.}
\end{figure}

A related issue is the methodological difference between explosion site and galaxy-nucleus spectroscopy.  In the combined sample (Section~\ref{sec:combsurv}), there are \compexpN{} SNe which have metallicity estimates at both the galaxy nucleus and explosion site.  Figure~\ref{fig:compesite} indicates that nuclear metallicity measurements introduce both a small additional scatter (rms discrepancy of $\compexprms$~dex) and bias (median difference of $-\compexpmeddif$~dex) relative to explosion site spectroscopy.  The sign of the effect is consistent with offset explosion sites being systematically lower-metallicity environments than galaxy nuclei due to galactic metallicity gradients, with the effect strongest among the highest-metallicity host galaxies (nuclear $Z>Z_\odot$).  In the median the discrepancy between the nuclear and explosion site metallicity measurements is only \compexpmederr{} times the uncertainties in the individual metallicity measurements.  We do not find evidence for a significant difference between the explosion site versus nuclear metallicity discrepancies of SN~Ib and Ic in the combined sample; their medians are different by only \compexpIbvIc~dex, which is much smaller than the rms.  

Simulations of the type illustrated in Figure~\ref{fig:kssim} indicate that the bias and scatter introduced by nuclear spectroscopy should be small for studies of SN~Ibc host environments.  Introducing an additional $0.1$~dex scatter for all measurements, the sample size necessary to recover a 0.1~dex metallicity difference between SN~Ib and Ic increases by only a small amount ($\sim20$\%).  Introducing a $0.1$~dex bias only for objects with super-solar metallicity (and assuming intrinsic median metallicities of log(O/H)+12$=8.5$ and 8.6 for SN~Ib and Ic, respectively), we find a negligible effect on the significance indicated by the KS test.  In summary, the combined sample demonstrates that nuclear metallicity measurements can be used as effectively unbiased tracers of SN~Ibc progenitor metallicity.

A third issue relating to the isolation of the SN explosion site is the use of photometry to estimate the nuclear galaxy metallicity \citep[e.g.][]{Arcavi10}, or additionally using a simple metallicity gradient model to estimate the explosion site metallicity \citep[e.g.][]{BP09}.  In either case, photometric metallicity estimates will carry the same scatter as spectroscopic nuclear metallicity measurements ($\sim0.1$~dex, see above) and an additional uncertainty $\gtrsim0.16$~dex due to the scatter in the $L-Z$ relation (see Section~\ref{sec:Iclow}).  Assuming a representative $0.2$~dex uncertainty for photometric metallicity estimates and performing the same simulations as in Figure~\ref{fig:kssim}, we find that a small (0.1~dex median) metallicity difference between SN~Ib and Ic host environments can only be detected with a sample $\sim2$ times as large as an equivalent sample of spectroscopic metallicity measurements.  Our conclusion that the combined sample of $>100$ SN~Ibc has not revealed a difference between the metallicity distributions of SN~Ib and Ic may therefore conflict with the suggestion by \cite{Arcavi10} that a sample of 18 photometric metallicity measurements for SN~Ibc host galaxies could reveal a significant difference.

In summary, the role of explosion site spectroscopy in SN host environment studies is complex.  The ability to isolate SN explosion sites is limited for smaller galaxies and at larger redshifts (e.g. those found by untargeted surveys).  However, given the significantly larger baseline for making measurements of metallicity distributions offered by untargeted SN searches (Section \ref{sec:targeffect}), the effect of physical resolution is likely to play a secondary role.  Substituting nuclear for explosion site spectroscopy can lead to discrepancies in metallicity of $\sim0.1$~dex, but we find that small differences in metallicity distributions could still be recovered by studies employing nuclear spectroscopy.  However, the additional uncertainty introduced by photometric metallicity measurements is larger and may limit the role of these investigations in discovering differences in metallicity distributions, unless sufficient statistical power can be gained by larger sample size.

\subsection{Uncertainties in SN classification}
\label{sec:classeffect}

The spectral classification of SNe could also potentially effect the metallicity distribution inferred for the host environments of each SN type (see also Section~\ref{sec:typing}).  

It is unlikely that a SN~Ic would be mistaken for a SN~Ib, if the type is designated based on a clear detection of \ion{He}{1} lines, but it is possible for a SN~Ib to be classified as an SN~Ic if the He lines are weak and S/N is poor, or if the He lines have not yet developed at the epoch of spectroscopy \citep{Hamuy02}.  Misidentifying SN~Ib as SN~Ic could act to blur any distinctions in the metallicity distribution of their host environments.   We evaluate the possibility that this blurring could hide a large ($0.2$~dex) difference in the metallicity distributions of SN~Ib and Ic host environments for a dataset as large as the combined sample using simulations.  We randomly draw SN~Ib and Ic metallicities from Gaussian distributions with widths of $0.2$~dex and then mis-attribute a certain fraction of the draws to SN~Ic.  We find that the KS~test indicates a statistically significant difference ($\ksp<0.05$ at $1\sigma$) unless a large percentage ($\gtrsim2/3$) of SN~Ib are mis-classified as SN~Ic.

It can be difficult to distinguish normal SN~Ib from SN~IIb, especially if spectra are not available near peak.  There is a diversity in the hydrogen mass surrounding SNe IIb \citep{Chevalier10} and, spectroscopically, SN~IIb seem to represent a transitional type between SN~Ib and SN~II \citep{Milisavljevic12}.  However, \cite{Arcavi10} have suggested that SN~IIb prefer low-mass, low-metallicity host galaxies.  It is difficult to reconcile these observations into a consistent progenitor model, but the conclusions of \cite{Arcavi10} imply that grouping SN~Ib and IIb together (as \citealt{Leloudas11} and \citealt{Modjaz11} have done explicitly) could bias downward the metallicity distribution inferred for SN~Ib environments.  

Similarly, if SN~Ic-BL are grouped together with SN~Ic, the SN~Ic metallicity distribution could also be biased downward (Section~\ref{sec:combIcBL}). There is no clear definition of the ejecta velocity or line width that distinguishes SN~Ic-BL from SN~Ic, and any distinction will strongly depend on the spectroscopy epoch due to the velocity evolution of the photosphere \citep[see e.g.][]{Pian06,nes2010ay}.

We have attempted to reduce the effects of SN classification errors by revisiting the SN spectroscopy on a case-by-case basis, rather than relying on the classification reported in the IAUCs, and by separating our dataset based on our degree of confidence in the classifications (Gold and Silver samples; Section~\ref{sec:typing}).  Of the previous studies in our combined sample (Section~\ref{sec:combsurv}), \cite{Leloudas11} and \cite{Modjaz11} made similar revisions to SN classifications, while \cite{Anderson10} and \cite{Kelly11} have relied on classifications reported in the circulars.

\subsection{Selection effects in spectroscopic follow-up}
\label{sec:selfoleffect}

The number of optical transients detected by SN searches often exceeds the resources available for spectroscopic follow-up.  For example, because SNe without host galaxies clearly detected in discovery images may be mistaken for non-SN optical transients, the spectroscopic follow-up for some SN searches will be biased against SNe with low-luminosity host galaxies \citep{CSS}.  Similarly, some untargeted SN searches such as ESSENCE and SDSS-II have focused on discovering SNe~Ia at the exclusion of core-collapse SNe \citep{Sako08,Foley09}.  Because the host galaxy properties of SN~Ia differ systematically from those of SN~Ibc \citep[see e.g.][]{Kelly08,Mannucci08}, this selection effect could influence resulting studies of SN~Ibc host environments.  Moreover, some optical surveys are less sensitive to SN detection in the central regions of galaxies.  However, because the photometric and host galaxy properties of SN~Ib and Ic are significantly more similar to other subclasses of SNe, the systematic effects of biases in SN search spectroscopic follow-up is likely to effect SN~Ib and Ic in similar ways.  Therefore we expect it to be a second order effect influencing the results of this study, likely much smaller than the previous effects discussed.  

In a magnitude limited survey, the ratio of SN~Ic to Ib has been estimated to be $\sim1.6$ \citep{Li11}.  In our study, this ratio is similar, \RatioIcIb{}, suggesting no significant bias for or against SN~Ib or Ic.  For comparison, this ratio is 1.4 for \cite{Anderson10}, 2.1 for \cite{Kelly11}, 0.5 for \cite{Leloudas11}, and 1.25 for \cite{Modjaz11}.  

SN~Ic-BL, comprising $\percentIcBL\%$ of our sample, are over-represented in our sample with respect to galaxy-targeted, volume-limited surveys (e.g. $<5\%$ of SN~Ibc in LOSS, not including SN~IIb, \citealt{Li11}).  This is a natural consequence of two factors.  First, SN~Ic-BL typically have brighter peak luminosities than other SNe~Ibc ($\sim1$~mag in $R$-band, \citealt{Drout11}) and can therefore be discovered over a larger volume in a magnitude-limited survey.  Indeed, the median redshift of SNe~Ic-BL in our sample ($ z=\medredIcBL$) is nearly twice that of the other SNe~Ibc in our sample ($z=\medredOther$).  Second, we have found that Ic-BL preferentially occur in low-metallicity galaxies that will have preferentially lower luminosities  \citep[see also][]{Arcavi10,Kelly11}.  These galaxies are therefore under-represented in galaxy-targeted SN searches, but are not excluded from the untargeted SN searches we draw our sample from.  We further note that the percentage of SN~IIb in our sample ($\percentIIb\%$) is similar to the value found by \cite{Smartt09b} among those SN classified and reported in the IAU circulars: $\sim16\%$ of SN~Ibc.  However, \cite{Li11} report an SN~IIb rate $>2$ times this value based on SNe found by LOSS.  \cite{Li11} that the LOSS classifications are based primarily on photometry (in particular, identification of the double-peaked SN~IIb lightcurve shape), while classifications reported in the circulars are primarily based on single-epoch spectroscopic observations.  Finally, we note that any effect due to the cosmic star formation history should be negligible over the modest redshift range ($z\lesssim0.3$) of our sample \citep{Grieco12}.

An additional systematic effect could act if the host galaxy properties of SN~Ibc are found to correlate with the explosion peak magnitude, as has been found for SN~Ia \citep{Hamuy96}.  For example, if it were the case that SN~Ic-BL in low-mass/metallicity galaxies are brighter than their counterparts in brighter galaxies, they would be over-represented in samples of host galaxy spectroscopy.  This potential effect could be evaluated more thoroughly with more extensive studies of SN~Ibc lightcurve properties, as the sample size of existing studies are small \citep{Richardson06,Drout11,Li11}, or by performing a volume-limited survey.  Moreover, dust obscuration could prevent the discovery of a significant fraction of SNe  \citep{Mattila12}, and could potentially be correlated with both host galaxy and explosion properties.

\subsection{Depth limits for host galaxy spectroscopy}
\label{sec:deptheffect}

Due to the galaxy $L-Z$ relation, the lowest metallicity galaxies will also have the lowest luminosities, and may therefore be under-represented in spectroscopic studies that require sufficient S/N in the nebular emission lines to derive metallicity.  Table~\ref{tab:noZ} indicates that we are potentially excluding 3 SN~Ic in low-metallicity hosts from our sample due to insufficient S/N, while we are likely only excluding one such SN~Ib.  The effect of these exclusions is significant: if we suppose that each of the excluded galaxies falls at the low-metallicity end of the observed distribution, than the median metallicities we measure for the SN~[Ib,Ic] in our sample become log(O/H)+12$=[\noexcIbmed,\noexcIcmed]$.  Considering these excluded galaxies indicates that the median metallicity for SN~Ic host environments could be \textit{lower} than for SN~Ib in our sample.  Such a revision would not conflict with the results of our statistical tests, which indicate that the difference we measure in the metallicity distributions of SN~Ib and Ic host environments is not significant (Section~\ref{sec:stat}).  This illustrates the role of small number statistics in spectroscopic studies of SN~Ibc host environments.  Moreover, it advocates for continued follow up of supernova host environments with facilities capable of measuring metallicity for the low-luminosity, relatively high-redshift host galaxies discovered by untargeted SN searches.

\section{DISCUSSION}
\label{sec:disc}

\subsection{SN~Ib and Ic progenitor models}
\label{sec:discIbIc}

The new observations presented in this work, and previous observations synthesized in our combined dataset, suggest and constrain differences in the progenitor star population of SNe~Ibc.  The modest difference observed in the median metallicity of SN~Ib and Ic host environments corresponds to a very small difference in mass-loss rates for single star progenitors.   In general, the mass loss rates of massive stars is taken to scale as $\dot{M}\sim(Z/Z_\odot)^{m}$ where $m=0.5$ \citep{Kudritzki87} or perhaps $m=0.86$ for WR winds \citep{Vink05}.  The difference in median metallicity we measure for SN~Ib and Ic from explosion site spectroscopy of Gold-classification SNe in our untargeted study (Section~\ref{sec:metaldist}) then corresponds to a difference in mass loss rates of a factor of $\dot{M}_{Ic}/\dot{M}_{Ib}\sim\MdotdiffIbIcEG~(\MdotdiffIbIcEGVink)$ for the power law slope $m=0.5~(0.86)$.  If we use instead the median difference inferred from the combined sample using only explosion-site spectroscopy ( Section~\ref{sec:combdist}), the mass loss rate differs by a factor of $\dot{M}_{Ic}/\dot{M}_{Ib}\sim\MdotdiffIbIccombEU~(\MdotdiffIbIccombEUVink)$.  In this simplistic analysis, it seems improbable that this small difference in mass loss rate is sufficient to strip the entire He layer from the progenitor star in order to produce the envelope composition indicated by the spectrum of the explosion.

In more detail, the difference in the SN~Ib and Ic metallicity distributions can be interpreted in terms of a metallicity-dependence for the critical initial-mass required for a progenitor star to explode as SN~Ib or Ic.   \cite{BP09} estimated how this critical mass depends on metallicity by comparing the observed difference in the rates of SN~Ib and Ic in different metallicity bins to a model that has an explicit dependence of the threshold mass for SN~Ib and Ic explosions.  Using photometric metallicity estimates, they find that the critical mass varies by a factor of $\sim2$ over a factor of $\sim3$ in metallicity.  However, our combined dataset indicates that a difference in the metallicity distribution of SN~Ib and Ic has yet to be measured robustly (Section~\ref{sec:combIbIc}), suggesting that the metallicity-dependence of this critical mass may be much more subtle.  

Moreover, the observed relation between galaxy mass, metallicity, and star formation rate implies that metal-poor galaxies typically have higher specific star formation rates \citep{LaraLopez10,Mannucci10}.  In a single star progenitor model where SN~Ic are produced by more massive stars, this could indicate an elevated rate of SN~Ic relative to SN~Ib in metal-poor galaxies if the star formation events are short-lived.  This could potentially mask the effect of metal-line dependent winds.  Our measurements of the ages of young stellar populations of SN~Ibc host galaxies do not indicate that SN~Ic come from younger stellar populations (Section~\ref{sec:Yage}), but a larger sample is needed to address this question in detail.  A metallicity-dependent slope for the initial mass function could also effect SN~Ibc rates \citep{BP03}.  Assuming SN~Ic progenitors are more massive than those of SN~Ib, a top-heavy IMF in low-metallicity environments would elevate the SN~Ic/Ib rate and could push the metallicity distribution of SN~Ic downward.

\cite{Smartt09b} suggest that luminosity limits from pre-explosion imaging and ejecta masses from light-curve modeling rule out massive WR progenitors for a sample of nearby SN~Ibc.  However, the SN~Ibc studied by \cite{Smartt09b} are primarily in nearby, high-mass host galaxies that are likely to reflect high-metallicity progenitor stars.  In essentially all progenitor models, SN~Ibc progenitors are expected to be less massive (and less luminous) at higher metallicities.  Indeed, the rate of SN~Ibc in nearby ($z\lesssim0.04$), low-metallicity environments is expected to be quite low (at best a few per year), making direct progenitor detection observationally challenging \citep{Young08}.  Moreover, local extinction may play a larger role in obscuring SN progenitor stars than previously recognized \citep{Walmswell12}.

\cite{Podsiadlowski92} suggested that Roche lobe overflow via binary interaction could be responsible for stripping significant amounts of material from the progenitor stars of SN~Ibc.  Observations of OB stars indicate that a significant percentage of potential Type~Ibc SN progenitors are likely to be in interacting binary systems \citep{Kobulnicky07,Kouwenhoven07,Sana12}.  \cite{Eldridge08} have shown that the observed metallicity-dependence of the relative rates of SN~II and Ibc can be reproduced using binary population synthesis models informed by the observed populations of red supergiants, Wolf-Rayet, and other massive stars.  \cite{Smith11} have argued based on SN rates that the majority of SN~Ibc progenitors may come from binary star systems.  However, additional modeling is needed to predict the relative metallicity-distribution of SN~Ib and Ic that would result from binary progenitor stars.  Moreover, the effect of mixing may further complicate the comparison of observations to SN~Ibc progenitor models for both single and binary stars \citep{Dessart12}.

\subsection{Comparison to nearby GRB-SNe}
\label{sec:combGRB}

In the ``collapsar'' model, the progenitors of LGRBs found in association with SN\,Ic-BL (GRB-SNe) are massive stars with high rates of core rotation, implying sub-solar metallicities ($Z\lesssim0.3~Z_\odot$) in order to minimize angular momentum losses due to line-driven winds \citep{woosley06}.  The observational result that most LGRBs and GRB-SNe are discovered in dwarf, sub-solar metallicity galaxies has been interpreted as evidence supporting this model \citep{Fruchter06,Stanek06,Levesque10}.  Because SN\,Ic-BL have traditionally been found in higher-metallicity environments, a ``cut-off'' metallicity has been proposed to distinguish stars which will produce GRB-SNe from those that will produce non-relativistic SN\,Ic-BL \citep{Modjaz08,Kocevski09}.

However, several discoveries have challenged the role of metallicity in the production of LGRBs.  LGRBs and relativistic, engine-driven SNe have been found in super-solar metallicity environments \citep{Berger07,Levesque102,Soderberg10,lkg+10,Graham09}.  Moreover, SN\,Ic-BL with strong limits on the association of relativistic ejecta have been found in sub-solar metallicity environments \citep[see e.g.][]{nes2010ay}.  This shift in the discovery environments of GRBs and SN\,Ic-BL can be explained in terms of the transient search strategy --- GRBs are discovered via their gamma-ray emission through untargeted searches, while past studies of SN\,Ibc host environments drew primarily from galaxy-targeted searches, which are heavily biased towards higher-metallicity host environments (see Section~\ref{sec:targeffect}).  The advent of wide-field, untargeted SN searches has enabled the discovery of SN\,Ic-BL in sub-solar metallicity environments.  Indeed, the results from this work and \cite{Kelly11} demonstrate that SN\,Ic-BL preferentially occur in lower-metallicity environments than SN\,Ib or Ic (see Section~\ref{sec:metaldist} and Section~\ref{sec:combIcBL}).

Alternatively, the observed metallicity distribution of LGRB host environments has been interpreted as a secondary manifestation of a preference for high-star formation rate environments \citep{Kocevski11,Mannucci11}.  Our results are consistent with this view.  In Section~\ref{sec:Yage} we found that SN\,Ic-BL host environments potentially have younger stellar populations than those of other SN\,Ibc and more similar to nearby LGRBs.  In Section~\ref{sec:metaldist} we find that the untargeted SN\,Ic-BL in our sample have a median metallicity $\sim0.2$~dex higher than that of the nearby LGRBs ($z<0.3$), but given the small size of both samples ($N=\NlgrbEL$ and 7, respectively), the KS test indicates no significant difference between the distributions ($\ksp=\KSlgrbIcBL$).

In summary, observations of SNe discovered by untargeted searches have substantially reduced the discrepancy reported between the metallicity distribution of SN\,Ic-BL with and without associated LGRBs, while verifying a shared preference for environments with high-star formation rates and/or very young stellar populations.  These findings are consistent with the view that massive stars ($M\gtrsim40~M_\odot$) are the common progenitor for both these types of explosions, but suggest that metallicity does not play the primary role in the formation of a central engine.  Similarly, \cite{Levesque10NoC} have found that host environment metallicity does not correlate to the gamma-ray energy release of LGRBs.  \cite{Georgy12} have suggested, using the models of \cite{Ekstrom12}, that differential rotation in LGRB progenitors could moderate the coupling between WR winds and the stellar core and may be responsible for reducing the role of metallicity in the explosion.

\subsection{Studies of SN~I\lowercase{bc} in the LSST era}
\label{sec:discLSST}

In the coming years, the discovery rate for SNe promises to grow dramatically as existing high-cadence, wide field surveys continue to operate and new surveys such as the Dark Energy Survey and the Large Synoptic Survey Telescope (LSST) come online.  As LSST could discover $\sim10^6$ SNe per year, it is anticipated that the SN discovery rate in this era will far outstrip the capacity to perform spectroscopic follow-up on individual objects \citep{LSSTbook}.  This significant increase in the SN discovery rate could permit studies of untargeted SN~Ibc host environments with sample sizes sufficient to detect even small ($\lesssim0.1$~dex) differences in the SN~Ib and Ic metallicity distributions without statistical ambiguity.  We suggest that a sample of $\gtrsim100$ such objects would be required for that purpose (see Section~\ref{sec:combIbIc}), but caution that attention must continue to be paid to the systematic effects discussed in Section~\ref{sec:SE}.

Moreover, such an SN discovery rate suggests that SN classification will become the limiting factor for certain science goals.  Investigating the GRB-SN connection will be more difficult if a sample of well-classified SN~Ic-BL cannot be identified.  In SN~Ia samples for precision cosmology that are assembled using light curve-based classification alone, SN~Ibc may be major contaminants ($\sim20\%$) because their light curve shape resembles that of SN~Ia \citep{Gong10,Kessler10,Gjergo12}.  The potential for ambiguity between SN~Ia and SN~Ibc light curves is especially great in the case of SN~Ic-BL, because the average absolute peak magnitude is more similar to SN~Ia than for normal SN~Ib or Ic \citep{Drout11}.

We therefore suggest that information about the distribution of host galaxy properties for SN~Ia and Ibc could be used to inform photometric SN classification in the LSST era.  Because SN~Ic-BL preferentially occur in lower luminosity, lower metallicity host galaxies than other SN~Ibc or SN~Ia \citep{Li11,Kelly11}, the difference in flux between the SN and the host galaxy could be used to distinguish SN~Ic-BL from other SN types for the dual purposes of 1) identifying SN~Ic-BL for follow-up SN studies and 2) filtering SN~Ibc from samples of SN~Ia for precision cosmology.  However, a study using information about the host galaxy distribution to inform SN classification would sacrifice the ability to make unbiased inferences about the properties of the host galaxy populations (similar to the case of galaxy-targeted SN surveys) and could introduce systematic effects if used to construct samples for SN~Ia cosmology.  Moreover, it would also exclude events with unexpected properties, such as the engine-driven Type~Ic-BL SN\,2009bb that exploded in a luminous, high-metallicity host galaxy \citep{lsf+10}.

\section{CONCLUSIONS}
\label{sec:conc}

We have presented the largest study to date of SN~Ibc host environments unbiased with respect to the galaxy $L-Z$ relation as imposed by galaxy-targeted SN searches.  By reporting metallicity measurements for \diagPPzerofourNtwoN{} objects, we more than double the number of host environment metallicity measurements for SN~Ibc discovered by untargeted surveys.

We conclude that:

\begin{enumerate}
\item SN~Ibc host environments discovered through targeted surveys are significantly biased towards higher metallicities, representing host environments that are typically more enriched by $\sim$\ZdifallNTPPzerofourNtwo\%.  The ratio of low-metallicity ($Z_\odot/3$) host environments probed by untargeted versus targeted SN searches is $N_U/N_T\approx\numTthirdsol$, and galaxy-targeted SN searches offer a smaller baseline for probing metallicity distribution differences.

\item In our own sample and combining observations from all spectroscopic studies of SN~Ibc host environments to date, we find no statistical difference between the metallicity distributions of SN~Ib and Ic or between SN~Ib and IIb.  We place a limit on the median metallicity difference between SN~Ib and Ic ($\lesssim0.1$~dex) and find that a sample $\gtrsim2\times$ as large would be required to unambiguously confirm a difference at that level.  This limit corresponds to a very small difference in the mass loss rate of metal line-driven winds, suggesting that it may not be the dominant factor distinguishing SN~Ib and Ic progenitors.  
\item SN~Ic-BL are found in host environments with substantially lower metallicity than SN~Ic, confirming the result of \cite{Kelly11}.  We show that the median metallicity of SN~Ic-BL found by untargeted SN searches is $\sim0.15$~dex lower than those found by targeted SN searches, yielding closer agreement to LGRB host environments.  Moreover, the young stellar populations of SN~Ic-BL appear to be lower than SN~Ib and Ic, but the sample is not large enough to be significant.
\item Evaluating the systematic effects afflicting studies of SN~Ibc host environment metallicities, we find that the bias introduced by galaxy-targeted SN searches is most significant.  Galaxy-nucleus spectroscopy can serve as a good proxy for explosion site metallicity measurements, but studies using photometric metallicity estimates would require sample sizes $\sim2$ times larger.  Biases in spectroscopic follow-up of SNe discovered by optical transient searches and uncertainties in SN classification play smaller roles.
\end{enumerate}

In the era of wide-field, untargeted SN searches, we anticipate that significant progress will be made towards unveiling the progenitors of SN~Ibc through the study of their host environments.  We advocate for continued spectroscopic studies of the host environments of SN~Ibc discovered by untargeted surveys to uncover or place stricter limits on the difference in the metallicity distribution of SN~Ib and Ic.  Optical facilities with large light-gathering power are required to measure metallicities in the low-luminosity, relatively high-redshift galaxies hosting SNe discovered by untargeted surveys.  To identify SN~Ic-BL in this era and make progress on the GRB-SN connection in the face of limited resources for spectroscopic follow up, we suggest that host galaxy properties will become a significant aid to future photometric SN classification schemes.

\acknowledgements
\label{sec:ackn}

We thank D. Eisenstein, W. Fong, A. Gal-Yam, P. Kelly, B. Kirshner, M. Modjaz, R. Narayan, and P. Podsiadlowski for helpful conversations.  We are grateful to I. Arcavi, S. Blondin, A. Drake, A. Gal-Yam, G. Leloudas, M. Stritzinger, R. Thomas, and S. Valenti for their help in refining supernova spectral classifications.  We thank the staffs of the MMT, Las Campanas, and Gemini observatories for their excellent support.  This work was supported by the National Science Foundation through Graduate Research Fellowships provided to I.C., M.R.D., and N.E.S; E.M.L. is supported by NASA through Einstein Postdoctoral Fellowship grant number PF0-110075 awarded by the Chandra X-ray Center, which is operated by the Smithsonian Astrophysical Observatory for NASA under contract NAS8-03060; and support for this work was provided by the David and Lucile Packard Foundation Fellowship for Science and Engineering awarded to A.M.S.

Based on observations obtained at the Gemini Observatory, which is operated by the Association of Universities for Research in Astronomy (AURA) under a cooperative agreement with the NSF on behalf of the Gemini partnership: the National Science Foundation (United States), the Science and Technology Facilities Council (United Kingdom), the National Research Council (Canada), CONICYT (Chile), the Australian Research Council (Australia), CNPq (Brazil) and CONICET (Argentina).

{\it Facilities:} \facility{Gemini:North} (GMOS-N), \facility{Magellan:Baade} (IMACS), \facility{Magellan:Clay} (LDSS3), \facility{MMT} (BlueChannel).

\bibliographystyle{fapj.bst}

\end{document}